# Observation of fractional edge excitations in nanographene spin chains

Shantanu Mishra[1,9#], Gonçalo Catarina[2,3#], Fupeng Wu[4], Ricardo Ortiz[3], David Jacob[5,6], Kristjan Eimre[1], Ji Ma[4], Carlo A. Pignedoli[1], Xinliang Feng[4,7*], Pascal Ruffieux[1*], Joaquín Fernández-Rossier[2*] and Roman Fasel[1,8]

[1]Empa – Swiss Federal Laboratories for Materials Science and Technology, Dübendorf, Switzerland
[2]International Iberian Nanotechnology Laboratory, Braga, Portugal
[3]University of Alicante, Sant Vicent del Raspeig, Spain
[4]Technical University of Dresden, Dresden, Germany
[5]University of the Basque Country, San Sebastián, Spain
[6]IKERBASQUE, Basque Foundation for Science, Bilbao, Spain
[7]Max Planck Institute of Microstructure Physics, Halle, Germany
[8]University of Bern, Bern, Switzerland
[9]*Present address:* IBM Research – Zurich, Rüschlikon, Switzerland

#These authors contributed equally to this work

*Correspondence to: xinliang.feng@tu-dresden.de (X.F.), pascal.ruffieux@empa.ch (P.R.) and joaquin.fernandez-rossier@inl.int (J.F.R.)

**Fractionalization is a phenomenon in which strong interactions in a quantum system drive the emergence of excitations with quantum numbers that are absent in the building blocks. Outstanding examples are excitations with charge $e/3$ in the fractional quantum Hall effect[1,2], solitons in one-dimensional conducting polymers[3,4] and Majorana states in topological superconductors[5]. Fractionalization is also predicted to manifest itself in low-dimensional quantum magnets, such as one-dimensional antiferromagnetic $S = 1$ chains. The fundamental features of this system are gapped excitations in the bulk[6] and, remarkably, $S = 1/2$ edge states at the chain termini[7–9], leading to a four-fold degenerate ground state that reflects the underlying symmetry-protected topological order[10]. Here, we use on-surface synthesis[11] to fabricate one-dimensional spin chains that contain the $S = 1$ polycyclic aromatic hydrocarbon triangulene as the building block. Using scanning tunneling microscopy and spectroscopy at 4.5 K, we probe length-dependent magnetic excitations at the atomic scale in both open-ended and cyclic spin chains, and directly observe gapped spin excitations and fractional edge states therein. Exact diagonalization calculations provide conclusive evidence that the spin chains are described by the $S = 1$ bilinear-biquadratic Hamiltonian in the Haldane symmetry-protected topological phase. Our results open a bottom-up approach to study strongly correlated quantum spin liquid[12] phases in purely organic materials, with the potential for the realization of measurement-based quantum computation[13].**

Quantum spin liquids are states of matter in which quantum fluctuations are sufficiently strong to quench long-range magnetic order, enabling the emergence of exotic phenomena such as topological order and fractionalization. The notion that spin chains with an antiferromagnetic Heisenberg exchange lack a classical magnetic order, and have a gapless excitation spectrum with a continuum of excited states above the ground state, goes back to the early theoretical work of Bethe performed almost a century ago for $S = 1/2$ chains[14] (where $S$ denotes the total spin quantum number of the elementary building block). In contrast to half-integer spin chains, Haldane predicted that integer spin chains with periodic boundary conditions should have a gapped excitation spectrum between a singlet ground state and the first excited state[6], known as the Haldane gap. It was later found that open-ended $S = 1$ chains additionally host fractional $S = 1/2$ edge states at the chain termini[7–9]. These edge states are coupled via an interedge effective exchange that gives rise to a singlet-triplet splitting, which decays exponentially with increasing chain length and results in a four-fold degeneracy of the ground state in the thermodynamic limit. The situation where the ground state degeneracy depends upon the open-ended or closed (cyclic) nature of the chains is a hallmark of topological order. In the case of $S = 1$ chains, topological order is associated to symmetries such as SO(3), time reversal and link inversion, and is known as symmetry-protected topological order[10].

In the past three decades, a plethora of experimental work has explored the existence of the Haldane gap and fractional edge excitations in materials containing quasi-one-dimensional (1D) $S = 1$ chains of transition metal ions[15], employing ensemble probes such as neutron scattering, electron spin resonance and thermodynamic property measurements. However, magnetic anisotropy of transition metal ions and a finite interchain magnetic exchange, inherently present in these materials, are detrimental for the emergence of the Haldane phase. An alternative approach to achieve physical realization of spin chains relies on the ability to image and manipulate individual atoms or molecules on solid surfaces by the scanning tunneling microscope (STM). Combined with the ability of STM to measure local electronic structure[16] and magnetic excitations[17] at the atomic scale, recent years have witnessed on-demand fabrication of atomic spin chains and demonstration of complex magnetic interactions and topological phenomena therein[18], including the realization of quantum $S = 1/2$ models[19,20]. However, the Haldane phase has so far not been realized using this approach, despite predictions to such effect[21].

Here, we use on-surface synthesis under ultra-high vacuum conditions to fabricate 1D spin chains on a Au(111) surface, where the elementary building block is triangulene − a diradical polycyclic aromatic hydrocarbon (hereafter, nanographene) with $S = 1$ ground state (Fig. 1a). Magnetism in triangulene arises due to an inherent sublattice imbalance in its bipartite honeycomb lattice, which translates to a net spin imbalance[22,23]. Triangulene and its homologues, although challenging to synthesize by solution chemical routes[24–26], have recently been synthesized on a range of metal and insulator surfaces[27–30], and are shown to retain their magnetic ground states on the relatively inert Au(111) surface. We have previously shown that triangulene dimers, which consist of two triangulene units connected by a single carbon-carbon bond through their minority sublattice atoms, exhibit a large intertriangulene antiferromagnetic exchange of 14 meV[31]. Furthermore, magnetic anisotropy in such carbon-based nanostructures is expected to be extremely weak[32]. Therefore, we expect triangulene spin chains (TSCs) to provide an ideal platform to explore the spin physics of $S = 1$ chains.

The fabrication of TSCs relies on the solution synthesis of dimethylphenyl-substituted anthracene precursors **1** and **2** (Fig. 1b, see Supplementary Information for solution synthesis and characterization data), which undergo surface-catalyzed Ullmann-like polymerization and subsequent oxidative cyclization upon thermal annealing on Au(111), thereby yielding the TSCs. We note that the use of only the dibrominated precursor **2** results in the growth of long TSCs with maximum length in excess of 100 nm (Supplementary Fig. 1). Therefore, we use a mixture of **2** and the monobrominated precursor **1** to limit the chain growth, resulting in short open-ended TSCs (oTSCs) with varying lengths, as shown in the overview STM image in Fig. 1c, which allows us to investigate the length-dependent magnetic structure of TSCs. As shown in the bond-resolved STM images in Fig. 1d,e, TSCs with both *cis* and *trans* intertriangulene bonding configurations are found, with long chains mostly containing a mixed *cis/trans* structure. Scanning tunneling spectroscopy (STS) measurements on TSCs over a wide bias range reveal an electronic band gap of 1.60 eV irrespective of the *cis/trans* structure (Extended Data Fig. 1 and Supplementary Fig. 2). Our STS results are in agreement with spin-polarized density functional theory (DFT) calculations, which show an antiferromagnetic exchange between nearest-neighbor triangulene units, and nearly dispersionless frontier bands indicative of a weak intertriangulene electronic hybridization (Extended Data Fig. 2). We also performed many-body perturbation theory *GW* calculations on TSCs (where $G$ and $W$ denote Green's function and screened Coulomb potential, respectively), including screening effects from the underlying surface, from which we obtain a theoretical electronic band gap of 1.43 eV that is consistent with the experimental band gap.

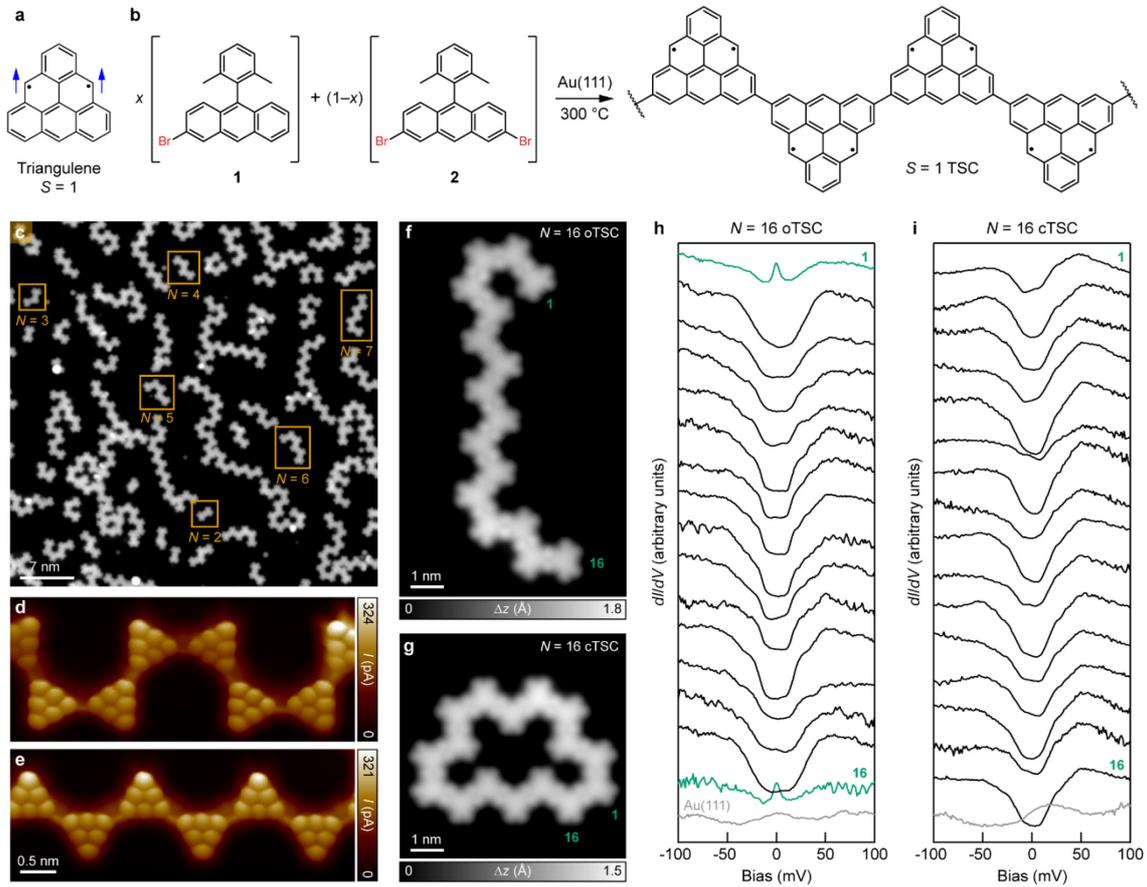

**Fig. 1 | On-surface synthesis of triangulene spin chains and observation of zero-energy edge excitations. a**, Chemical structure of triangulene. **b**, On-surface synthesis of TSCs using precursor mixture **1**+**2**. **c**, Overview STM image after annealing the precursor mixture ($x = 0.2$) on Au(111) at 300 °C ($V = -0.7$ V, $I = 70$ pA). The image is acquired with a carbon monoxide (CO) functionalized tip. oTSCs with $N = 2–7$ are highlighted. **d,e**, Bond-resolved STM images of TSCs with *cis* (**d**) and *trans* (**e**) intertriangulene bonding configurations (open feedback parameters: $V = -5$ mV, $I = 50$ pA; $\Delta h = -0.7$ Å). $\Delta h$ denotes the offset applied to the tip-sample distance with respect to the STM setpoint above the TSCs. **f,g**, High-resolution STM images of $N = 16$ oTSC (**f**) and cTSC (**g**). Scanning parameters: $V = -0.6$ V, $I = 200$ pA (**f**) and $V = -0.7$ V, $I = 500$

pA (**g**). **h,i**, $dI/dV$ spectra acquired on every unit of the $N = 16$ oTSC (**h**) and cTSC (**i**), revealing zero-energy excitations exclusively at the terminal units of the oTSC (green curves). Numerals near the curves indicate the unit number, marked in the high-resolution STM images, on which the corresponding spectrum was acquired. The $dI/dV$ spectra in the panels are offset vertically for visual clarity. Open feedback parameters for the $dI/dV$ spectra: $V = -100$ mV, $I = 1.4$ nA; root mean squared modulation voltage $V_{rms} = 1$ mV.

Figure 1f,g presents high-resolution STM images of $N = 16$ oTSC (Fig. 1f) and cyclic TSC (cTSC, Fig. 1g) (where $N$ denotes the number of triangulene units in a TSC). $dI/dV$ spectroscopy (where $I$ and $V$ correspond to the tunneling current and bias voltage, respectively) performed on these TSCs in the low-bias regime ($|V| \leq 100$ mV; Fig. 1h,i) reveals two salient features. First, terminal units in the oTSC show peaks at zero bias (Fig. 1h), which exhibit an anomalous linewidth broadening with increasing temperature that is characteristic of a Kondo resonance[33] (Supplementary Figs. 3–5). These Kondo resonances are absent both in the non-terminal units of the oTSC and throughout the cTSC (Fig. 1i) and, as is shown later, they are indicative of the emergence of $S = 1/2$ edge states. Second, several conductance steps symmetric with respect to zero bias and with energies below 50 meV are found throughout the oTSC and cTSC, corresponding to inelastic excitations. We ascribe these inelastic spectral features to spin excitations[34–36] in the TSCs, as has been previously observed in spin chains of magnetic adatoms on surfaces[17]. The spin excitation energies, which reflect the energy difference between the magnetic ground state and the excited states, show a marked dependence on both $N$ and the open-ended/cyclic topology of the TSCs. In addition, the spin excitation amplitudes exhibit a unit-to-unit modulation across a TSC that is linked to the spin spectral weight[21] (see Methods), which is the probability of exciting the final state by means of spin-dependent electron tunneling across a given location.

A natural starting point to account for our experimental observations is the 1D Heisenberg model, $\hat{H}_{Heisenberg} = J \sum_i \vec{S}_i \cdot \vec{S}_{i+1}$ (here, $\vec{S}_i$ denotes the spin-1 operator at site $i$ and $J > 0$ the exchange coupling), where individual triangulene units are described as $S = 1$ spins with a nearest-neighbor antiferromagnetic exchange. However, the Heisenberg model, with $J$ taken to be 14 meV from STS measurements on an $N = 2$ TSC[31], fails to provide a quantitative agreement with the observed spin excitation energies for oTSCs (Extended Data Fig. 3). We therefore conducted extensive Hubbard model calculations using configuration interaction in the complete active space (CAS) approximation, exact diagonalization (ED) and density matrix renormalization group (DMRG), and compared them with model spin Hamiltonians solved by ED (see Methods). The results of these calculations and their comparison with both the energies and the modulation of the spin excitation steps (Extended Data Fig. 4) show that the TSCs are well described by the spin-1 Hamiltonian

$$\hat{H}_{BLBQ} = J \sum_i \left[ \vec{S}_i \cdot \vec{S}_{i+1} + \beta \left( \vec{S}_i \cdot \vec{S}_{i+1} \right)^2 \right] \tag{1}$$

that includes both bilinear and biquadratic exchange terms, and is referred to as the bilinear-biquadratic (BLBQ) model (here, $\beta$ is a parameter that determines the strength of the biquadratic term relative to the bilinear term). From a comparison of the BLBQ and Hubbard model calculations for an $N = 2$ TSC, we obtain $J = 18$ meV and $\beta = 0.09$, which, hereafter, we adopt for all values of $N$. The emerging physical picture is that cTSCs have a unique $S = 0$ ground state, which is qualitatively similar[8] to the analytical solution obtained for $\beta = 1/3$ – the Affleck-Kennedy-Lieb-Tasaki (AKLT) limit[7] – whose ground state is the valence bond solid given by the concatenation of singlets formed between two $S = 1/2$ virtual spins located at adjacent triangulene units (Fig. 2a). For oTSCs, the valence bond solid picture naturally accounts for the existence of fractional edge states with $S = 1/2$, which can be Kondo screened on a metal surface, and gapped bulk excitations. Since the terminal $S = 1$ units in an oTSC only have a single neighbor, one of their constituent $S = 1/2$ spins is excluded from the valence bond solid, thus generating unpaired spins (Fig. 2a). An effective interedge exchange couples these unpaired spins, leading to a singlet-triplet splitting whose magnitude decays exponentially with increasing $N$. In contrast, complete pairing of spins is achieved in a cTSC, and therefore no edge states are to be expected.

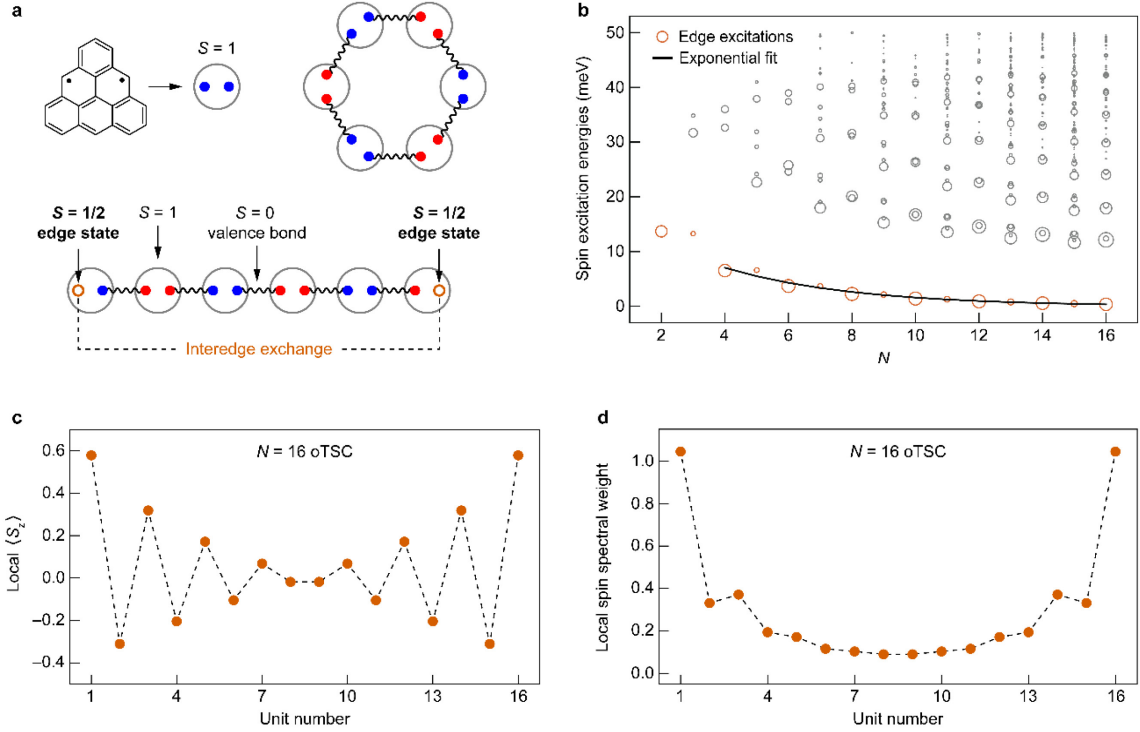

**Fig. 2 | The valence bond solid picture and theoretical calculations of spin excitations in open-ended triangulene spin chains. a**, Representation of triangulene as two virtual $S = 1/2$ spins (smaller filled circles) projected over the $S = 1$ triplet state (larger circle). Also shown is the valence bond solid spin state for $N = 6$ oTSC and cTSC, accounting for $S = 1/2$ edge states in the oTSC and their absence in the cTSC. Wavy lines denote valence bonds, which couple $S = 1/2$ spins from neighboring triangulene units into an $S = 0$ singlet state. Blue and red filled circles denote spin up and spin down electrons, respectively. **b**, Spin excitation energies calculated by ED of the BLBQ model ($J = 18$ meV and $\beta = 0.09$) for oTSCs with $N = 2$–16. Size of the circles represents the spin spectral weight. Orange circles correspond to edge excitations, while gray circles represent all other excitations predicted by the BLBQ model up to 50 meV, which constitute more than 96% of the spin spectral weight for each $N$. The solid line is an exponential fit to the edge excitation energies, $A e^{-N/\zeta}$, with $A = 19$ meV and spin correlation length $\zeta = 4$. **c,d**, Average magnetization (**c**) and spin spectral weight (**d**) of the edge state with $|S, S_z\rangle = |1, +1\rangle$ obtained with the BLBQ model for an $N = 16$ oTSC.

In addition to the low-energy edge excitations for oTSCs, the BLBQ model features multiple spin excitations at higher energies for both oTSCs and cTSCs. Some of them are spin waves spread across the entire TSC, while others are spin waves hybridized with the edge states (Extended Data Fig. 5). In Fig. 2b, we present the BLBQ spin excitation energies of oTSCs with $N = 2$–16, calculated with ED, where the size of the symbols accounts for the spin spectral weight of the corresponding spin excitation, with a larger weight leading to a more prominent step amplitude in $dI/dV$ spectroscopy. Our calculations show that (1) the edge excitation energy exponentially decreases with increasing $N$ and (2) the lowest energy bulk excitation extrapolates toward the Haldane gap with increasing $N$ (Extended Data Fig. 6), in agreement with the experimental results (Supplementary Fig. 6). Figure 2c,d shows the average magnetization (Fig. 2c) and the spin spectral weight (Fig. 2d) of the edge state with the quantum numbers $|S, S_z\rangle = |1, +1\rangle$ for an $N = 16$ oTSC, revealing a strong localization of this state at the terminal triangulene units.

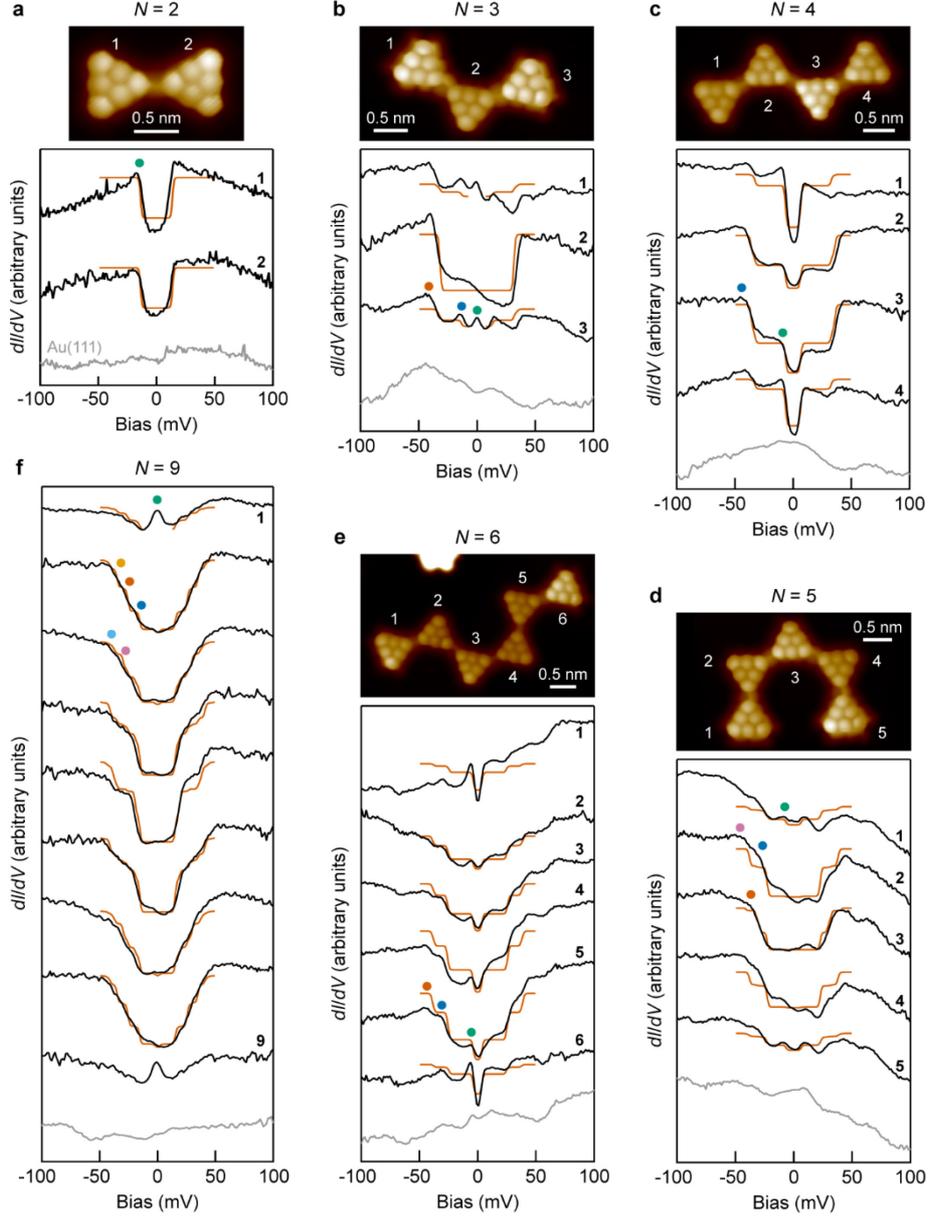

**Fig. 3 | Magnetic excitations in selected open-ended triangulene spin chains and comparison with the bilinear-biquadratic model. a–f**, $dI/dV$ spectroscopy on oTSCs with $N = 2$–6 and 9 (black curves). Representative bond-resolved STM images of oTSCs with $N = 2$–6 are shown (open feedback parameters: $V = -5$ mV, $I = 50$ pA; $\Delta b = -0.6$ or $-0.7$ Å). Also shown are the unit-resolved fits to the $dI/dV$ spectra between ±50 mV, obtained with the BLBQ model (orange curves; $J = 18$ meV, $\beta = 0.09$ and effective temperature $T_{eff} = 5$ K). Since the BLBQ model does not account for the underlying surface, it does not capture the Kondo exchange phenomena. Therefore, for the terminal units of $N = 3$ and 9 oTSCs, no fits are performed near the

Kondo resonances. Colored filled circles indicate the unique spin excitations experimentally observed for each $N$ ($N = 2$: 14 mV; $N = 3$: 0, 11 and 35 mV; $N = 4$: 6 and 37 mV; $N = 5$: 5, 25, 30 and 40 mV; $N = 6$: 3, 27 and 40 mV; $N = 9$: 0, 18, 28, 30, 36 and 40 mV). Open feedback parameters for the $dI/dV$ spectra: $V = -100$ mV, $I = 600$ pA (**a**) and $I = 1.4$ nA (**b–f**); $V_{rms} = 1$ mV.

We performed a systematic experimental study of spin excitations in seventeen oTSCs with $N$ between 2 and 20 (Fig. 3 and Supplementary Figs. 7–18), and eight cTSCs with $N = 5$, 6, 12, 13, 14, 15, 16 and 47 (Fig. 4 and Supplementary Figs. 19–24) that validate our theoretical picture. Figure 3 shows $dI/dV$ spectroscopy performed on oTSCs with $N = 2$–6 (Fig. 3a–e) and 9 (Fig. 3f), which reveals three principal features. First, all TSCs exhibit multiple spin excitations, with the exception of the $N = 2$ TSC, which shows a single spin excitation (singlet-triplet) at 14 meV. It is notable that the BLBQ model accurately accounts for both the energies and amplitude modulation of the spin excitation steps across the triangulene units for these chain lengths. The spin excitation energies calculated by ED of the BLBQ model for TSCs with $N \leq 16$ exhibit a good agreement with the corresponding experimental spin excitation energies (Extended Data Fig. 6 and Supplementary Fig. 6). Deviations between theory and experiments can be partially accounted for by the renormalization of excitation energies due to interactions with the metal surface[37,38]. Second, with the exception of the $N = 3$ TSC, the energy of the lowest energy spin excitation progressively decreases with increasing $N$, as predicted by the BLBQ model (Fig. 2b). Third, TSCs with $N \geq 9$ exhibit Kondo resonances at the terminal units, which are a hallmark of topological degeneracy and fractionalization – Kondo resonances arise at the edges due to screening of the emergent $S = 1/2$ edge states by the underlying metal surface. The Kondo exchange competes with the interedge magnetic exchange, whose magnitude decays exponentially with increasing $N$, but overcomes the Kondo exchange for a small enough $N$ (experimentally, for $N \leq 8$). We note that the zero-bias resonances observed at the terminal units of the $N = 3$ oTSC do not correspond to the emergent $S = 1/2$ edge states. It is observed that the amplitude of the zero-bias resonance for the $N = 3$ oTSC is considerably lower than that of the Kondo resonances for oTSCs with $N \geq 9$. We calculated the spectral function for the $N = 3$ oTSC with a non-perturbative treatment of a multi-orbital Anderson model (MOAM), including coupling to the surface (see Methods and Extended Data Fig. 7). These calculations show that the zero-bias resonance in the $N = 3$ oTSC can be associated to a Kondo resonance of an $S = 1$ ground state, in agreement with previous works[39]. Our calculations also account for the spin excitation steps that are experimentally observed for the $N = 3$ oTSC. Given the large computational cost of such calculations, we presently cannot employ them for TSCs with $N > 3$.

A final confirmation of the validity of the BLBQ model to describe TSCs comes from STS measurements on cTSCs. Figure 4a,b shows high-resolution STM images of $N = 6$ and 13 cTSCs. $dI/dV$ spectroscopy on these cTSCs (Fig. 4c–e) reveals spin excitations that are in agreement with the prediction of the BLBQ model using the same parameters as for the oTSCs. Expectedly, no Kondo resonances are observed in cTSCs given the absence of terminal units. Moreover, the spin excitation spectra for all units of a cTSC are roughly identical, reflecting the equivalence of units in a cyclic structure.

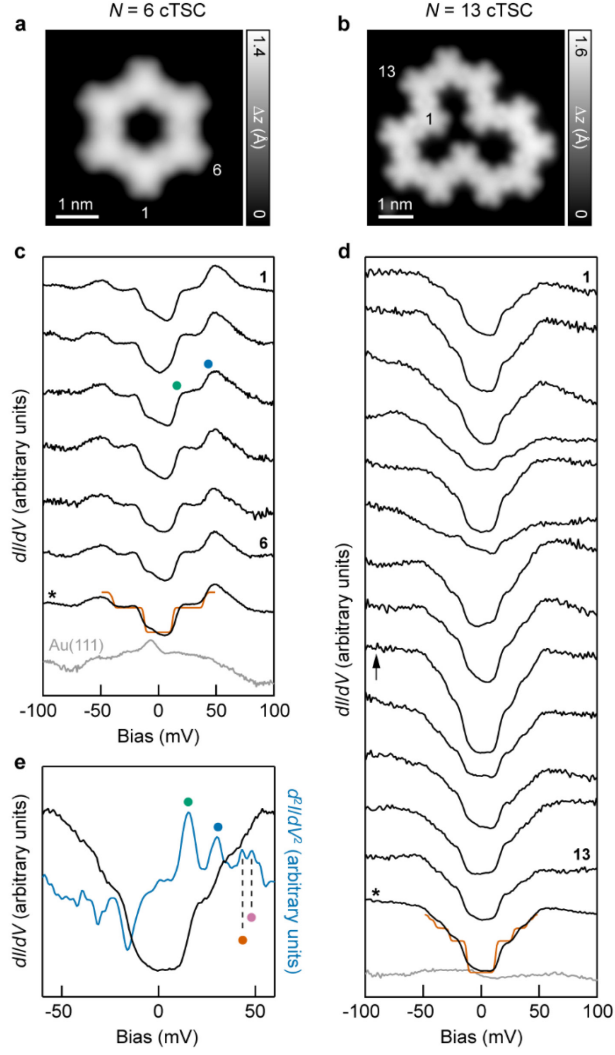

**Fig. 4 | Magnetic excitations in $N = 6$ and 13 cyclic triangulene spin chains and comparison with the bilinear-biquadratic model. a–d,** High-resolution STM images (**a, b**), and $dI/dV$ spectroscopy (black curves) on every unit of $N = 6$ (**c**) and 13 (**d**) cTSCs. The curves marked with an asterisk in **c** and **d** denote the corresponding averaged $dI/dV$ spectrum of all six and thirteen units, respectively. Also shown are the fits to the averaged $dI/dV$ spectra between ±50 mV, obtained with the BLBQ model (orange curves; $J = 18$ meV, $\beta = 0.09$ and $T_{eff} = 5$ K). **e,** High-resolution $dI/dV$ spectrum (black curve) for the curve indicated by an arrow in **d**, and the corresponding $d^2I/dV^2$ spectrum (blue curve) obtained from numerical differentiation. Colored filled circles indicate the unique spin excitations experimentally observed for each chain length ($N = 6$: 15 and 42 mV, and $N = 13$: 15, 30, 43 and 48 mV). Scanning parameters for the STM images: $V = -0.4$ V, $I = 350$ pA (**a**) and $V = -0.7$ V, $I = 210$ pA (**b**). Open feedback parameters for the $dI/dV$ spectra: $V = -100$ mV, $I = 1.3$ nA (**c**) and $I = 1.4$ nA (**d**); $V = -60$ mV, $I = 1.4$ nA (**e**); $V_{rms} = 1$ mV (**c, d**) and 400 µV (**e**).

The ground state of the BLBQ model in the AKLT limit, as well as its generalization in two dimensions, are known to be a universal resource for measurement-based quantum computation[13]. Our results should therefore motivate future work addressing the possibility to tune $\beta$, so that these non-trivial quantum states naturally occur as the ground state of coupled magnetic nanographenes. On a general note, our on-surface synthetic protocol demonstrated here for TSCs can be extended to afford scalable fabrication of purely organic quantum spin chains, lattices and networks – thus opening exciting opportunities in the realization of non-trivial spin liquid phases[12], quantum simulators[40] and nanoscale spintronic devices.

**Acknowledgements.** We thank O. Gröning and J.C. Sancho-García for fruitful discussions. This work was supported by the Swiss National Science Foundation (grant numbers 200020-182015 and IZLCZ2-170184), the NCCR MARVEL funded by the Swiss National Science Foundation (grant number 51NF40-182892), the European Union's Horizon 2020 research and innovation program (grant number 881603, Graphene Flagship Core 3), the Office of Naval Research (N00014-18-1-2708), ERC Consolidator grant (T2DCP, grant number 819698), the German Research Foundation within the Cluster of Excellence Center for Advancing Electronics Dresden (cfaed) and EnhanceNano (grant number 391979941), the Basque Government (Grant No. IT1249-19), the Generalitat Valenciana (Prometeo2017/139), the Spanish Government (Grant PID2019-109539GB-C41), and the Portuguese FCT (grant number SFRH/BD/138806/2018). Computational support from the Swiss Supercomputing Center (CSCS) under project ID s904 is gratefully acknowledged.

**Author contributions**. X.F., P.R. and R.F. conceived the project. F.W. and J.M. synthesized and characterized the precursor molecules. S.M. performed the on-surface synthesis and, STM and STS measurements. G.C., R.O. and J.F.R. performed the tight-binding, CAS, ED and DMRG calculations. D.J. performed the MOAM-NCA calculations. K.E. and C.A.P. performed the DFT and $GW$ calculations. All authors contributed toward writing the manuscript.

**Competing interests.** The authors declare no competing interests.

**Methods**

**Synthesis of molecular precursors.** The synthesis of molecular precursors **1** and **2**, and associated characterization data, are reported in Supplementary Figs. 25–49.

**Sample preparation and STM/STS measurements.** STM measurements were performed with a low-temperature STM from Scienta Omicron operating at a temperature of 4.5 K and base pressure below $5\times10^{-11}$ mbar. Au(111) single-crystal surfaces were prepared through cycles of $Ar^+$ sputtering and subsequent annealing at 723 K. Powder samples of precursors **1** and **2** were contained in quartz crucibles and sublimed from a home-built evaporator at 323 K and 343 K, respectively, onto Au(111) surface held at room temperature. STM images and $dI/dV$ maps were recorded either in constant-current or constant-height modes, while $dI/dV$ spectra were recorded in constant-height mode. For constant-height $dI/dV$ mapping, feedback was opened above the TSC. Bias voltages are provided with respect to the sample. All $dI/dV$ measurements were obtained using a lock-in amplifier (SR830, Stanford Research Systems) operating at a frequency of 860 Hz. Modulation voltages for each measurement are reported as root mean squared amplitude ($V_{rms}$). $d^2I/dV^2$ spectra were obtained by numerical differentiation of the corresponding $dI/dV$ curves, with a binomial smoothing (1–5 iterations) applied to the $dI/dV$ curves. Unless otherwise noted, STM and STS measurements were performed with gold-coated tungsten tips. Bond-resolved STM images were acquired by scanning the TSCs with CO functionalized tips in constant-height mode. CO molecules were deposited onto a cold sample (with a maximum sample temperature of 13 K) containing the reaction products. Analysis of Kondo resonances was performed following the procedure in ref.[35] The data reported in this study were processed with WaveMetrics Igor Pro software.

**DFT and $GW$ calculations.** DFT band structure calculations of TSCs were performed with the Quantum Espresso[41] software package using the PBE exchange-correlation functional.[42] A plane wave basis with an energy cut-off of 400 Ry for the charge density was used together with PAW pseudopotentials (SSSP[43]). Monkhorst k-meshes of $12 \times 1 \times 1$ and $10 \times 1 \times 1$ were used for TSCs with two (*trans* TSC) and four (*cis* TSC) triangulene units in the periodic cell, respectively. The cell and atomic geometries were relaxed until forces were smaller than 0.001 a.u.

The adsorption geometry of an $N = 6$ oTSC on Au(111) was calculated with the CP2K[44] software package using the PBE exchange-correlation functional together with the DFT-D3 van der Waals scheme proposed by Grimme et al.[45] and norm-conserving GTH pseudopotentials.[46] A TZV2P MOLOPT basis set[47] was used for C and H species, and a DZVP MOLOPT basis set for the Au species, together with a cut-off of 600 Ry for the plane wave basis set. Unrestricted Kohn-Sham approach was used together with an anti-ferromagnetic spin guess to the triangulene chain to model the magnetic ground state. The surface/adsorbate system was modeled within the repeated slab scheme, with a simulation cell containing 4 atomic layers of Au along the [111] direction and a layer of hydrogen atoms to suppress one of the two Au(111) surface states. 40 Å of vacuum was included in the simulation cell to decouple the system from its periodic replicas in the direction perpendicular to the surface. The gold surface was modeled by a supercell of $67.80 \times 35.74$ Å$^2$ corresponding to 322 surface units. The adsorption geometry was optimized by keeping the positions of the two bottom layers of the slab fixed to the ideal bulk coordinates, while all the other atoms were relaxed until forces were lower than 0.005 eV/Å.

The eigenvalue self-consistent $GW$ calculations[48] were performed on an $N = 6$ oTSC with the CP2K code on the isolated geometry corresponding to the adsorption conformation. The calculations were performed based on the unrestricted DFT PBE wave functions using the GTH pseudopotentials and analytic continuation with a two-pole model. The aug-DZVP basis set from Wilhelm et al.[48] was used. To account for screening by the Au(111) surface, we applied the image charge model by Neaton et al.[49], and to determine the image plane position with respect to the molecular geometry, we used a distance of 1.42 Å between the image plane and the first surface layer, as reported by Kharche et al.[50]

The calculations were performed using the AiiDAlab platform.[51]

**Derivation of the BLBQ model.** Our starting point to describe the TSCs is a tight-binding model where we only consider $p_z$ orbitals from carbon[23,52], which we refer to as the complete tight-binding model. The resulting single-particle spectrum for a TSC with $N$ triangulenes features $2N$ zero-energy states, each hosting one electron, which arise due to the inherent sublattice imbalance in triangulene. Strict zero-energy states occur within the nearest-neighbor tight-binding approximation, whereas the presence of third-nearest-neighbor hopping leads to hybridization of the zero-energy states. In order to describe the formation of local magnetic moments and their exchange interaction, we include electron-electron interactions in the Hubbard approximation, where only intra-atomic Coulomb repulsion ($U > 0$) is considered. Comparison of the Hubbard model with DFT calculations justifies this approximation[23,52]. Further, we employ the CAS approximation, where we consider a subset of many-body states: the occupation of the set of molecular orbitals that correspond to the $2N$ hybridized zero-energy states is allowed to vary, whereas the occupation of orbitals lower or higher in energy is frozen. The Hubbard model is represented in this restricted space and diagonalized numerically. The CAS approximation for a single triangulene and an $N = 2$ TSC predicts $S = 1$ and $S = 0$ ground states, respectively[31,52]. The CAS approximation allows us to obtain the spin excitation energies as a function of $U$ for oTSCs with $N \leq 4$ ($t_1 = -2.70$ eV, $t_2 = 0$ eV and $t_3 = -0.35$ eV; where $t_1$, $t_2$ and $t_3$ denote the first-, second- and third-nearest-neighbor hopping parameters, respectively[53]). By comparing the calculated spin excitation energies with the corresponding experimental values (Extended Data Fig. 3), we infer $U \cong 2 |t_1|$.

To address oTSCs with $N > 4$, which are beyond our current computational capabilities using the CAS approximation, we instead use a simpler tight-binding toy model that captures the salient features of triangulene, that is (1) $C_3$ symmetry and, (2) a sublattice imbalance of two, resulting in two zero-energy states[52]. We refer to this model as the four-site model. This model has two parameters $t$ and $t'$ that describe intratriangulene and intertriangulene hopping, respectively, along a TSC (Extended Data Fig. 3). We choose $t = -1.11$ eV and $t' = -0.20$ eV, such that the low-energy single-particle spectra of both the complete and the four-site tight-binding models are the same for arbitrary chain lengths. Importantly, comparison of the low-energy many-body spectra of an $N = 3$ oTSC for both the complete and the four-site Hubbard models, as a function of $U$, validates this approach (Extended Data Fig. 3).

We can model oTSCs with $N \leq 6$, described with the four-site model, using DMRG as implemented in the ITensor[54] library. For a fixed $U = 1.45 |t|$, DMRG calculations are in agreement with both the CAS approximation and experiments (Extended Data Fig. 3). Importantly, for oTSCs with $N = 2$–6, the DMRG calculations match not only the experimental spin excitation energies, but also the unit-to-unit modulation of the spin excitation amplitudes (Extended Data Fig. 4).

The excitation energies computed with the Hubbard model, both for the four-site and complete versions, can be compared with those of the BLBQ model to derive the parameters $J$ and $\beta$. Specifically, using the four-site Hubbard model results for the $N = 2$ TSC, we determine $J = 18$ meV and $\beta = 0.09$ (Extended Data Fig. 3). We then extend the comparison of the four-site Hubbard and BLBQ models for oTSCs with $N = 3$–6, while using the aforementioned values of $J$ and $\beta$. We find that the calculated spin excitation energies exhibit an excellent match, with differences smaller than 3 meV. Additionally, we obtain the same pattern of spin degeneracies and identical spin spectral weights (Extended Data Fig. 4) using both models.

Finally, using ED of the BLBQ model, we could extend our calculations for both oTSCs and cTSCs with $N \leq 16$. Comparison with experimental data (Figs. 3 and 4) provides the final evidence that the BLBQ model describes the TSCs.

**Modeling of low-bias experimental $dI/dV$ spectra.** The calculated $dI/dV$ spectra in the main text are obtained using the following expression, which treats coupling to the substrate to the lowest order[55]

$$\frac{dI}{dV}\Big|_n = g_0 \sum_M P_M \sum_{M',a=x,y,z} |\langle M|S_a(n)|M'\rangle|^2 \, \Theta_{MM'}(eV) \tag{2}$$

where $n$ denotes the triangulene unit on which the $dI/dV$ spectrum is recorded, $M$ and $M'$ denote the many-body states of triangulene, $g_0$ is a constant prefactor, $P_M$ denotes the equilibrium occupation of $M$, $\Theta_{MM'}(eV)$ is a thermally broadened step function centered around $eV \pm (E_{M'} - E_M)$ ($e$ is the elementary charge), where $E_{M'} - E_M$ is the excitation energy for a transition from state $M$ to $M'$, and $S_a(n)$ are the $S = 1$ spin operators acting on the $n^{\text{th}}$ triangulene unit. The expression for $dI/dV$ contains the spin spectral weight, defined for the state $M'$ and the $n^{\text{th}}$ triangulene unit as

$$\mathcal{S}_{M'}(n) \equiv \sum_M P_M \sum_{a=x,y,z} |\langle M|S_a(n)|M'\rangle|^2 \tag{3}$$

Equation (2) relates the $dI/dV$ spectra to the many-body wave functions and excitation energies. Specifically, it yields stepwise $dI/dV$ curves, with steps at $eV = \pm(E_{M'} - E_M)$ and relative heights determined by the spin spectral weights. Importantly, for a given pair of states $M$ and $M'$, the height of the inelastic step can change for different triangulene units $n$. Thus, the theory predicts both the energies of the inelastic $dI/dV$ steps and the modulation of their heights across a TSC.

The matrix elements in $\mathcal{S}_{M'}(n)$ are only non-zero for states $M$ and $M'$ whose total spin quantum number $S$ differs by zero or one, reflecting the conservation of the total spin of the system formed by the tunneling electron and triangulene. In addition, equation (2) contains the following sum rule for spin-1 models: for very large $eV$, the unit-resolved $dI/dV$ saturates to $S(S+1) \times g_0 = 2g_0$. We have verified that by considering transition energies up to 50 meV, we have, for all TSCs described by the BLBQ model, more than 92% of the spin spectral weight in each unit, and more than 96% of the total spin spectral weight (that is, the spin spectral weight summed over all units).

In order to compare the experimental $dI/dV$ spectra, which is in arbitrary units, to the theoretical predictions given by equation (2), we make a fit to set the constant of proportionality $g_0$ (we also allow for a vertical shift that has no physical relevance). For cTSCs, where all the triangulene units are equivalent, we average the experimental $dI/dV$ spectra of all the units, and we perform a single fit. In the case of oTSCs, variations of the heights of the spin excitation steps are expected across different units[21,56], so that we perform one fit for each experimental curve, using the expression $m(n) \times dI/dV(n) + b(n)$, where $m$ and $b$ are fitting parameters. This fit assumes that the constant of proportionality may change when the tip is moved laterally to scan across the structure, which can occur due to surface variations or minor vertical tip deviations. It must be noted that these fitting parameters do not change the relative height of the steps in $dI/dV$. Thus, only the spin spectral weight matrix elements control the relative heights in a given unit.

For the $N = 3$ oTSC, we have also calculated the $dI/dV$ spectra, including the coupling to the surface, non-perturbatively for a MOAM formed by the zero-energy states of triangulene. The $dI/dV$ spectra are calculated as the spectral function of the zero-energy states in the non-crossing approximation (NCA), which is capable of modeling Kondo resonances. However, the computational cost of these calculations is too high for $N > 3$.

The starting point for the MOAM-NCA calculations is the complete Hubbard model, with 22 states per triangulene, for an $N = 3$ oTSC, taking into account nearest-neighbor and third-nearest-neighbor hopping ($t_1 = -2.70$ eV, $t_2 = 0$ eV, $t_3 = -0.35$ eV) and $U = 1.90 |t_1|$. With $t_3 = 0$, the single-particle spectrum would have six zero-energy states. $t_3$ partially lifts this degeneracy, leaving two zero-energy states and four low-energy states, all well separated from the other molecular levels. These six single-particle states, which we label with index $k$, form the localized states of the MOAM. We assume the single-particle broadening (hybridization) $\Gamma$ to the bath to be equal for all local levels and energy independent. In addition, the single-particle energy levels can be shifted rigidly with respect to the Fermi energy of the surface by an amount $\delta\varepsilon - V_{dc}$, which controls the occupation of the zero-energy states and the particle-hole symmetry. $U = 1.90 |t_1| = 5.13$ eV and $V_{dc} = 0.47$ eV ensure charge neutrality and particle-hole symmetry for $\delta\varepsilon = 0$. Finite values of $\delta\varepsilon$ allow charge fluctuations and lift particle-hole symmetry (Extended Data Fig. 7).

In order to solve the MOAM, NCA expands the eigenstates of the *isolated* impurity in the coupling ($\Gamma$) to the bath[57]. The first step is thus an exact diagonalization of the impurity Hamiltonian. The eigenstates are simultaneously eigenstates of the total number of electrons $N_e$ and the total spin. At half-filling ($N_e = 6$), the ground state is an $S = 1$ spin triplet, and the first and second excited states are $S = 0$ and $S = 2$, respectively. Coupling to the surface leads to fluctuations of electrons in the impurity, and thus requires the charged sectors with $N_e \pm 1$ electrons. The solution yields the orbital-resolved spectral function $A_k(\omega)$ from which the atom-resolved spectral function $A_{loc}(\omega)$ can be calculated, which is proportional to $dI/dV$[58,59]. More details on the application of NCA to nanoscale quantum magnets can be found in ref.[60]

**Data availability.** Analytical solutions of Heisenberg and BLBQ models, additional STM and STS data, materials and methods in solution synthesis and characterization, solution synthetic procedures, and characterization data of chemical compounds (NMR spectroscopy and high-resolution mass spectrometry) are available in the Supplementary Information.

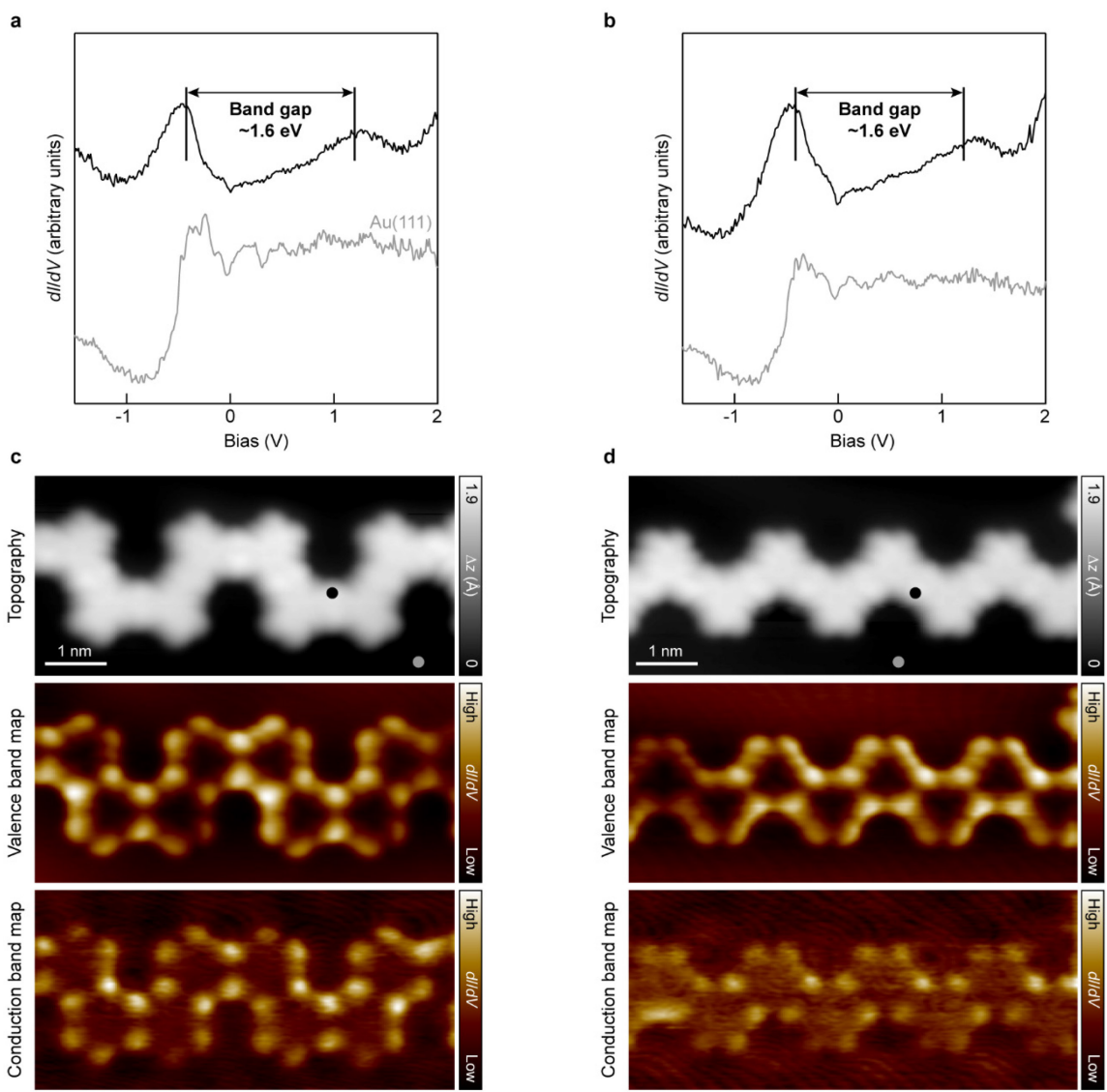

**Extended Data Fig. 1 │ Scanning tunneling spectroscopy measurements of the frontier bands of TSCs. a,b,** $dI/dV$ spectroscopy on TSCs with *cis* (**a**) and *trans* (**b**) intertriangulene bonding configurations (open feedback parameters: $V = -1.5$ V, $I = 250$ pA; $V_{rms} = 16$ mV). Acquisition positions are marked with filled circles in **c** and **d**. Irrespective of the bonding configuration, TSCs exhibit an electronic band gap of 1.6 eV. **c,d,** High-resolution STM images (top panels), and constant-current $dI/dV$ maps of the valence (middle panels) and conduction (bottom panels) bands of *cis* (**c**) and *trans* (**d**) TSCs. Scanning parameters: $V = -0.4$ V, $I = 250$ pA (top and middle panels, **c** and **d**) and $V = 1.1$ V, $I = 280$ pA (bottom panels, **c** and **d**); $V_{rms} = 30$ mV. All measurements were performed with a CO functionalized tip.

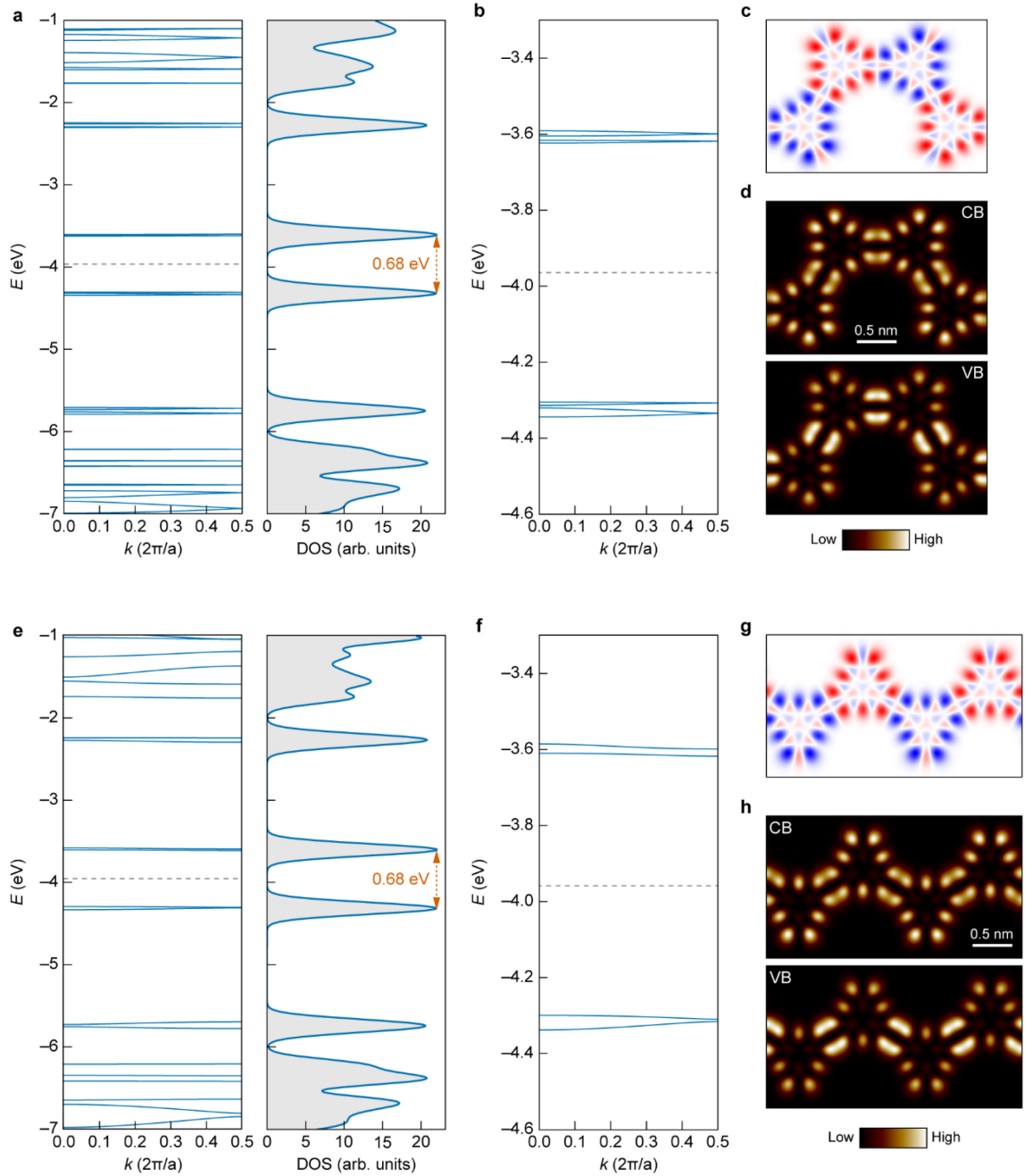

**Extended Data Fig. 2 | Density functional theory calculations on triangulene spin chains. a,e,** DFT band structure and density of states (DOS) plots of TSCs with *cis* (**a**) and *trans* (**e**) intertriangulene bonding configurations in their antiferromagnetic ground state. Energies are given with respect to the vacuum level. A Gaussian broadening of 100 meV has been applied to the DOS plots. Note that spin up and spin down bands are energetically degenerate. **b,f,** Corresponding band structure plots around the frontier bands. The unit cells for the band structure calculations contain four and two triangulene units for *cis* and *trans* TSCs, respectively, with the lattice periodicities $a = 30.0$ Å (*cis* TSC) and 17.4 Å (*trans* TSC). The dashed lines indicate the middle of the band gap. The calculations reveal nearly dispersionless frontier bands due to a weak intertriangulene electronic hybridization. In addition, TSCs exhibit a band gap of 0.68 eV irrespective of the intertriangulene bonding configurations. **c,g,** Ground state spin density distributions for *cis* (**c**) and *trans* (**g**) TSCs. Spin up and spin down densities are denoted in blue and red, respectively. **d,h,** Local DOS maps of the valence (VB) and conduction (CB) bands of *cis* (**d**) and *trans* (**h**) TSCs. Spin density distributions and local DOS maps were calculated at a height of 3 Å above the TSCs.

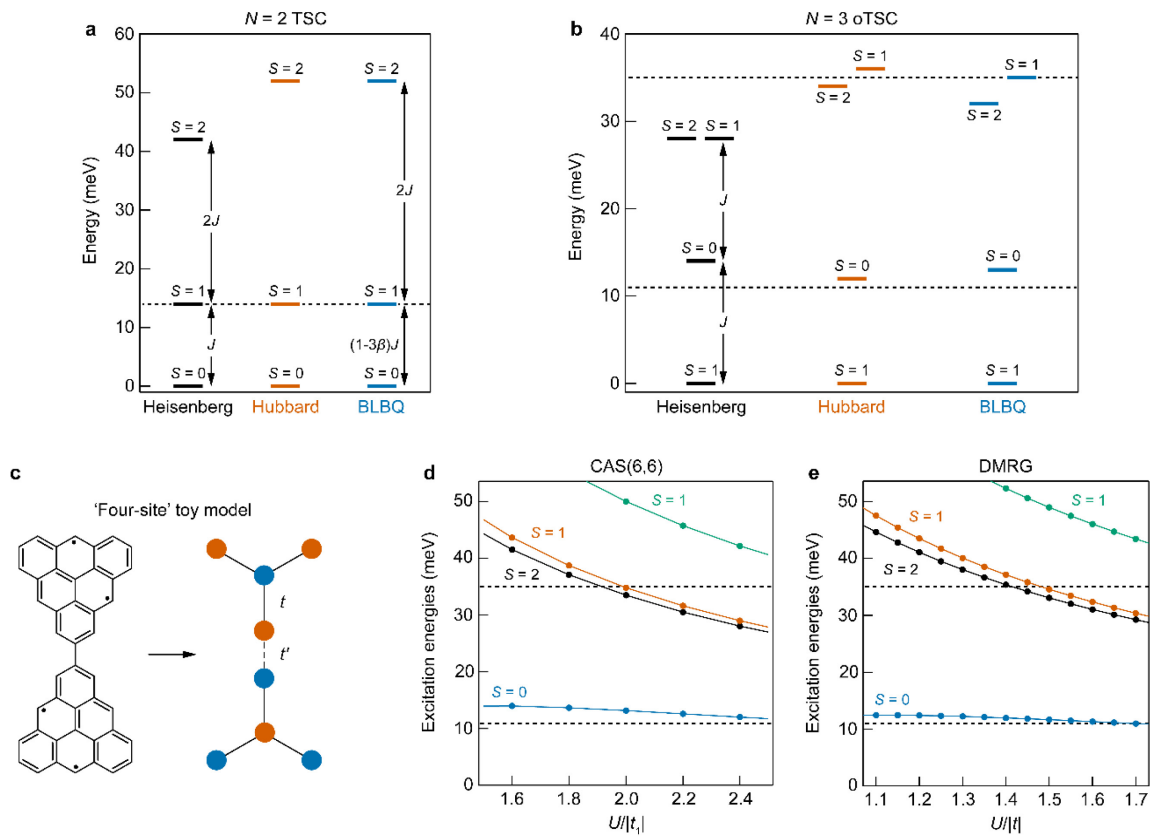

**Extended Data Fig. 3 │ Derivation of the bilinear-biquadratic model. a,b,** Schematic energy level diagram of $N = 2$ (**a**) and 3 (**b**) oTSCs for Heisenberg, Hubbard and BLBQ models. Analytical expressions for the spin models are provided in the Supplementary Information. The Hubbard model is defined such that each triangulene unit is represented by a four-site lattice (**c**) and the many-body energy levels are computed with DMRG, taking $t = -1.11$ eV, $t' = -0.20$ eV and $U = 1.45 |t|$. The parameters of the BLBQ model ($J = 18$ meV and $\beta = 0.09$) are obtained by matching its excitation energies to those of the Hubbard model for the $N = 2$ TSC. **c,** Description of the four-site toy model with the intra- and intertriangulene hopping, $t$ and $t'$, respectively, indicated. The colored filled circles denote the two sublattices. **d,e,** Comparison of the excitation energies for an $N = 3$ oTSC computed with CAS(6,6) for the complete Hubbard model with $t_1 = -2.70$ eV, $t_2 = 0$ eV and $t_3 = -0.35$ eV (**d**), and with DMRG for the four-site Hubbard model (**e**), as the atomic Hubbard $U$ is varied. Dashed lines indicate the experimental spin excitation energies of 14 meV for $N = 2$ TSC (**a**) and, 11 and 35 meV for $N = 3$ oTSC (**b**, **d** and **e**). Note that the Heisenberg model fails to capture both the experimental spin excitation energies for the $N = 3$ oTSC (**b**), and the Hubbard results for the $N = 2$ (**a**) and $N = 3$ (**b**) oTSCs.

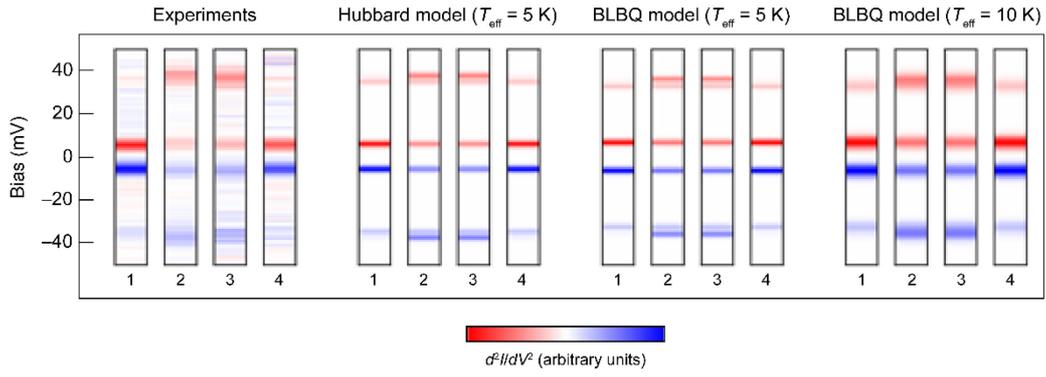

**Extended Data Fig. 4 | Experimental and theoretical spectroscopic signatures of spin excitations in an $N = 4$ open-ended triangulene spin chain.** Comparison between experimental and theoretical (using the four-site Hubbard and BLBQ models) $d^2I/dV^2$ spectra of an $N = 4$ oTSC shows a good agreement in both the energies and the modulation of the spin spectral weight across the different units in the TSC. Numerals along the abscissa denote the unit number of the TSC. BLBQ model calculations are performed with two different $T_{\text{eff}}$ values for the tunneling quasiparticle, which determine the linewidth of the $d^2I/dV^2$ profile. Model parameters are as in Extended Data Fig. 3.

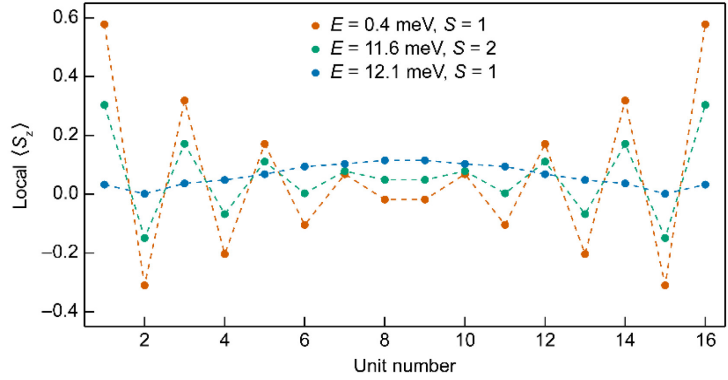

**Extended Data Fig. 5 | Average magnetization for the first three $S_z = +1$ states of an $N = 16$ open-ended triangulene spin chain, obtained with the bilinear-biquadratic model.** Calculations were performed with $J = 18$ meV and $\beta = 0.09$. Orange filled circles denote the magnetization profile of the state with the lowest excitation energy $E = 0.4$ meV, much smaller than the Haldane gap (9 meV), and $|S, S_z\rangle = |1, +1\rangle$. The average magnetization is clearly the largest at the terminal units, and is strongly depleted at the central units, as expected for an edge state. Blue and green filled circles denote spin excitations with energies larger the Haldane gap. Blue filled circles correspond to a state with $E = 12.1$ meV and $|S, S_z\rangle = |1, +1\rangle$, where the magnetization profile forms a nodeless standing wave with maximum average magnetization at the central units. This can be identified as a spin wave state, except for the minor upturn at the terminal units. Green filled circles are associated to a state with $E = 11.6$ meV and $|S, S_z\rangle = |2, +1\rangle$, where the average magnetization shares similarities with both the edge and nodeless spin wave states.

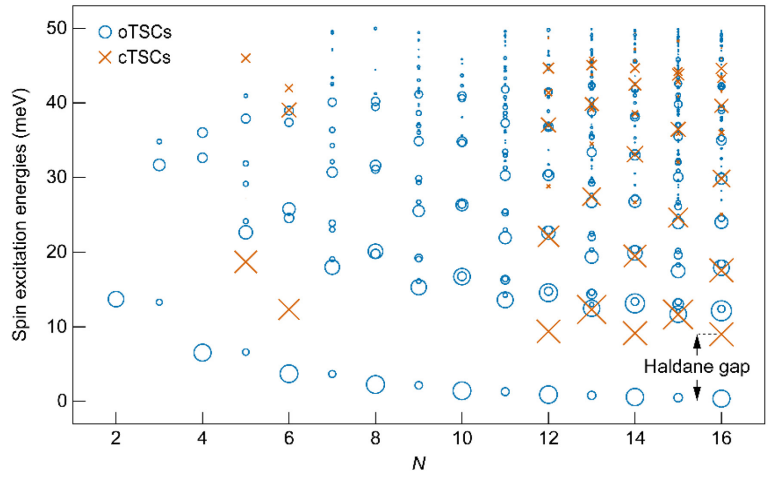

**Extended Data Fig. 6 │ Theoretical spin excitation spectrum of open-ended and cyclic triangulene spin chains.** Spin excitation energies calculated by ED of the BLBQ model ($J$ = 18 meV and $\beta$ = 0.09) for oTSCs with $N$ = 2–16 (circles) and cTSCs with $N$ = 5, 6, 12, 13, 14, 15 and 16 (crosses) up to 50 meV. Size of the symbols accounts for the spin spectral weight of the corresponding spin excitation. The lowest energy bulk excitation, as indicated for the $N$ = 16 cTSC, converges to the Haldane gap (9 meV) with increasing $N$. Note the odd-even effect observed for the lowest energy excitation of cTSCs, which is also seen in the experiments (Supplementary Fig. 6).

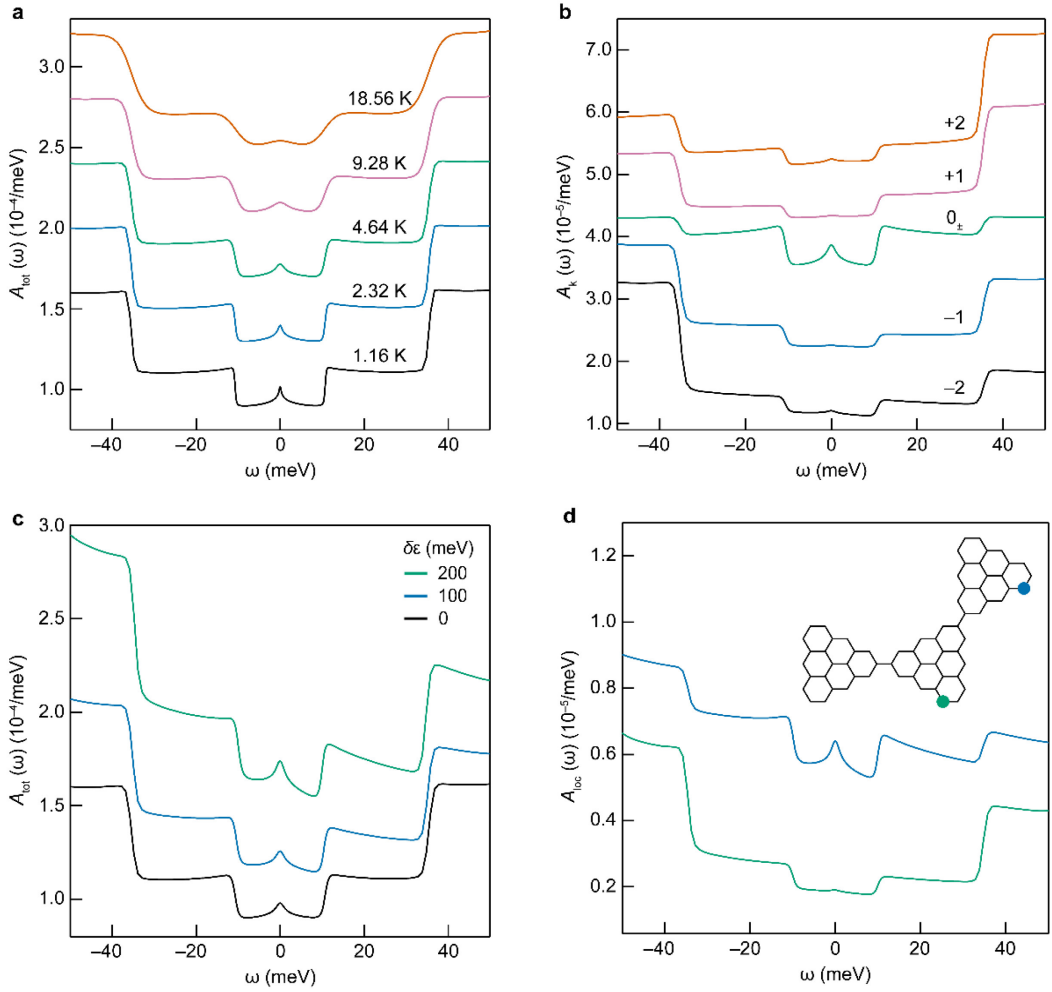

**Extended Data Fig. 7 │ Non-crossing approximation results for the multi-orbital Anderson model of an $N = 3$ open-ended triangulene spin chain ($t_1 = -2.70$ eV, $t_2 = 0$ eV, $t_3 = -0.35$ eV and $U = 1.90\,|t_1|$) coupled to the surface ($\Gamma/\pi = 13$ meV). a,** Total spectral function of CAS(6,6) at different temperatures $T$ for the case of particle-hole symmetry. **b,** Orbital-resolved spectral function of CAS(6,6) for $T = 4.64$ K and for the particle-hole symmetric case. **c,** Detuning from particle-hole symmetry: total spectral function of CAS(6,6) for different values of $\delta\varepsilon$ and $T = 4.64$ K. **d,** Local spectral functions at $T = 4.64$ K for carbon sites of one of the outer triangulene units and the central triangulene unit ($\delta\varepsilon = 200$ meV). The inset shows a sketch of the $N = 3$ oTSC with the two carbon sites marked with the corresponding colored filled circles. The spectral functions in individual panels are offset vertically for visual clarity.

## Supplementary Information

# Observation of fractional edge excitations in nanographene spin chains

Shantanu Mishra, Gonçalo Catarina, Fupeng Wu, Ricardo Ortiz, David Jacob, Kristjan Eimre, Ji Ma, Carlo A. Pignedoli, Xinliang Feng, Pascal Ruffieux, Joaquín Fernández-Rossier and Roman Fasel

Contents:

# 1. Analytical solutions of Heisenberg and BLBQ models

## (1) BLBQ dimer

Here we provide the derivation of the excitation energies of the BLBQ model for the $N$ = 2 TSC (dimer). The model Hamiltonian reads as

$$\hat{H}_{BLBQ} = J\left[\vec{S}_1 \cdot \vec{S}_2 + \beta\left(\vec{S}_1 \cdot \vec{S}_2\right)^2\right] \tag{1}$$

where $\vec{S}_1$ and $\vec{S}_2$ correspond to the individual spin operators of the triangulene units, which we take as $S$ = 1 in the following. We define $\vec{S}_T = \vec{S}_1 + \vec{S}_2$, and use the fact that the eigenvalues of $S_T{}^2 = \vec{S}_T \cdot \vec{S}_T$ are $S_T(S_T + 1)$, where $S_T$ covers the range $|S_1 - S_2|, |S_1 - S_2| + 1, \ldots, S_1 + S_2$. In this case, $S_T$ = 0, 1 and 2. We now write

$$S_T{}^2 = \left(\vec{S}_1 + \vec{S}_2\right)^2 = S_1{}^2 + S_2{}^2 + 2\vec{S}_1 \cdot \vec{S}_2 \tag{2}$$

Equation 2 allows to replace the spin operator $\vec{S}_1 \cdot \vec{S}_2$ by scalars, using $S_1{}^2 = S_2{}^2 = S(S + 1) = 2$. We can thus express the BLBQ Hamiltonian for the dimer as

$$\hat{H}_{BLBQ} = J\left[\left(\frac{S_T(S_T+1)-4}{2}\right) + \beta\left(\frac{S_T(S_T+1)-4}{2}\right)^2\right] \tag{3}$$

This expression yields the eigenvalues in terms of the total spin quantum number $S_T$

$$E(S_T) = J\left[\left(\frac{S_T(S_T+1)-4}{2}\right) + \beta\left(\frac{S_T(S_T+1)-4}{2}\right)^2\right] \tag{4}$$

From equation (4), we obtain the excitation energies for the BLBQ dimer as

$$E(S_T = 1) - E(S_T = 0) = J(1 - 3\beta) \tag{5}$$

and

$$E(S_T = 2) - \mathrm{E}(S_T = 0) = 3J(1 - \beta) \tag{6}$$

The Heisenberg model limit may be obtained by taking $\beta$ = 0.

## (2) Heisenberg trimer

We now derive the expression for the excitation energies of the Heisenberg model for the $N$ = 3 oTSC (trimer). We first write the model Hamiltonian as

$$\hat{H}_{Heisenberg} = J\vec{S}_2 \cdot \left(\vec{S}_1 + \vec{S}_3\right) = J\vec{S}_2 \cdot \vec{S}_{edge} \tag{7}$$

where we have defined $\vec{S}_{edge} = \vec{S}_1 + \vec{S}_3$, which can take the values $S_{edge}$ = 0, 1 and 2, assuming that all individual spins are $S$ = 1 objects. We now define $\vec{S}_T = \vec{S}_2 + \vec{S}_{edge}$, which can take the values $S_T$ = 1 for $S_{edge}$ = 0, $S_T$ = 0, 1 and 2 for $S_{edge}$ = 1, and $S_T$ = 1, 2 and 3 for $S_{edge}$ = 2. Using equation (2), replacing $S_1$ by $S_{edge}$, we write the eigenvalues of the Heisenberg trimer Hamiltonian as

$$E(S_T, \ S_{edge}) = \frac{J}{2}\left[S_T(S_T + 1) - S_{edge}\left(S_{edge} + 1\right) - 2\right] \tag{8}$$

For $J > 0$, the ground state has $S_T$ = 1 and $S_{edge}$ = 2. The first excited state, with excitation energy $J$, has $S_T$ = 0 and $S_{edge}$ = 1. Two degenerate multiplets, with $S_T = S_{edge} = 1$ and $S_T = S_{edge} = 2$, give the second excited state, with excitation energy $2J$.

## 2. STM and STS data

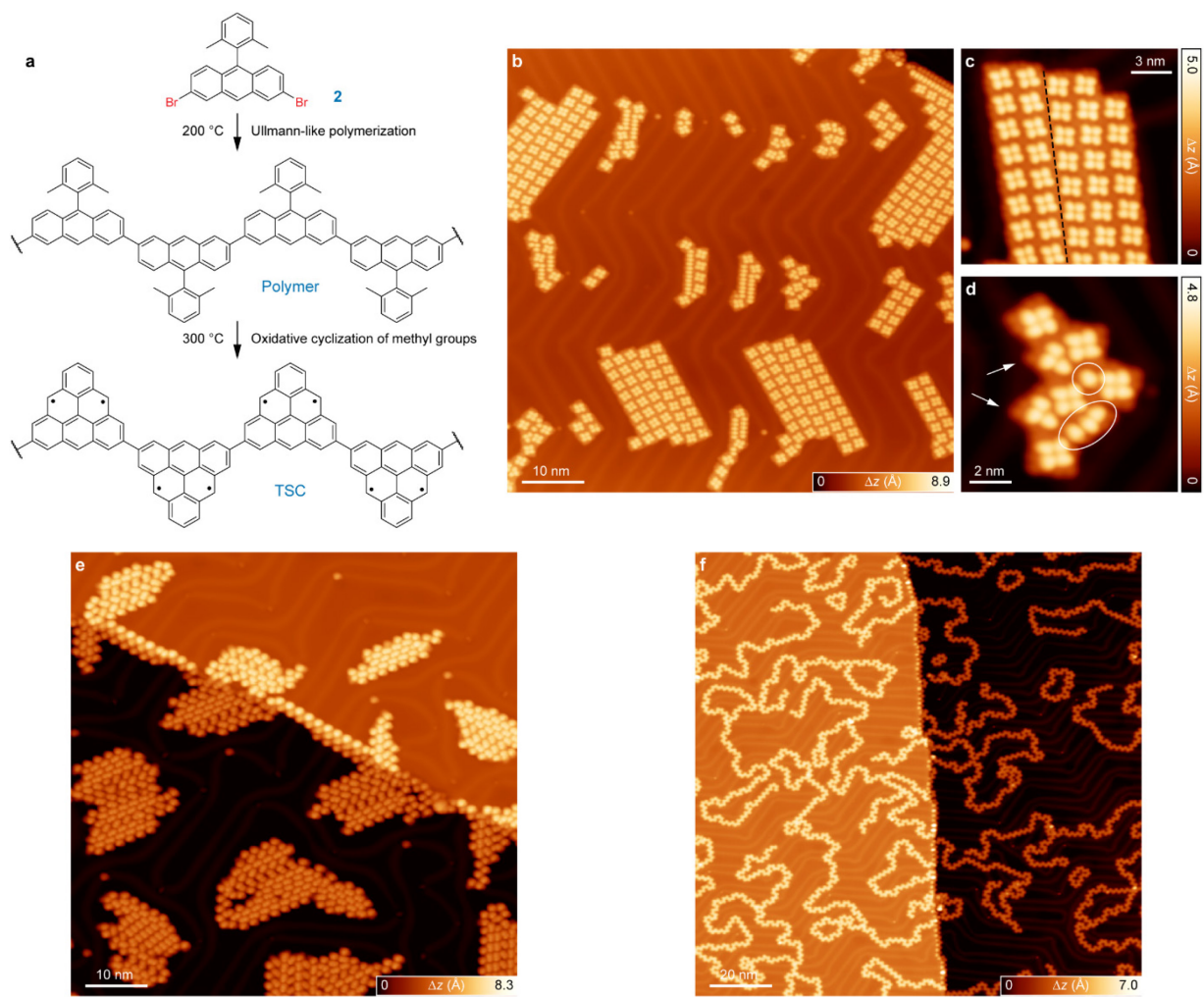

**Supplementary Fig. 1 | On-surface synthesis of TSCs using precursor 2. a**, Schematic illustration of the on-surface reactions of **2** leading to the synthesis of TSCs. **b**, Overview STM image after deposition of **2** on a Au(111) surface held at room temperature ($V = -1.2$ V, $I = 30$ pA). The precursor molecules self-assemble into both ordered and disordered islands. **c**, High-resolution STM image of an ordered island ($V = -1.0$ V, $I = 40$ pA). We assign each bright lobe to the dimethylphenyl moiety of **2**, which is bent out of plane due to steric repulsion with the dibromoanthracene moiety that likely adopts a planar conformation on the surface. Evidently, the constituent building block of the island is a self-assembled tetramer of **2**. A single island usually contains domains with different chirality of the tetramers. A domain boundary is highlighted with a dashed line. **d**, High-resolution STM image of a disordered island ($V = -1.0$ V, $I = 40$ pA). The island contains five tetramers, two trimers (highlighted with arrows) and four isolated molecules/monomers (highlighted with ellipses). **e**, Overview STM image after successively annealing the sample in **a** to 150 °C and 200 °C for 5 minutes each ($V = -1.2$ V, $I = 30$ pA). The surface consists of polymer islands. **f**, Overview STM image after annealing the sample in **e** to 300 °C for 5 minutes ($V = -1.2$ V, $I = 30$ pA). The surface predominantly consists of long and mostly interconnected TSCs, with a minority of isolated cTSCs.

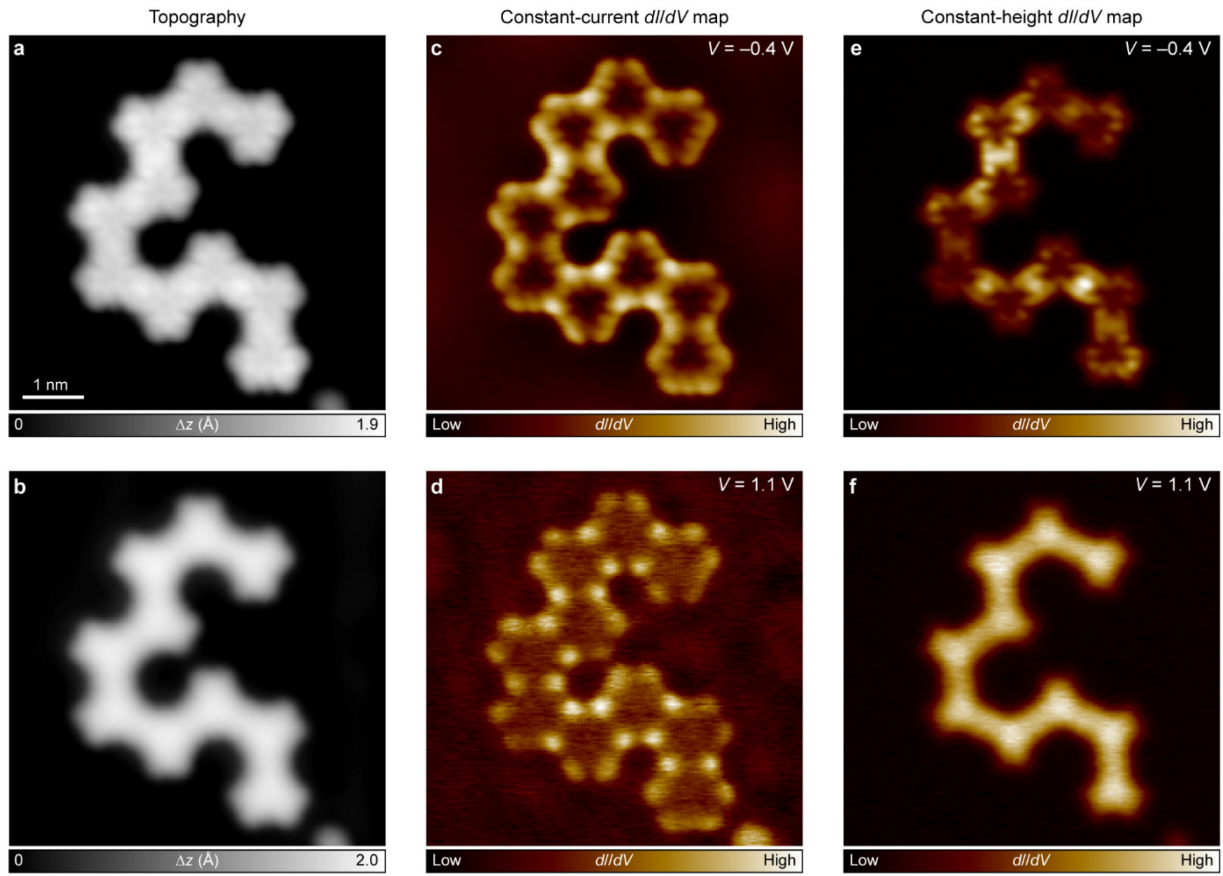

Topography | Constant-current *dI/dV* map | Constant-height *dI/dV* map

**a** | **c** $V = -0.4$ V | **e** $V = -0.4$ V

1 nm

0      $\Delta z$ (Å)      1.9 | Low    *dI/dV*    High | Low    *dI/dV*    High

**b** | **d** $V = 1.1$ V | **f** $V = 1.1$ V

0      $\Delta z$ (Å)      2.0 | Low    *dI/dV*    High | Low    *dI/dV*    High

**Supplementary Fig. 2 | $dI/dV$ mapping of the frontier bands of an $N = 10$ oTSC. a,b,** High-resolution STM images of an $N = 10$ oTSC at the valence (**a**) and conduction (**b**) band resonances. **c–f,** Constant-current (**c, d**) and constant-height (**e, f**) $dI/dV$ maps of the valence (**c, e**) and conduction (**d, f**) bands of the TSC. Scanning parameters: $V = -0.4$ V, $I = 300$ pA (**a, c**) and $V = 1.1$ V, $I = 350$ pA (**b, d**). Open feedback parameters: $V = -0.4$ V, $I = 320$ pA (**e**) and $V = 1.1$ V, $I = 350$ pA (**f**). For all measurements, $V_{rms} = 30$ mV. All measurements were performed with a CO functionalized tip.

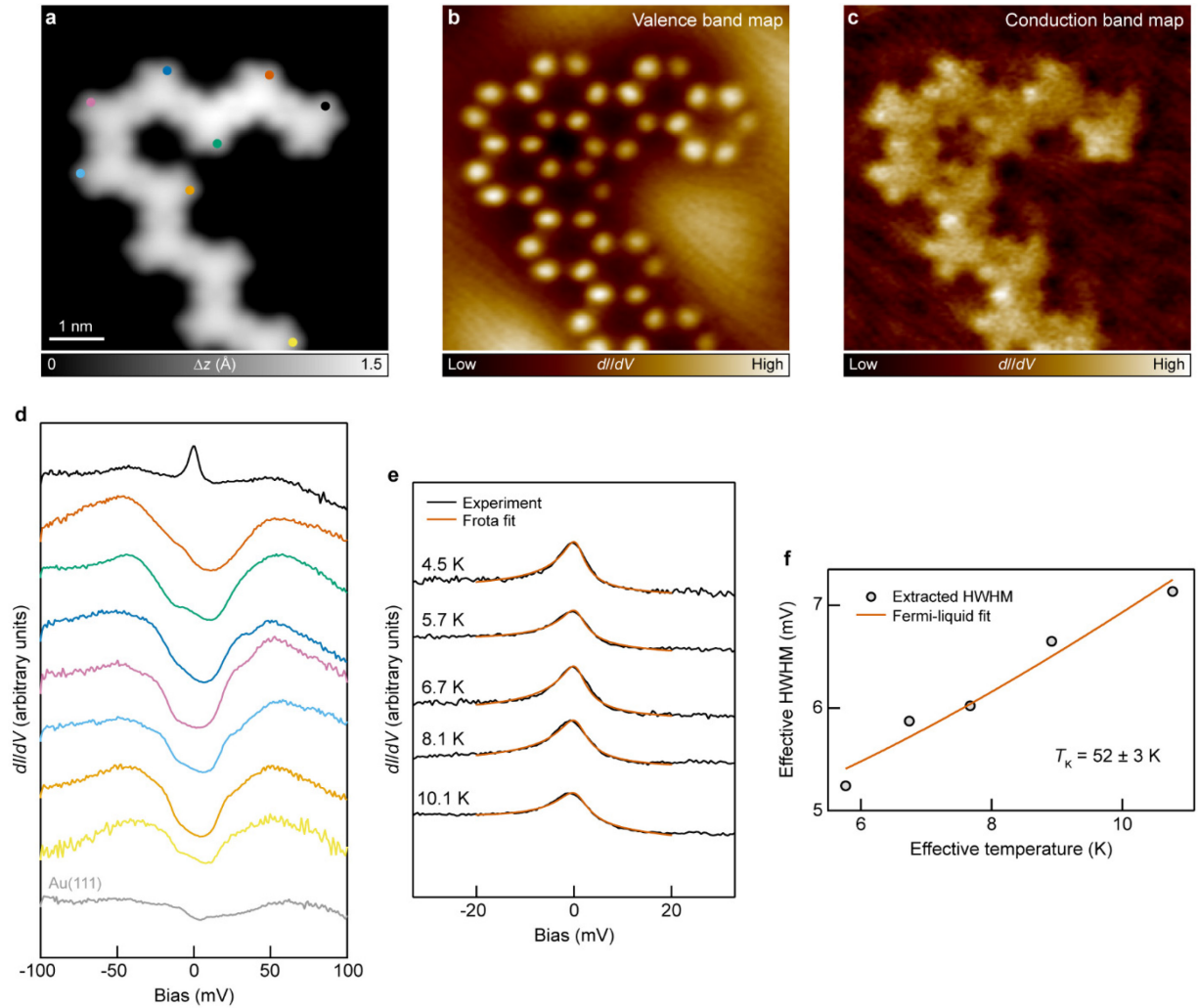

**Supplementary Fig. 3 | Temperature-dependent spectroscopy of zero bias peaks in oTSCs.**
**a−c**, High-resolution STM image ($V = -0.4$ V, $I = 350$ pA) (**a**), and constant-current $dI/dV$ maps of the valence ($V = -0.4$ V, $I = 350$ pA; $V_{rms} = 30$ mV) (**b**) and conduction ($V = 1.1$ V, $I = 400$ pA; $V_{rms} = 30$ mV) (**c**) bands of an oTSC with only one free terminus. The second terminus is connected to another TSC. **d**, $dI/dV$ spectroscopy on the TSC at the positions indicated with the corresponding colored filled circles in **a**, revealing the typical zero energy peak at the terminus and spin excitations at the non-terminal units. **e**, Temperature evolution of the zero bias peak, with fit to the experimental data with the Frota function.[1,2] **f**, Extracted half-width at half-maximum (HWHM) of the Kondo resonance as a function of temperature, with corresponding fit using the Fermi-liquid model.[3] We have taken into account the experimental broadening of the zero bias peaks due to the finite modulation of the lock-in amplifier[4] (corresponding to an effective temperature) and the temperature of the tip[5] (corresponding to an effective HWHM). The HWHM of the zero bias peak exhibits an anomalous broadening with increasing temperature, and follows the characteristic trend of a Kondo-screened state with the Kondo temperature $T_K = 52 \pm 3$ K. Open feedback parameters for the $dI/dV$ spectra: $V = -100$ mV (**d**) and $V = -35$ mV (**e**), $I = 1.3$ nA; $V_{rms} = 1$ mV.

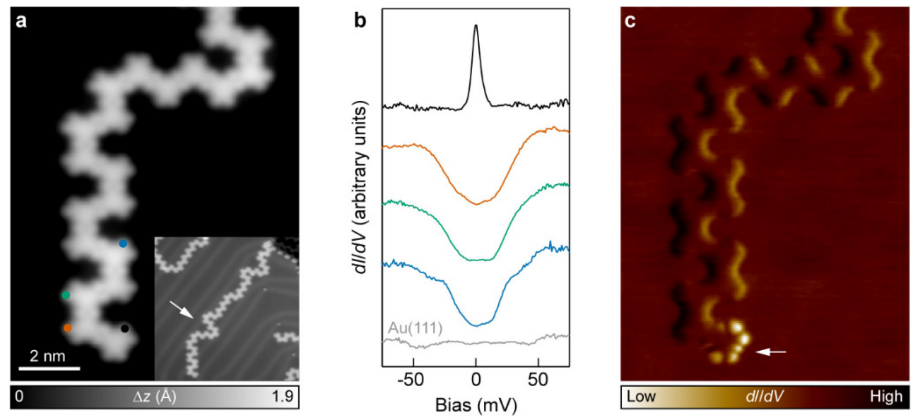

**Supplementary Fig. 4 | $dI/dV$ mapping of Kondo resonance. a**, High-resolution STM image of a long oTSC with only one free terminus ($V = -0.4$ V, $I = 400$ pA). The second terminus is connected to another species near a Au(111) step edge, as seen in the large-scale inset STM image. The arrow highlights the TSC. **b**, $dI/dV$ spectroscopy on the TSC at the positions indicated with the corresponding colored filled circles in **a**. A Kondo resonance is present at the terminal triangulene unit. **c**, Constant-current $dI/dV$ map near zero bias, revealing the spatial distribution of the Kondo resonance over the terminal triangulene unit, as highlighted by an arrow ($V = 3$ mV, $I = 150$ pA; $V_{rms} = 1.6$ mV).

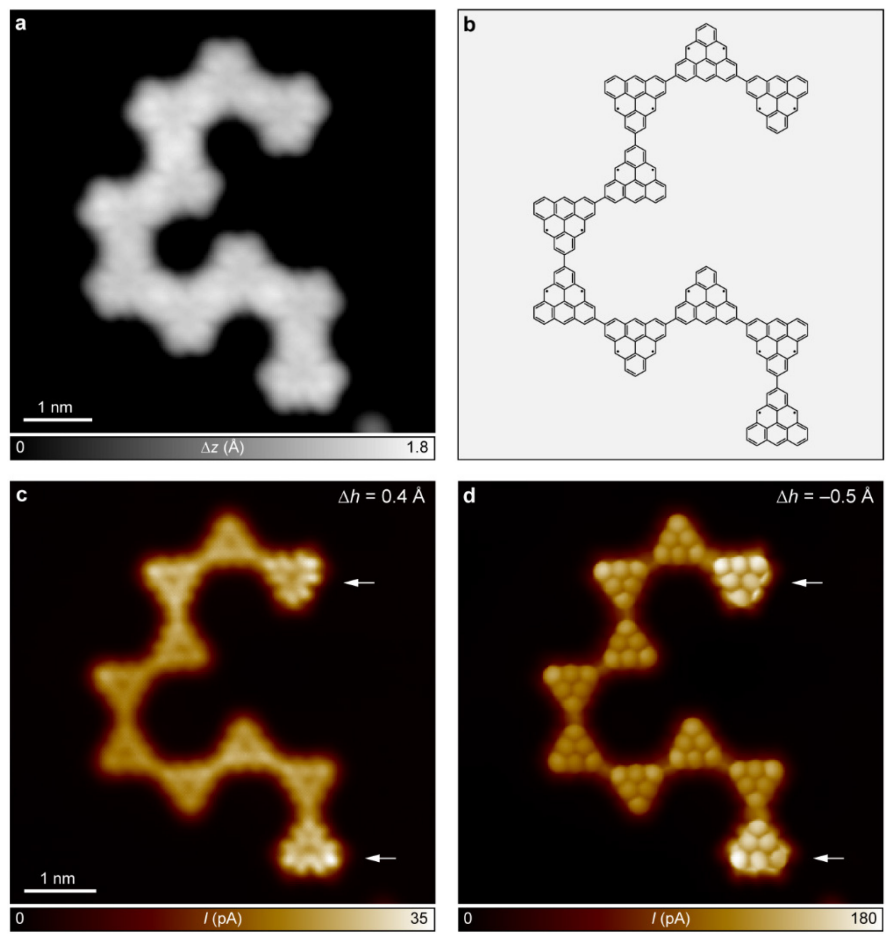

**Supplementary Fig. 5 | Visualization of Kondo resonance through bond-resolved STM imaging. a**, High-resolution STM image of an $N = 10$ oTSC ($V = -0.4$ V, $I = 300$ pA). **b**, Chemical structure of the TSC. **c,d**, Bond-resolved STM images of the TSC at $\Delta h = 0.4$ Å (**c**) and $-0.5$ Å (**d**). At a large tip-sample distance as in **c**, structural contrast due to Pauli repulsion between the tip and TSC is minimized, and the image is dominated by the local density of states (LDOS) around zero bias. As shown by the

arrows, significant LDOS, which is uniformly distributed over the terminal triangulene units, is observed. Given that there are no resonances at or near zero bias in the TSC apart from a Kondo resonance, the zero-bias LDOS thus depicts the spatial distribution of the Kondo resonance. At a small tip-sample distance, as in **d**, clear structural contrast is obtained over the TSC. In addition, a convolution of structural contrast with the zero-bias LDOS occurs, which leads to a larger tunneling current over the terminal triangulene units (indicated by arrows). Open feedback parameters for the bond-resolved STM images: $V = -5$ mV, $I = 50$ pA.

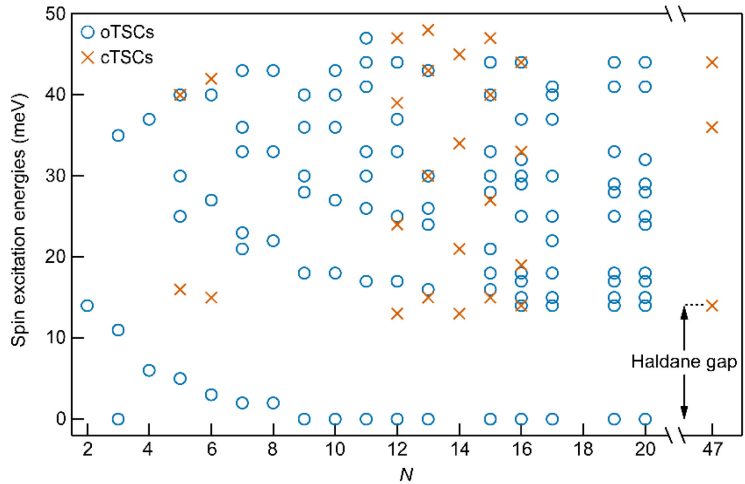

**Supplementary Fig. 6 | Experimental spin excitation spectrum of open-ended and cyclic TSCs.** Experimental spin excitation energies up to 50 meV for seventeen oTSCs with $N$ between 2 and 20 (circles), and eight cTSCs with $N = 5$, 6, 12, 13, 14, 15, 16 and 47 (crosses). The lowest energy bulk excitation, indicated for the $N = 47$ cTSC, converges to the Haldane gap (14 meV) with increasing $N$. In fact, starting from both oTSC and cTSC with $N = 16$, convergence to the Haldane gap is observed.

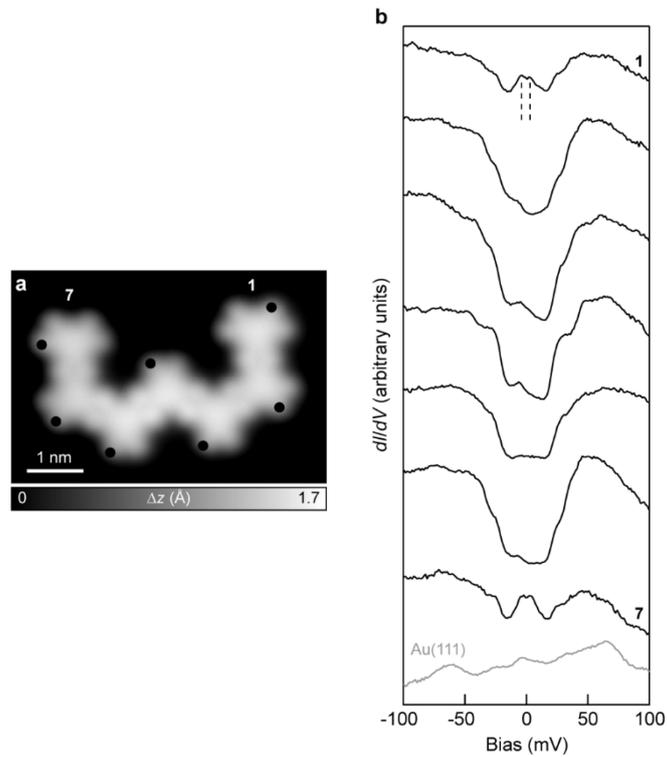

**Supplementary Fig. 7 | Low-bias *dI/dV* spectroscopy on an *N* = 7 oTSC. a**, High-resolution STM image of an *N* = 7 oTSC (*V* = −0.7 V, *I* = 320 pA). **b**, *dI/dV* spectroscopy on every unit of the TSC (open feedback parameters: *V* = −100 mV, *I* = 1.4 nA; $V_{rms}$ = 1 mV). Acquisition positions are marked with a filled circle in **a**. Numerals indicate the unit number, marked in **a**, on which the corresponding spectrum was acquired. Six unique spin excitations are observed – units 1 and 7: 2, 21, 33 and 43 mV; units 2 and 6: 21, 33 and 43 mV; units 3 and 5: 23 and 36 mV; unit 4: 21 and 43 mV. The dashed lines in **b** indicate the splitting of the apparent peak on units 1 and 7 into spin excitations at ±2 mV.

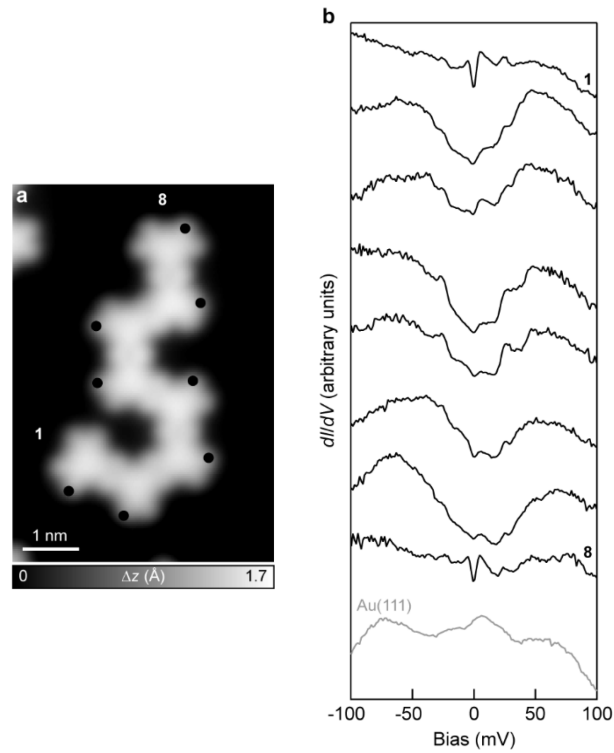

**Supplementary Fig. 8 | Low-bias *dI/dV* spectroscopy on an *N* = 8 oTSC. a**, High-resolution STM image of an *N* = 8 oTSC (*V* = −0.4 V, *I* = 300 pA). **b**, *dI/dV* spectroscopy on every unit of the TSC (open feedback parameters: *V* = −100 mV, *I* = 1.4 nA; $V_{rms}$ = 1 mV). Acquisition positions are marked with a filled circle in **a**. Four unique spin excitations are observed: units 1, 2, 3, 6, 7 and 8: 2, 22 and 33 meV; units 4 and 5: 2, 22 and 43 mV.

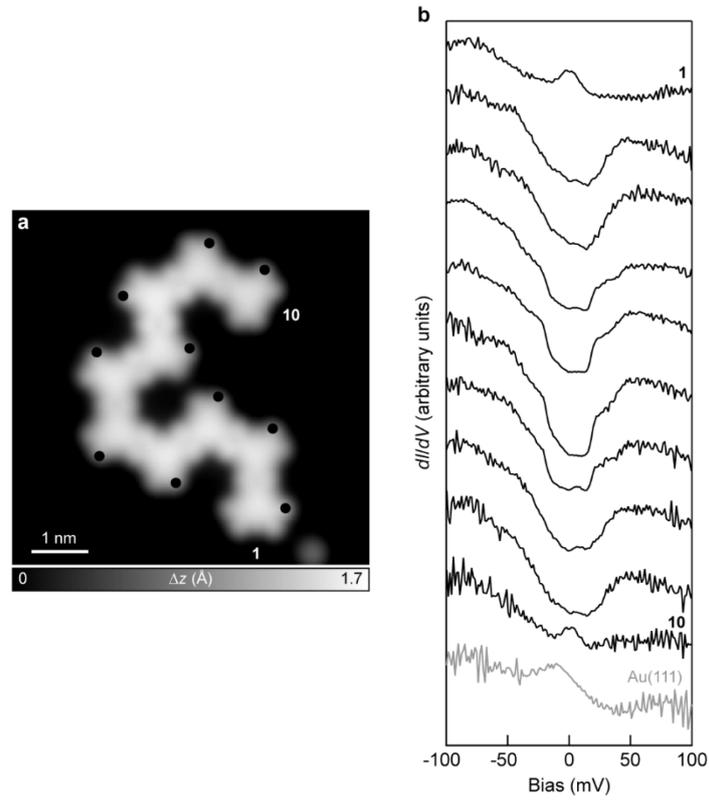

**Supplementary Fig. 9 | Low-bias $dI/dV$ spectroscopy on an $N$ = 10 oTSC. a**, High-resolution STM image of an $N$ = 10 oTSC ($V$ = −0.4 V, $I$ = 320 pA). **b**, $dI/dV$ spectroscopy on every unit of the TSC (open feedback parameters: $V$ = −100 mV, $I$ = 1.4 nA; $V_{rms}$ = 1 mV). Acquisition positions are marked with a filled circle in **a**. A Kondo resonance on units 1 and 10, and five unique spin excitations on units 2−9, are observed − units 2, 3, 8 and 9: 18, 27 and 36 mV; units 4 and 7: 18 and 43 mV; units 5 and 6: 18 and 40 mV.

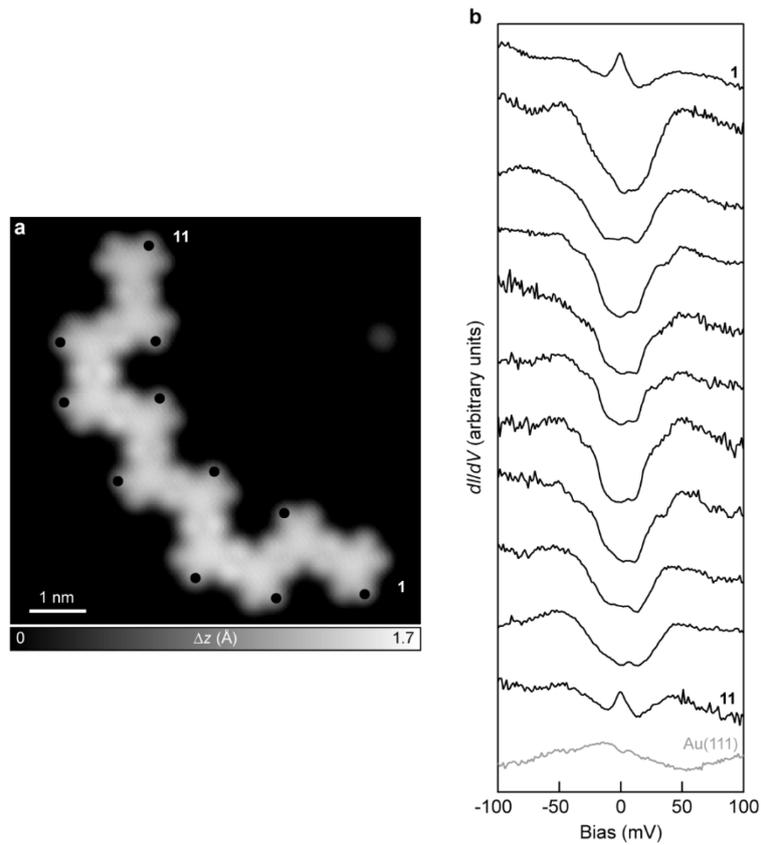

**Supplementary Fig. 10 | Low-bias *dI/dV* spectroscopy on an *N* = 11 oTSC. a**, High-resolution STM image of an *N* = 11 oTSC (*V* = −0.7 V, *I* = 200 pA). **b**, *dI/dV* spectroscopy on every unit of the TSC (open feedback parameters: *V* = −100 mV, *I* = 1.4 nA; $V_{rms}$ = 1 mV). Acquisition positions are marked with a filled circle in **a**. A Kondo resonance on units 1 and 11, and seven unique spin excitations on units 2−10, are observed − units 2 and 10: 30 mV; units 3 and 9: 17, 26 and 33 mV; units 4 and 8: 17, 26, 41 and 47 mV; units 5−7: 17, 33 and 44 mV.

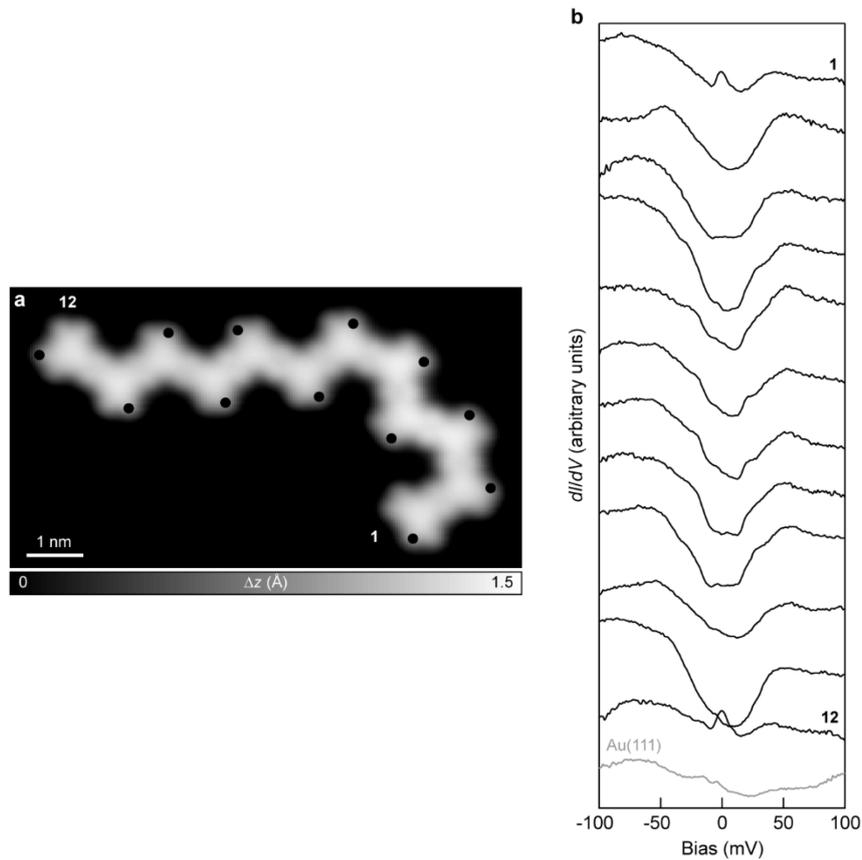

**Supplementary Fig. 11 | Low-bias *dI/dV* spectroscopy on an *N* = 12 oTSC. a**, High-resolution STM image of an *N* = 12 oTSC (*V* = −0.4 V, *I* = 350 pA). **b**, *dI/dV* spectroscopy on every unit of the TSC (open feedback parameters: *V* = −100 mV, *I* = 1.4 nA; *V*rms = 1 mV). Acquisition positions are marked with a filled circle in **a**. A Kondo resonance on units 1 and 12, and five unique spin excitations on units 2−11, are observed − units 2, 3, 10 and 11: 17, 25 and 33 mV; units 4 and 9: 17, 25, 37 and 44 mV; units 5−8: 17, 33 and 44 mV.

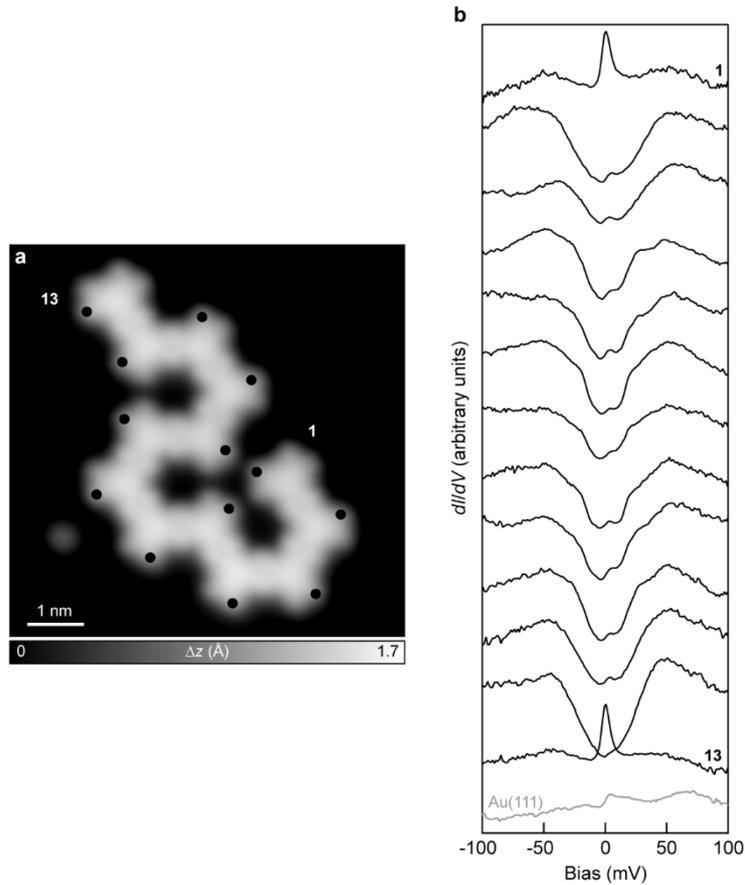

**Supplementary Fig. 12 | Low-bias *dI/dV* spectroscopy on an *N* = 13 oTSC. a**, High-resolution STM image of an *N* = 13 oTSC (*V* = −0.6 V, *I* = 300 pA). **b**, *dI/dV* spectroscopy on every unit of the TSC (open feedback parameters: *V* = −100 mV, *I* = 1.4 nA; $V_{rms}$ = 1 mV). Acquisition positions are marked with a filled circle in **a**. A Kondo resonance on units 1 and 13, and four unique spin excitations on units 2−12, are observed − units 2, 3, 11 and 12: single broad excitation at 26 mV, likely consisting of multiple closely spaced excitations; units 4, 5, 9 and 10: 16, 24 and 43 mV; units 6−8: 16, 30 and 43 mV.

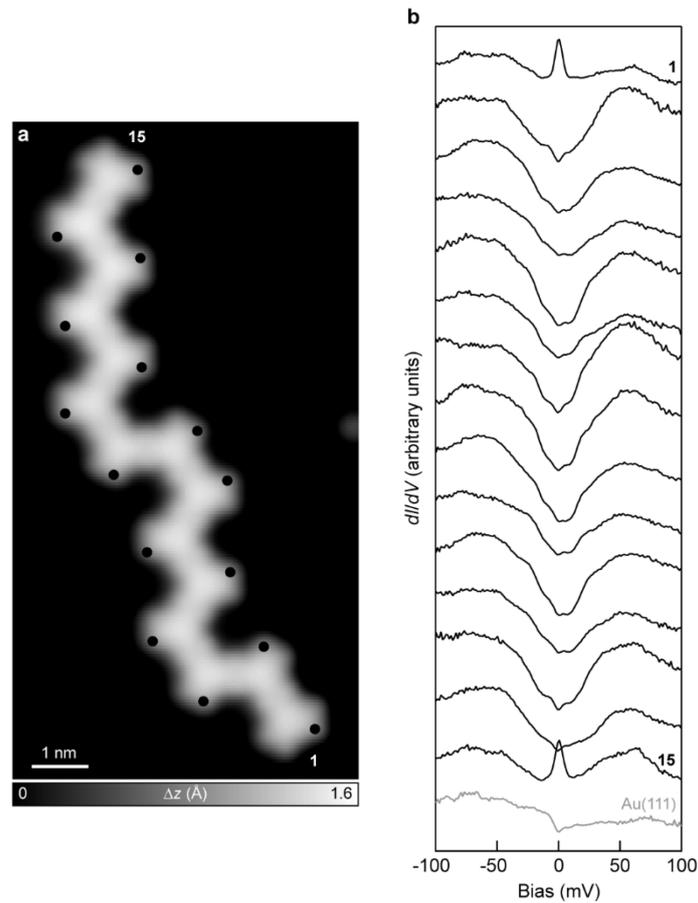

**Supplementary Fig. 13 | Low-bias $dI/dV$ spectroscopy on an $N = 15$ oTSC. a**, High-resolution STM image of an $N = 15$ oTSC ($V = -0.4$ V, $I = 350$ pA). **b**, $dI/dV$ spectroscopy on every unit of the TSC (open feedback parameters: $V = -100$ mV, $I = 1.4$ nA; $V_{rms} = 1$ mV). Acquisition positions are marked with a filled circle in **a**. A Kondo resonance on units 1 and 15, and eight unique spin excitations on units 2–14, are observed − units 2 and 14: single broad excitation at 30 mV; units 3 and 13: single broad excitation at 21 mV; units 4 and 12: 18 and 44 mV; units 5 and 11: 18 and 40 mV; units 6 and 10: 16, 33 and 44 mV; units 7 and 9: 16, 28 and 44 mV; unit 8: 16, 28 and 40 mV.

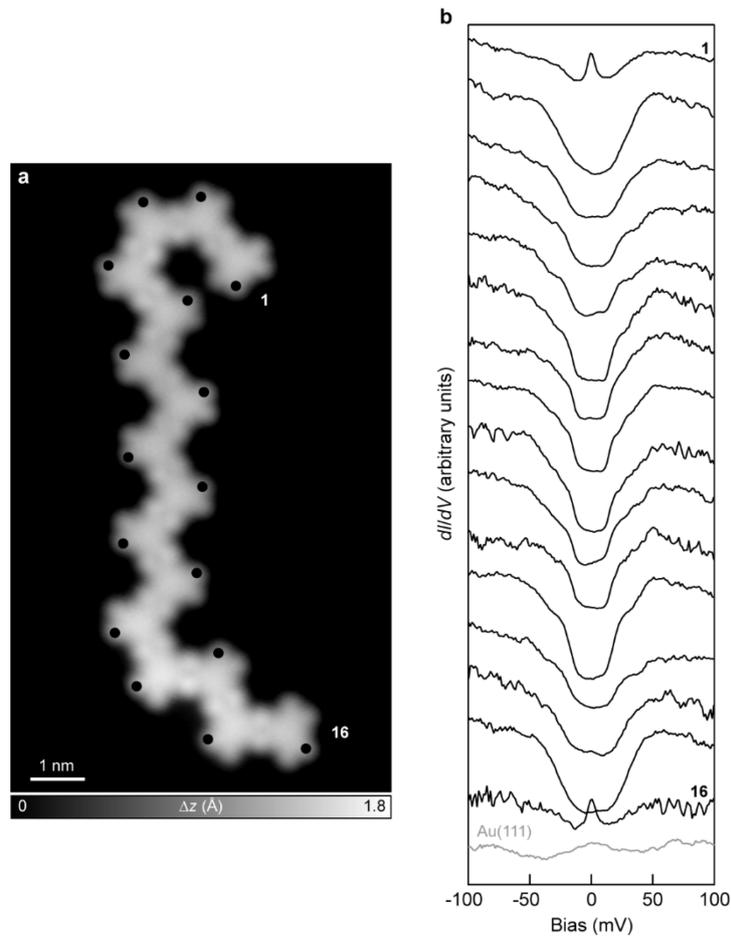

**Supplementary Fig. 14 | Low-bias *dI/dV* spectroscopy on an *N* = 16 oTSC. a**, High-resolution STM image of an *N* = 16 oTSC (*V* = −0.6 V, *I* = 200 pA). **b**, *dI/dV* spectroscopy on every unit of the TSC (open feedback parameters: *V* = −100 mV, *I* = 1.4 nA; $V_{rms}$ = 1 mV). Acquisition positions are marked with a filled circle in **a**. A Kondo resonance on units 1 and 16, and ten unique spin excitations on units 2−15, are observed − units 2 and 15: single broad excitation at 29 mV; units 3 and 14: single broad excitation at 25 mV; units 4 and 13: 18 and 44 mV; units 5 and 12: 17, 37 and 44 mV; units 6 and 11: 15, 32 and 44 mV; units 7 and 10: 14, 30 and 44 mV; units 8 and 9: 14, 25 and 44 mV.

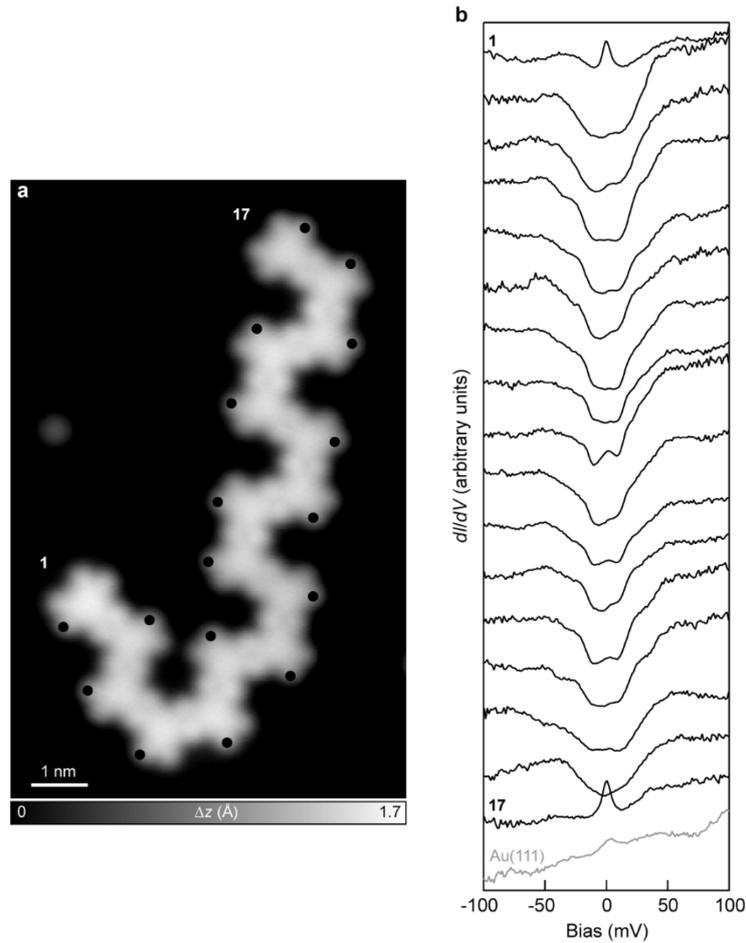

**Supplementary Fig. 15 | Low-bias** *dI/dV* **spectroscopy on an** *N* = **17 oTSC. a**, High-resolution STM image of an *N* = 17 oTSC (*V* = −0.7 V, *I* = 150 pA). **b**, *dI/dV* spectroscopy on every unit of the TSC (open feedback parameters: *V* = −100 mV, *I* = 1.4 nA; $V_{rms}$ = 1 mV). Acquisition positions are marked with a filled circle in **a**. A Kondo resonance on units 1 and 17, and nine unique spin excitations on units 2−16, are observed − units 2 and 16: single broad excitation at 25 mV; units 3 and 15: single broad excitation at 22 mV; units 4, 5, 13 and 14: 18 and 40 mV; units 6, 7, 11 and 12: 15, 30 and 41 mV; units 8 and 10: 14, 25 and 40 mV; unit 9: 14, 25 and 37 mV.

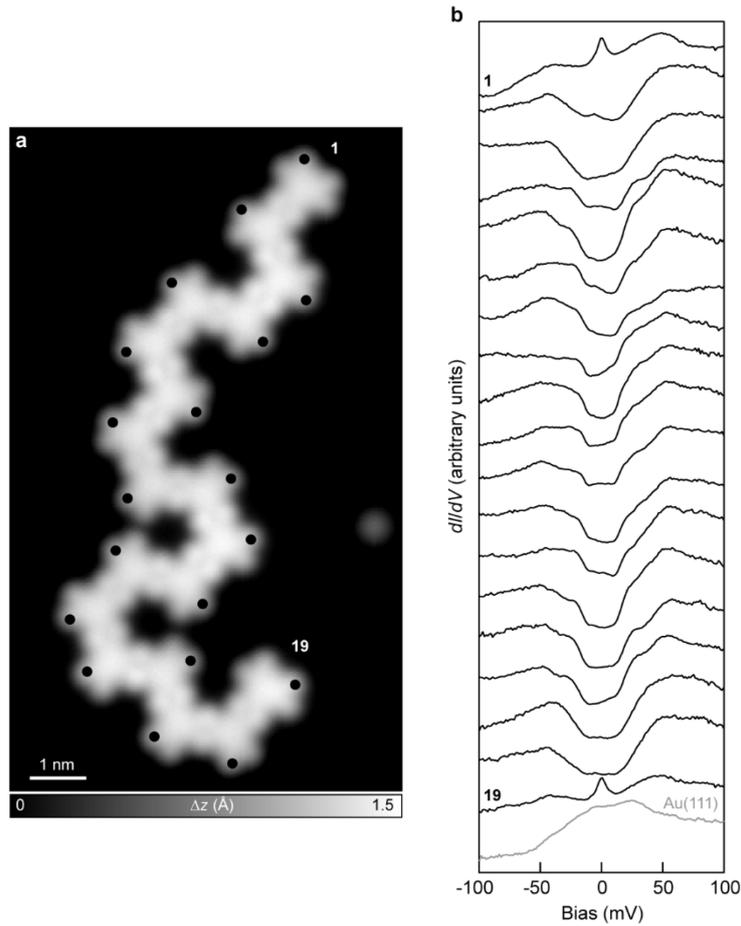

**Supplementary Fig. 16 | Low-bias *dI/dV* spectroscopy on an *N* = 19 oTSC. a**, High-resolution STM image of an *N* = 19 oTSC (*V* = −0.7 V, *I* = 200 pA). **b**, *dI/dV* spectroscopy on every unit of the TSC (open feedback parameters: *V* = −100 mV, *I* = 1.4 nA; $V_{rms}$ = 1 mV). Acquisition positions are marked with a filled circle in **a**. A Kondo resonance on units 1 and 19, and ten unique spin excitations on units 2–18, are observed – units 2 and 18: single broad excitation at 29 mV; units 3 and 17: single broad excitation at 25 mV; units 4, 5, 15 and 16: 18 and 41 mV; units 6, 7, 13 and 14: 15, 33 and 44 mV; units 8 and 12: 15, 28 and 41 mV; units 9 and 11: 14 and 44 mV; unit 10: 14 and 41 mV.

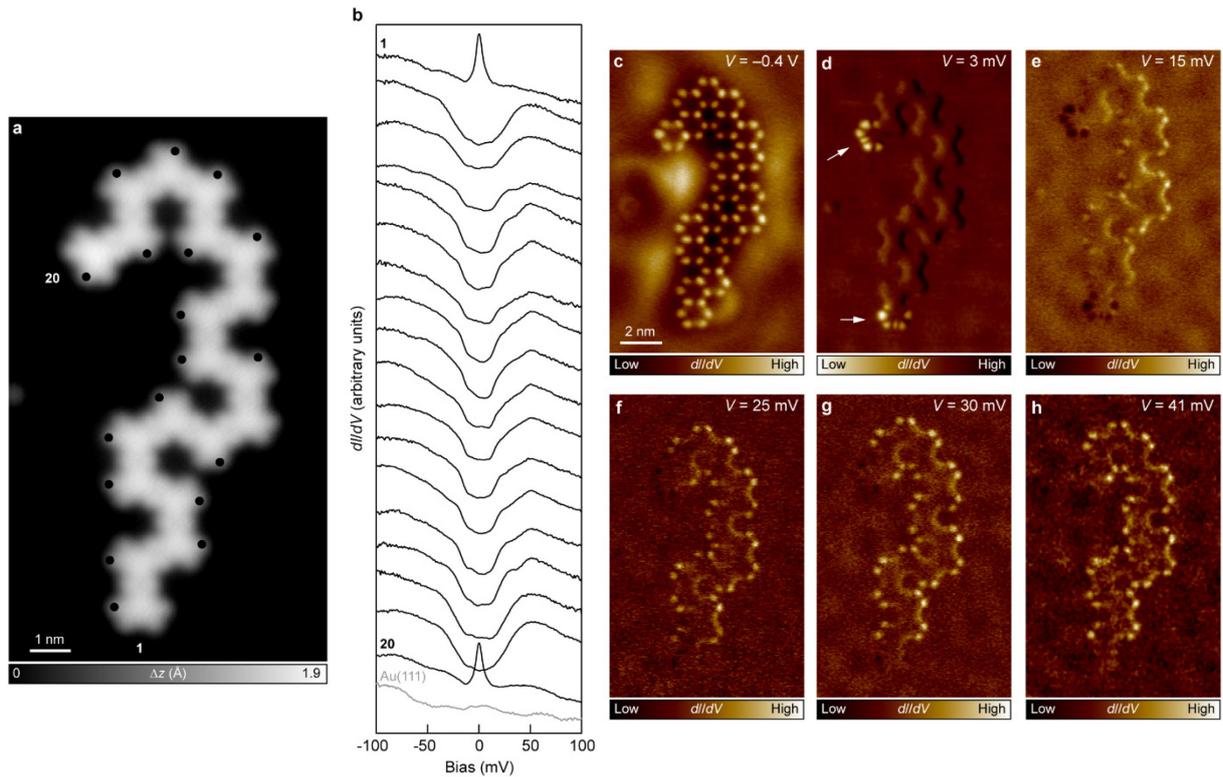

**Supplementary Fig. 17 | Low-bias d$I$/d$V$ spectroscopy on an $N$ = 20 oTSC. a**, High-resolution STM image of an $N$ = 20 oTSC ($V$ = −0.6 V, $I$ = 50 pA). **b**, d$I$/d$V$ spectroscopy on eighteen units of the TSC (open feedback parameters: $V$ = −100 mV, $I$ = 1.4 nA; $V_{rms}$ = 1 mV). Acquisition positions are marked with a filled circle in **a**. A Kondo resonance on units 1 and 20, and eleven unique spin excitations on units 2−19, are observed − units 2 and 19: single broad excitation at 29 mV; units 3 and 18: single broad excitation at 25 mV; units 4 and 17: 18 and 44 mV; units 5 and 16: 18 and 41 mV; units 6, 7, 14 and 15: 17, 32 and 44 mV; units 8 and 12: 15, 28 and 44 mV; units 10 and 11: 14, 24, 32 and 44 mV. **c**, Constant-current d$I$/d$V$ map of the valence band of the TSC ($V$ = −0.4 V, $I$ = 500 pA; $V_{rms}$ = 30 mV). **d−h**, Constant-current d$I$/d$V$ maps acquired near the Kondo resonance and some spin excitations, at biases indicated in the corresponding figures ($I$ = 200 pA (**d, e**), 250 pA (**f, h**) and 220 pA (**g**) and 320 pA (**i**); $V_{rms}$ = 1 mV). The d$I$/d$V$ maps clearly reveal the spatial distribution of the Kondo resonance over the terminal triangulene units (**d**), and distribution of spin excitation signals over multiple triangulene units of the TSC.

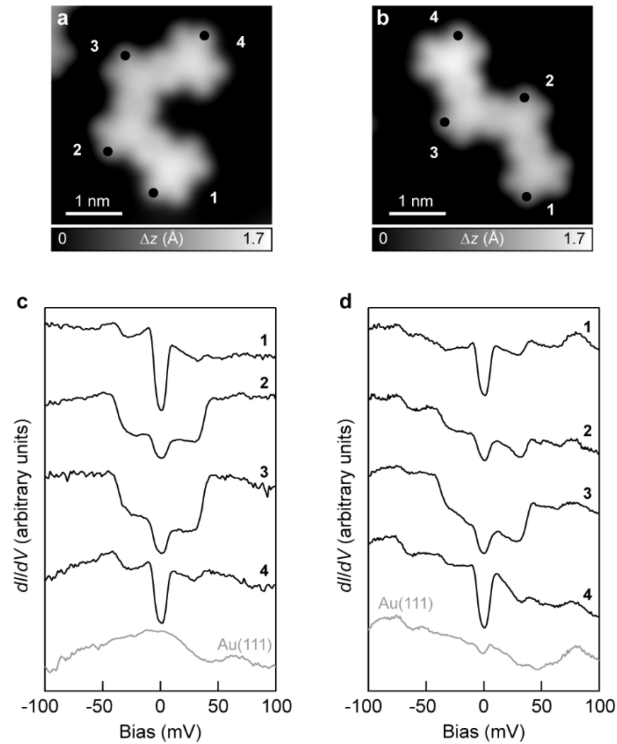

**Supplementary Fig. 18 | Low-bias *dI/dV* spectroscopy on *cis* and *trans* *N* = 4 oTSCs. a,b,** High-resolution STM images of *N* = 4 oTSCs with *cis* (**a**) and *trans* (**b**) inter-triangulene bonding configurations. **c,d,** *dI/dV* spectroscopy on every unit of the *cis* (**c**) and *trans* (**d**) TSCs (open feedback parameters: $V = -100$ mV, $I = 1.4$ nA; $V_{rms} = 1$ mV). Acquisition positions are marked with a filled circle in **a** and **b**. Irrespective of the bonding configuration, both TSCs exhibit spin excitations at 6 and 37 mV, which proves that differences in bonding configurations do not affect the magnetic coupling. Scanning parameters for the STM images: $V = -0.7$ V, $I = 200$ pA (**a**) and $V = -0.6$ V, $I = 100$ pA (**b**).

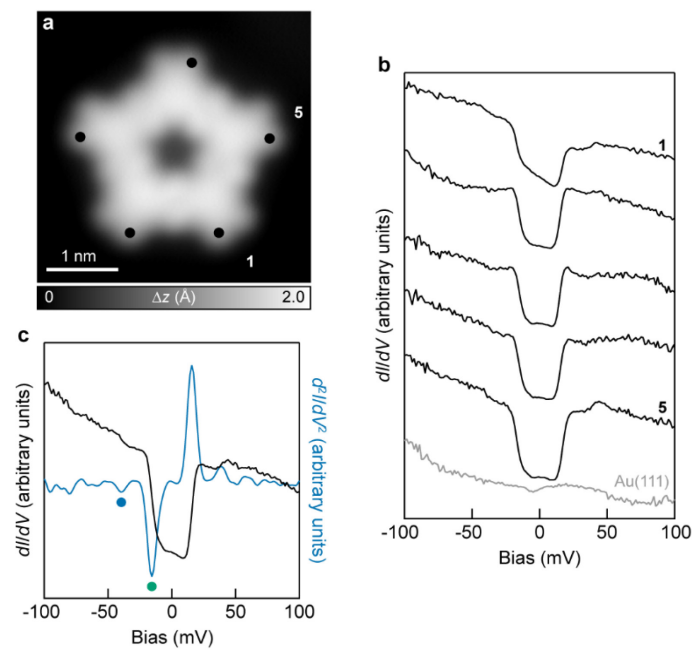

**Supplementary Fig. 19 | Low-bias *dI/dV* spectroscopy on an *N* = 5 cTSC. a,** High-resolution STM image of an *N* = 5 cTSC ($V = -0.7$ V, $I = 300$ pA). **b,** *dI/dV* spectroscopy on every unit of the TSC (open feedback parameters: $V = -100$ mV, $I = 1.4$ nA; $V_{rms} = 1$ mV). Acquisition positions are marked with a filled circle in **a**. **c,** Averaged *dI/dV* spectrum of all five units of the TSC shown in **b**, and the

corresponding $d^2I/dV^2$ spectrum obtained from numerical differentiation. The filled circles denote spin excitations at 16 and 40 mV.

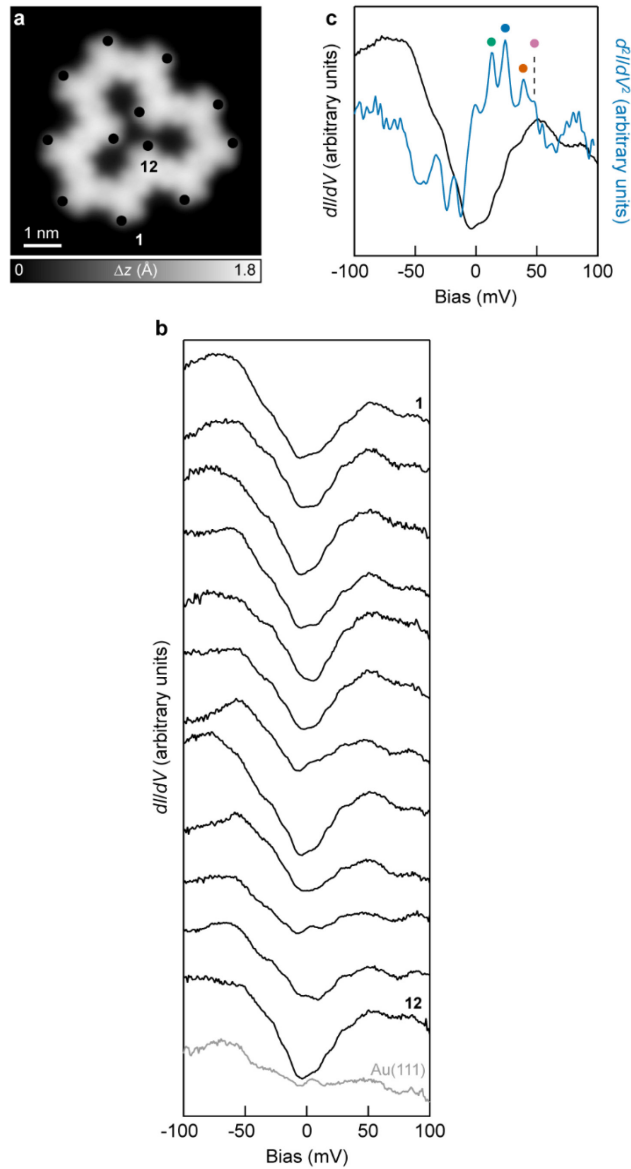

**Supplementary Fig. 20 | Low-bias $dI/dV$ spectroscopy on an $N$ = 12 cTSC. a**, High-resolution STM image of an $N$ = 12 cTSC ($V$ = −0.6 V, $I$ = 100 pA). **b**, $dI/dV$ spectroscopy on every unit of the TSC (open feedback parameters: $V$ = −100 mV, $I$ = 1.4 nA; $V_{rms}$ = 1 mV). Acquisition positions are marked with a filled circle in **a**. **c**, Averaged $dI/dV$ spectrum of all twelve units of the TSC shown in **b**, and the corresponding $d^2I/dV^2$ spectrum obtained from numerical differentiation. The filled circles denote spin excitations at 13, 24, 39 and 47 mV.

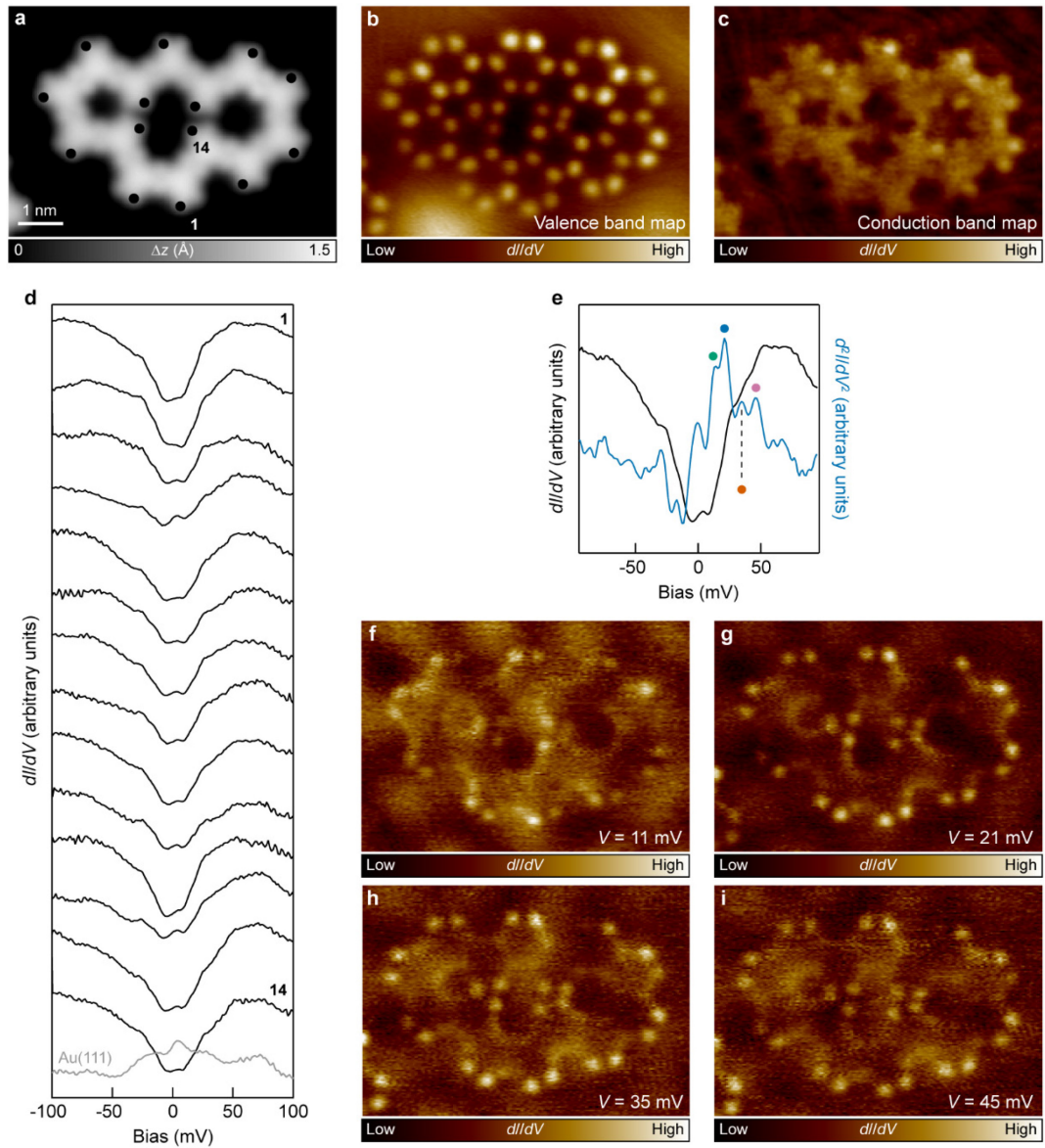

**Supplementary Fig. 21 | Low-bias *dI/dV* spectroscopy on an *N* = 14 cTSC. a–c**, High-resolution STM image (*V* = −0.4 V, *I* = 350 pA) (**a**), and constant-current *dI/dV* maps of the valence (*V* = −0.4 V, *I* = 350 pA; *V*_rms = 30 mV) (**b**) and conduction (*V* = 1.1 V, *I* = 400 pA; *V*_rms = 30 mV) (**c**) bands of an *N* = 14 cTSC. **d**, *dI/dV* spectroscopy on every unit of the TSC (open feedback parameters: *V* = −100 mV, *I* = 1.4 nA; *V*_rms = 1 mV). Acquisition positions are marked with filled circles in **a**. **e**, Averaged *dI/dV* spectrum of all fourteen units of the TSC shown in **d**, and the corresponding *d²I/dV²* spectrum obtained from numerical differentiation. The filled circles denote spin excitations at 13, 21, 34 and 45 mV. **f–i**, Constant-current *dI/dV* maps acquired near the spin excitations shown in **e**, at biases indicated in the corresponding figures (*I* = 250 pA (**f**), 280 pA (**g**), 300 pA (**h**) and 320 pA (**i**); *V*_rms = 2 mV). Spin excitation signals are uniformly distributed over all fourteen units of the TSC.

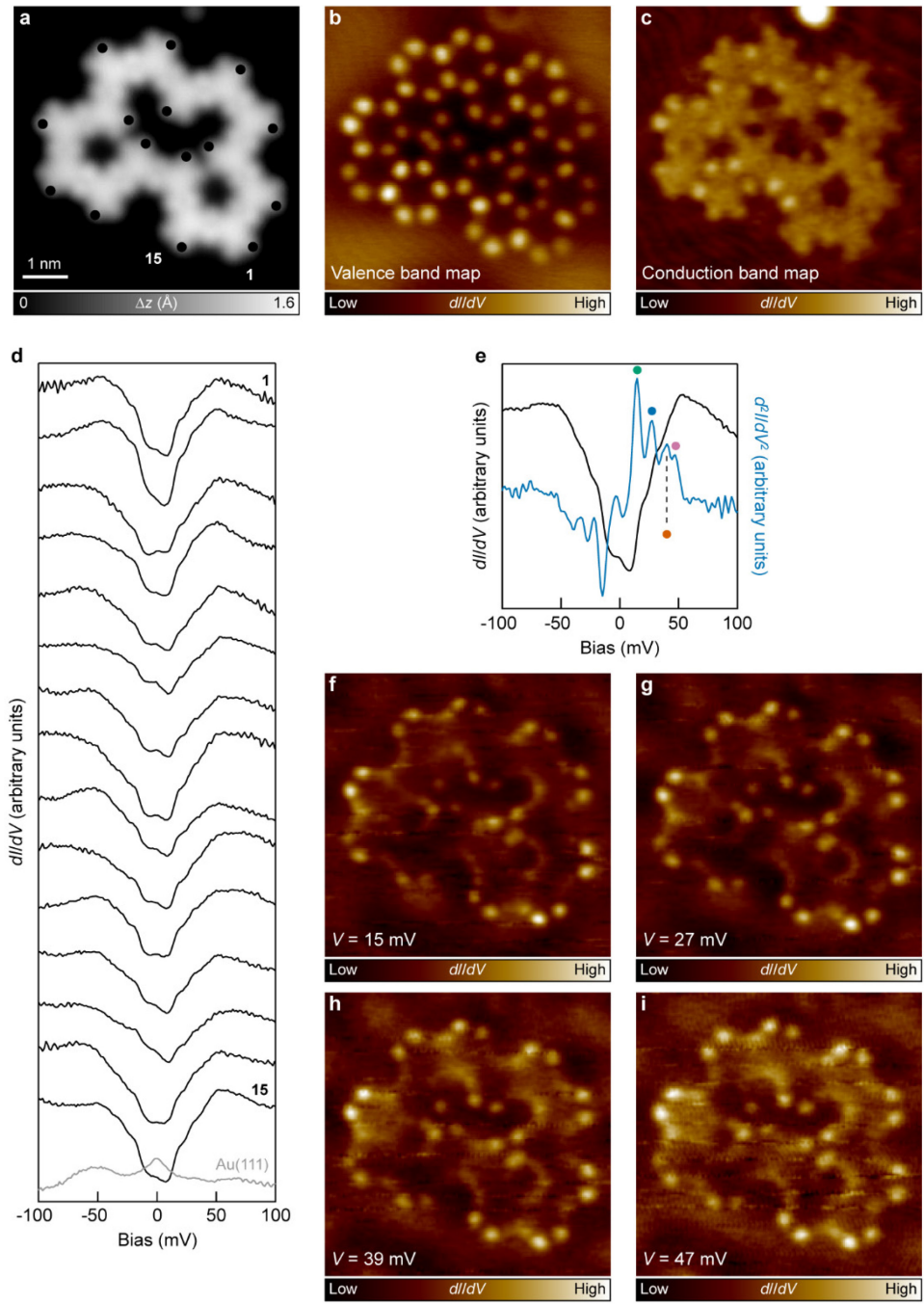

**Supplementary Fig. 22 | Low-bias $dI/dV$ spectroscopy on an $N = 15$ cTSC. a–c**, High-resolution STM image ($V = -0.7$ V, $I = 200$ pA) (**a**), and constant-current $dI/dV$ maps of the valence ($V = -0.4$ V, $I = 400$ pA; $V_{rms} = 32$ mV) (**b**) and conduction ($V = 1.1$ V, $I = 470$ pA; $V_{rms} = 32$ mV) (**c**) bands of an $N = 15$ cTSC. **d**, $dI/dV$ spectroscopy on every unit of the TSC (open feedback parameters: $V = -100$ mV, $I = 1.4$ nA; $V_{rms} = 1$ mV). Acquisition positions are marked with filled circles in **a**. **e**, Averaged $dI/dV$ spectrum of all fifteen units of the TSC shown in **d**, and the corresponding $d^2I/dV^2$ spectrum obtained from numerical differentiation. The filled circles denote spin excitations at 15, 27, 40 and 47 mV. **f–i**, Constant-current $dI/dV$ maps acquired near the spin excitations shown in **e**, at biases indicated in the corresponding figures ($I = 280$ pA (**f**), 320 pA (**g**) and 350 pA (**h, i**); $V_{rms} = 2$ mV). Spin excitation signals are uniformly distributed over all fifteen units of the TSC.

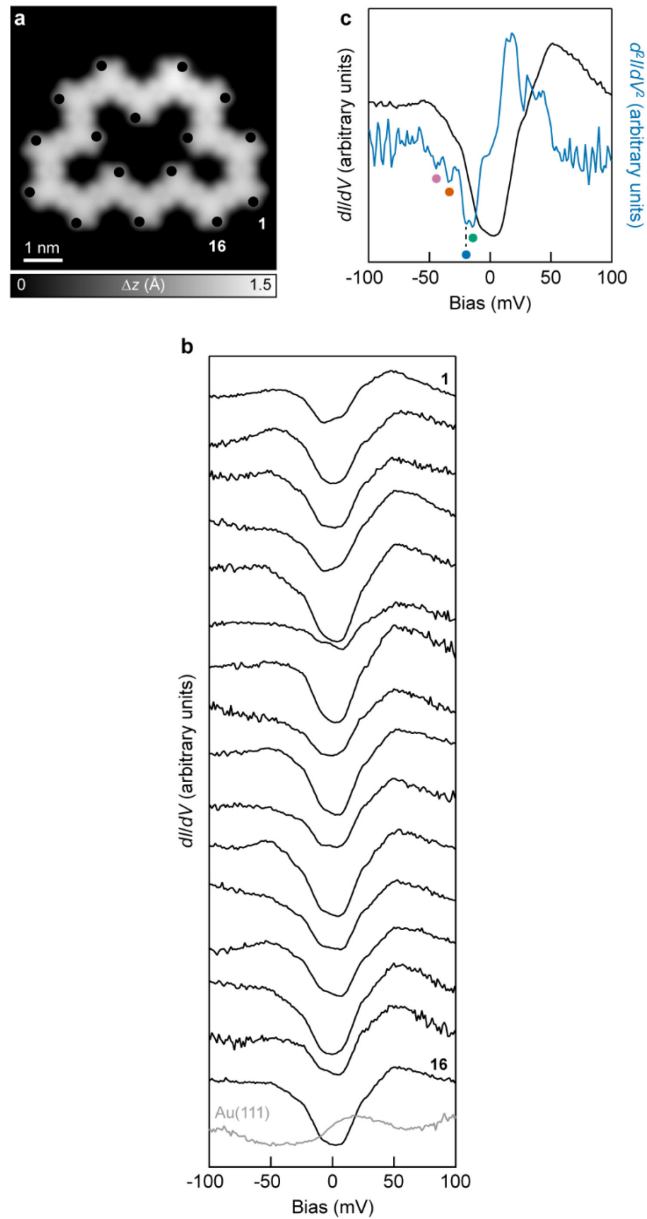

**Supplementary Fig. 23 | Low-bias $dI/dV$ spectroscopy on an $N$ = 16 cTSC. a**, High-resolution STM image of an $N$ = 16 cTSC ($V$ = −0.7 V, $I$ = 500 pA). **b**, $dI/dV$ spectroscopy on every unit of the TSC (open feedback parameters: $V$ = −100 mV, $I$ = 1.4 nA; $V_{rms}$ = 1 mV). Acquisition positions are marked with a filled circle in **a**. **c**, Averaged $dI/dV$ spectrum of all sixteen units of the TSC shown in **b**, and the corresponding $d^2I/dV^2$ spectrum obtained from numerical differentiation. The filled circles denote spin excitations at 14, 19, 33 and 44 mV.

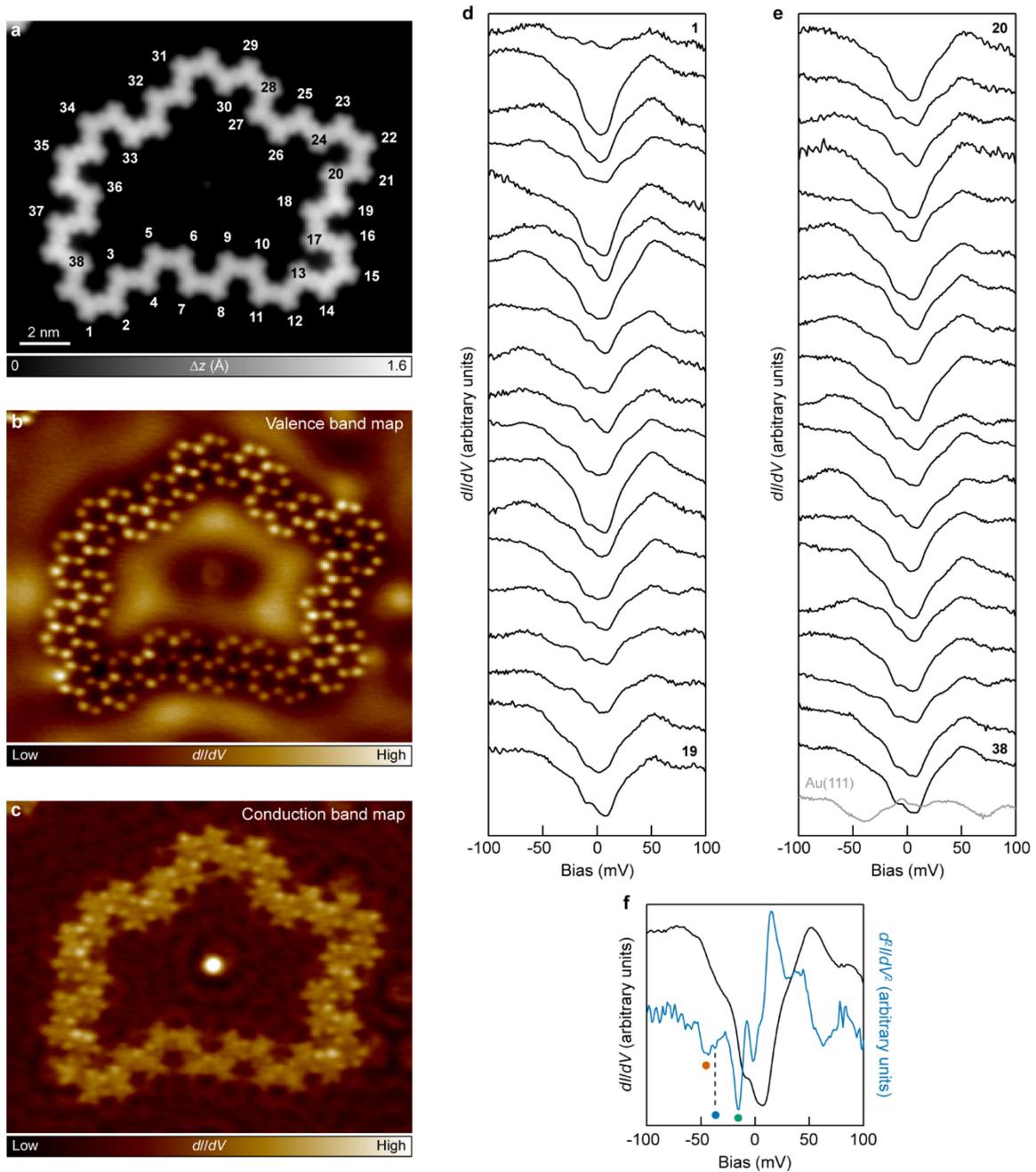

**Supplementary Fig. 24 | Low-bias $dI/dV$ spectroscopy on an $N = 47$ cTSC. a–c**, High-resolution STM image ($V = -0.4$ V, $I = 400$ pA) (**a**), and constant-current $dI/dV$ maps of the valence ($V = -0.4$ V, $I = 400$ pA; $V_{rms} = 32$ mV) (**b**) and conduction ($V = 1.1$ V, $I = 500$ pA; $V_{rms} = 32$ mV) (**c**) bands of an $N = 47$ cTSC. **d,e**, $dI/dV$ spectroscopy on thirty-eight units of the TSC (open feedback parameters: $V = -100$ mV, $I = 1.4$ nA; $V_{rms} = 1$ mV). **f**, Averaged $dI/dV$ spectrum of thirty-eight units of the TSC shown in **d** and **e**, and the corresponding $d^2I/dV^2$ spectrum obtained from numerical differentiation. The filled circles denote spin excitations at 14, 36 and 44 mV.

## 3. General methods and materials in solution synthesis and characterization

Unless otherwise noted, commercially available starting materials, reagents, catalysts, and dry solvents were used without further purification. Reactions were performed using standard vacuum-line and Schlenk techniques. All starting materials were obtained from TCI, Sigma Aldrich, abcr, Alfa Aesar, Acros Organics, or Fluorochem. Catalysts were purchased from Strem. Column chromatography was performed on silica ($SiO_2$, particle size $0.063-0.200$ mm, purchased from VWR). Silica-coated aluminum sheets with a fluorescence indicator (TLC silica gel 60 F254, purchased from Merck KGaA) were used for thin layer chromatography.

NMR data were recorded on a Bruker AV-II 300 spectrometer operating at 300 MHz for [1]H and 75 MHz for [13]C with standard Bruker pulse programs at room temperature (296 K). Chemical shifts ($\delta$) are reported in ppm. Coupling constants ($J$) are reported in Hz. Dichloromethane-$d_2$ ($\delta$([1]H) = 5.32 ppm, $\delta$([13]C) = 53.8 ppm) was used as solvent. The following abbreviations are used to describe peak patterns as appropriate: s = singlet, d = doublet, t = triplet, q = quartet, and m = multiplet. Dichloromethane-$d_2$ (99.9 atom % D) was purchased from Eurisotop.

High-resolution electrospray ionization mass spectrometry (HR-ESI-MS) was recorded with an Agilent 6538 Ultra High Definition (UHD) Accurate-Mass Q-TOF LC/MC system. Matrix-assisted laser desorption/ionization time-of-flight (MALDI-TOF) MS was recorded on a Bruker Autoflex Speed MALDI-TOF MS (Bruker Daltonics, Bremen, Germany). All of the samples, were prepared by mixing the analyte and the matrix, 1,8-dihydroxyanthracen-9(10H)-one (dithranol, purchased from Fluka Analytical, purity > 98%) or trans-2-[3-(4-tert-butylphenyl)-2-methyl-2-propenylidene]malononitrile (DCTB, purchased from Sigma Aldrich, purity > 99%) in the solid state.

## 4. Solution synthesis of **1** and **2**

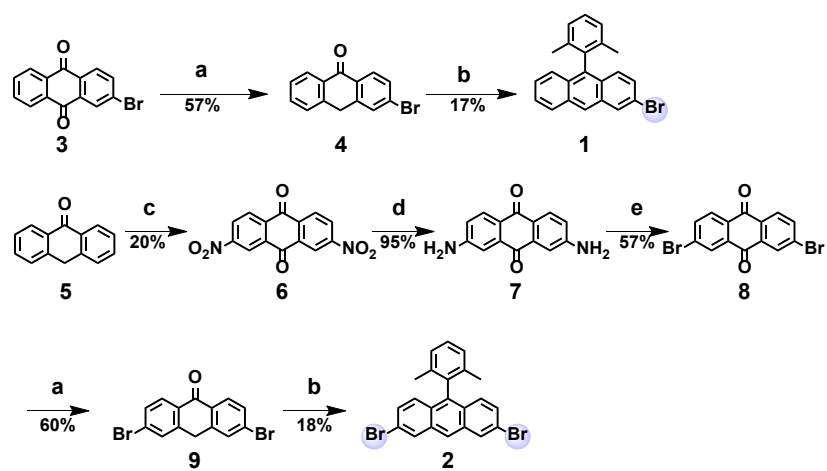

**Supplementary Fig. 25 | Synthetic route towards 1 and 2.** Reagents and conditions: **a**, Al, H$_2$SO$_4$, 30 ºC, overnight; **b**, (i) 2,6-dimethylphenylmagnesium bromide, toluene, 60 ºC, 1 h; (ii) conc. HCl. **c**, (i) Fuming HNO$_3$, 5 ºC, 1.5 h; (ii) CH$_3$COOH, rt, 1 week; **d**, Na$_2$S, NaOH, EtOH/H$_2$O, reflux, 6 h; **e**, CuBr$_2$, t-BuNO$_2$, 65 ºC, 2 h, then aq. HCl.

Precursors **1** and **2** were synthesized as shown in Supplementary Fig. 25. The reduction of commercially available 2-bromoanthracene-9,10-dione (**3**) gave 3-bromo-9,10-dihydro-9-oxoanthracene (**4**) in 57% yield. Subsequently, **4** was treated with 2,6-dimethylphenylmagnesium bromide, followed by dehydroxylation with concentrated HCl, to afford 2-bromo-10-(2,6-dimethylphenyl)anthracene (**1**) in 17% yield. **2** was synthesized in five steps from anthrone (**5**). 2,7-dinitro-9,10-anthraquinone (**6**) was initially prepared from **5** by nitration in 20% yield, which was converted into 2,7-diamino-9,10-anthraquinone (**7**) with a yield of 95% by Zinin reduction. Afterwards, diazotization and bromination of **7** afforded 2,7-dibromo-9,10-anthraquinone (**8**) in 57% yield. Following a similar reduction, nucleophilic addition and dehydroxylation reactions as for **3**, 2,7-dibromo-10-(2,6-dimethylphenyl)anthracene (**2**) was obtained from **8** in 6% yield.

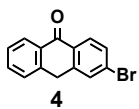

**Supplementary Fig. 26 |** Chemical structure of **4**.

**3-bromo-9,10-dihydro-9-oxoanthracene (4)**: The synthesis of **4** was performed in accordance with literature protocol.[6] Commercially available **3** (1.0 g, 3.48 mmol) was dissolved in 50 mL of hot concentrated H$_2$SO$_4$. After cooling to 30 ºC, Al powder (0.56 g, 20.9 mmol, and pretreatment with dilute HCl) was added within 10 min. The mixture was stirred at 30 ºC overnight. The color changed from orange to black and green and finally to bright yellow. After the reaction was complete and cooled to room temperature, the mixture was poured on ice. The precipitate and aqueous phase was extracted three times with CH$_2$Cl$_2$, and the combined organic layer was washed with brine and dried over magnesium sulfate (MgSO$_4$). The crude product **4** was then purified by silica gel chromatography using CH$_2$Cl$_2$/iso-hexane (1:3, v/v) as eluent to afford **4** as a pale-yellow solid (0.54 g, 57%).

¹H NMR (300 MHz, CD₂Cl₂): δ 8.29 (dd, *J* = 7.8, 1.5 Hz, 1H), 8.17 (d, *J* = 8.4 Hz, 1H), 7.72−7.57 (m, 3H), 7.54−7.44 (m, 2H), 4.35 (s, 2H). ¹³C NMR (75 MHz, CD₂Cl₂): δ 183.74, 142.94, 140.64, 133.54, 132.27, 132.01, 131.53, 131.00, 129.66, 129.16, 128.31, 127.82, 127.73, 32.51. HR-ESI-MS (negative mode): calc. for [M−H]⁻: 270.9764, found for [M−H]⁻: 270.9761 (deviation: −1.1 ppm).

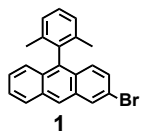

**Supplementary Fig. 27** | Chemical structure of **1**.

**2-bromo-10-(2,6-dimethylphenyl)anthracene (1)**: To a solution of **4** (250 mg, 0.91 mmol) in an-hydrous toluene (20 ml) was added 2,6-dimethylphenylmagnesium bromide solution (2.29 mL, 1 M in THF, 2.29 mmol) at room temperature under Ar. Then, the reaction mixture was heated to 60 °C. After the reaction mixture was stirred at 60 °C for 1 h, conc. HCl (2.0 mL) was added dropwise. The aqueous phase was extracted three times with CH₂Cl₂, and the combined organic layer was washed with brine and dried over magnesium sulfate (MgSO₄). The crude reaction product was then purified by silica gel chromatography using iso-hexane as eluent to afford **1** as a pale-yellow solid (57 mg, 17%). Further pu-rification by rGPC (eluent: chloroform) and recrystallization from dichloromethane/methanol affords white solid **1** as the precursor for on-surface synthesis.

¹H NMR (300 MHz, CD₂Cl₂) δ 8.43 (s, 1H), 8.26 (d, *J* = 1.9 Hz, 1H), 8.08 (d, *J* = 8.5 Hz, 1H), 7.26−7.54 (m, 8H), 1.72 (s, 6H). ¹³C NMR (75 MHz, CD₂Cl₂) δ 138.15, 137.42, 136.84, 132.90, 132.83, 130.79, 130.32, 129.59, 129.14, 128.51, 128.37, 128.34, 128.12, 126.75, 126.52, 126.39, 125.90, 119.99, 20.25. HR-ESI-MS (positive mode): calc. for [M]⁺: 360.0508, found for [M]⁺:360.04990 (deviation: −2.4 ppm).

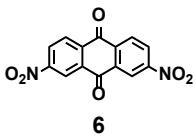

**Supplementary Fig. 28** | Chemical structure of **6**.

**2,7-dinitro-9,10-anthraquinone (6)**: The synthesis of **6** was performed in accordance with litera-ture protocol.[7] Commercially available anthrone **5** (5 g, 26.0 mmol) was added slowly with stirring to fuming nitric acid (40 mL) at 5 °C. Subsequently, 100 mL of glacial acetic acid was added to the reaction mixture with cooling. The resulting solution was stirred at room temperature for 1 week. Then, the yel-low precipitate was collected by filtration, washed with acetic acid (3 × 10 mL) and hexane (3 × 10 mL), and dried under vacuum (1.54 g, 20%). The crude product was used directly for the next step. Recrystal-lization from nitrobenzene/glacial acetic acid (1:1, v/v) afforded a pure sample of **6** and used for struc-tural characterization.

¹H NMR (300 MHz, DMSO-*d₆*): δ 8.84 (d, *J* = 2.4 Hz, 2H), 8.69 (dd, *J* = 8.5 Hz, *J* = 2.4 Hz, 2H), 8.48 (d, *J* = 8.5 Hz, 2H). The spectroscopic data are consistent with those described in literature.[7]

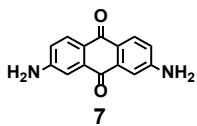

**Supplementary Fig. 29** | Chemical structure of **7**.

**2,7-diamino-9,10-anthraquinone (7)**: The synthesis of compound **7** was performed in accordance with literature protocol.[7] An aqueous solution (110 mL) of $Na_2S$ (2.39 g, 30.7 mmol) and NaOH (2.95 g, 73.6 mmol) was added to a stirred suspension of **6** (1.83 g, 6.1 mmol) in ethanol (60 mL). The mixture was refluxed for 6 h and allowed to stand overnight. Ethanol was removed by rotary evaporation. The resulting red precipitate was collected and washed with water and ethanol, dried under vacuum (1.38 g, 95%). The crude product was used directly for the next step. Recrystallization from ethanol/water (3:1, v/v) afforded a pure red solid of **7**.

[1]H NMR (300 MHz, DMSO-$d_6$): $\delta$ 7.84 (d, $J$ = 8.5 Hz, 2H), 7.23 (d, $J$ = 2.3 Hz, 2H), 6.89 (dd, $J$ = 8.5 Hz, $J$ = 2.4 Hz, 2H), 6.39 (s, 4H). [13]C NMR (75 MHz, DMSO-$d_6$): $\delta$ 184.28, 179.30, 153.70, 134.88, 128.92, 121.95, 117.97, 109.62. The spectroscopic data are consistent with those described in literature.[7]

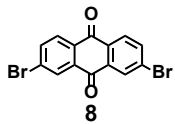

**Supplementary Fig. 30** | Chemical structure of **8**.

**2,7-dibromo-9,10-anthraquinone (8)**: The synthesis of **8** was performed in accordance with literature protocol.[7] To a three-necked round-bottom flask, anhydrous copper (II) bromide (2.12 g, 9.5 mmol), *tert*-butyl nitrite (1.3 mL, 10.9 mmol), and anhydrous acetonitrile (50 mL) were added, and the mixture was heated to 65 °C. **7** (0.86 g, 3.6 mmol) was added slowly over a period of 5 min to the reaction mixture. Nitrogen was evolving during the reaction. After nitrogen evolution had subsided, the reaction mixture was cooled to room temperature and poured into an aqueous HCl solution (100 mL, 20% w/v). The aqueous phase was extracted three times with $CH_2Cl_2$, and the combined organic layer was washed with brine and dried over magnesium sulfate ($MgSO_4$). The crude product was then purified by silica gel chromatography using $CH_2Cl_2$/iso-hexane (1:3, v/v) as eluent to afford **8** as a pale-yellow solid (0.75 g, 57%).

[1]H NMR (300 MHz, CDCl$_3$): $\delta$ 8.43 (d, $J$ = 2.0 Hz, 2H), 8.17 (d, $J$ = 8.3 Hz, 2H), 7.94 (dd, $J$ = 8.3 Hz, $J$ = 2.0 Hz, 2H). [13]C NMR (75 MHz, CDCl$_3$): $\delta$ 181.79, 181.11, 137.67, 134.33, 132.04, 130.53, 130.18, 129.27. The spectroscopic data are consistent with those described in literature.[7]

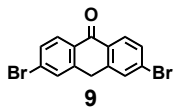

**Supplementary Fig. 31** | Chemical structure of **9**.

**3,6-dibromo-9,10-dihydro-9-oxoanthracene (9)**: The synthesis of **9** was performed in accordance with literature protocol.[6] **8** (0.74 g, 2.0 mmol) was dissolved in 40 mL of hot concentrated $H_2SO_4$. After cooling to 30 °C, Al powder (0.33 g, 12.0 mmol, pretreatment with dilute HCl) was added within 10 min. The mixture was stirred at 30 °C overnight. The color changed from orange to black and green and finally to bright yellow. After the reaction was complete and cooled to room temperature, the mixture was poured on ice. The precipitate and aqueous phase was extracted three times with CH$_2$Cl$_2$, and the combined organic layer was washed with brine and dried over magnesium sulfate (MgSO$_4$). The crude product was then purified by silica gel chromatography using CH$_2$Cl$_2$/iso-hexane (1:3, v/v) as eluent to afford **9** as a pale-yellow solid (0.42 g, 60%).

[1]H NMR (300 MHz, CDCl$_3$): $\delta$ 8.20 (d, $J$ = 8.4 Hz, 2H), 7.64 (s, 2H), 7.61 (dd, $J$ = 8.4 Hz, $J$ = 1.8 Hz, 2H), 4.31 (s, 2H). [13]C NMR (75 MHz, CDCl$_3$): $\delta$ 182.82, 141.51, 131.46, 131.01, 130.72, 129.55, 128.51, 31.74. The spectroscopic data are consistent with those described in literature.[6]

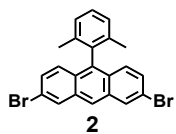

**Supplementary Fig. 32** | Chemical structure of **2**.

**2,7-dibromo-10-(2,6-dimethylphenyl)anthracene (2)**: To a solution of **9** (200 mg, 0.6 mmol) in anhydrous toluene (20 ml) was added 2,6-Dimethylphenylmagnesium bromide solution (2.84 mL, 1 M in THF, 2.8 mmol) at room temperature under Ar. Then, the reaction mixture was heated to 60 ºC. After the reaction mixture was stirred at 60 ºC for 1 h, conc. HCl (2.0 mL) was added dropwise. The aqueous phase was extracted three times with CH$_2$Cl$_2$, and the combined organic layer was washed with brine and dried over magnesium sulfate (MgSO$_4$). The crude reaction product was then purified by silica gel chromatography using iso-hexane as eluent to afford **2** as a pale-yellow solid (45 mg, 18%). Further purification by rGPC (eluent: chloroform) and recrystallization from dichloromethane/methanol affords **2** with higher purity as the precursor for on-surface synthesis.

[1]H NMR (300 MHz, CD$_2$Cl$_2$) $\delta$ 8.34 (s, 1H), 8.25 (d, $J$ = 1.8 Hz, 2H), 7.26−7.43 (m, 7H), 1.70 (s, 6H); [13]C NMR (75 MHz, CD$_2$Cl$_2$) $\delta$ 138.09, 137.55, 136.82, 133.50, 130.75, 130.05, 128.73, 128.58, 128.45, 128.20, 124.99, 120.81, 20.21. HR-MALDI-TOF (matrix: dithranol): calc. [M]$^+$: 437.9613, found for [M]$^+$: 437.9611 (deviation: 0.4 ppm).

## 5. Solution characterization data

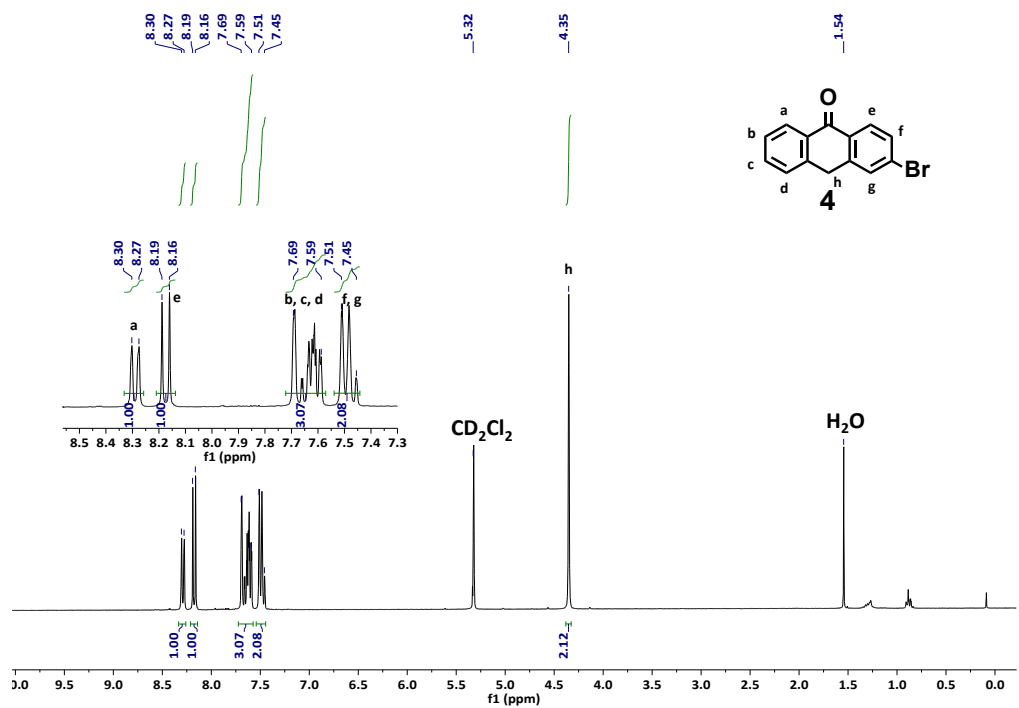

**Supplementary Fig. 33** | ¹H-NMR spectrum of **4** dissolved in CD₂Cl₂, 300 MHz, 296 K.

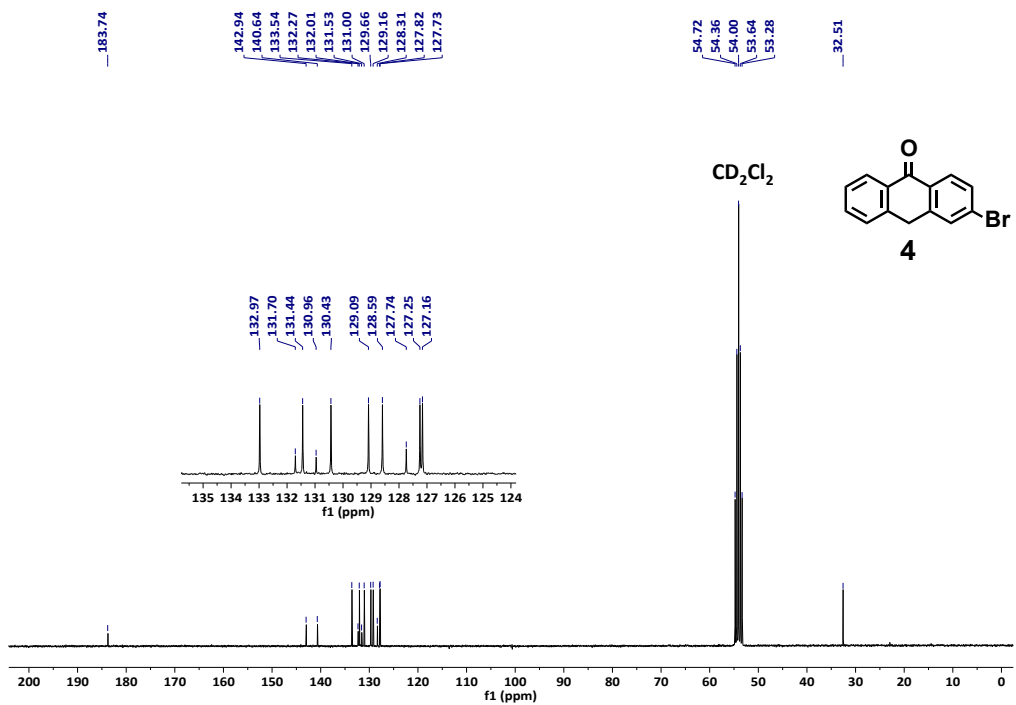

**Supplementary Fig. 34** | ¹³C-NMR spectrum of **4** dissolved in CD₂Cl₂, 75 MHz, 296 K.

**Qualitative Compound Report**

| Data File | F6017-10.d | Compound 4 | Sample Name | F6017 |
|---|---|---|---|---|
| Sample Type | Sample | | Position | Vial 33 |
| Instrument Name | Instrument 1 | | User Name | |
| Acq Method | ESI-neg-1-2-AcNi.m | | Acquired Time | 2/10/2021 8:35:17 PM |
| IRM Calibration Status | Success | | DA Method | default.m |
| Comment | C14H9BrO | | | |
| Sample Group | | | Info. | 271.9837 |
| Stream Name | LC 1 | | Acquisition SW Version | 6200 series TOF/6500 series Q-TOF B.06.01 (B6172 SP1) |

**Compound Table**

| Compound Label | RT | Mass | Abund | Formula | Tgt Mass | Diff (ppm) |
|---|---|---|---|---|---|---|
| Cpd 1: C14 H9 Br O | 1.171 | 271.9833 | 63564 | C14 H9 Br O | 271.98368 | -1.4 |

MS Zoomed Spectrum

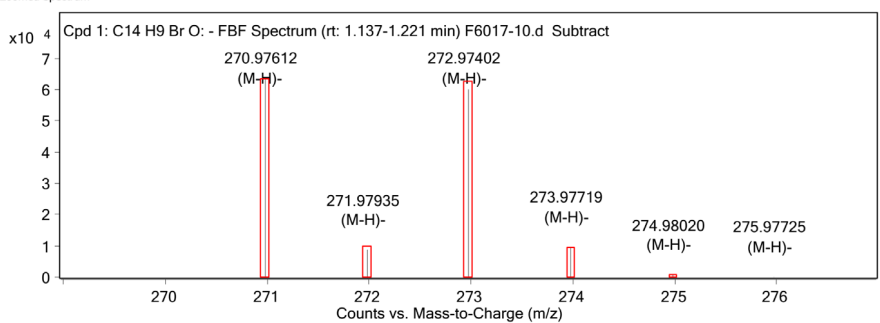

**MS Spectrum Peak List**

| m/z | z | Abund | Formula | Ion |
|---|---|---|---|---|
| 270.97612 | 1 | 63563.69 | C14H9BrO | (M-H)- |
| 271.97935 | 1 | 8694.76 | C14H9BrO | (M-H)- |
| 272.97402 | 1 | 60134.14 | C14H9BrO | (M-H)- |
| 273.97719 | 1 | 9623.55 | C14H9BrO | (M-H)- |
| 274.9802 | 1 | 673.23 | C14H9BrO | (M-H)- |
| 275.97725 | 1 | 51.68 | C14H9BrO | (M-H)- |

--- End Of Report ---

**Supplementary Fig. 35** | Qualitative compound report of HR-ESI-MS (negative mode) for **4**.

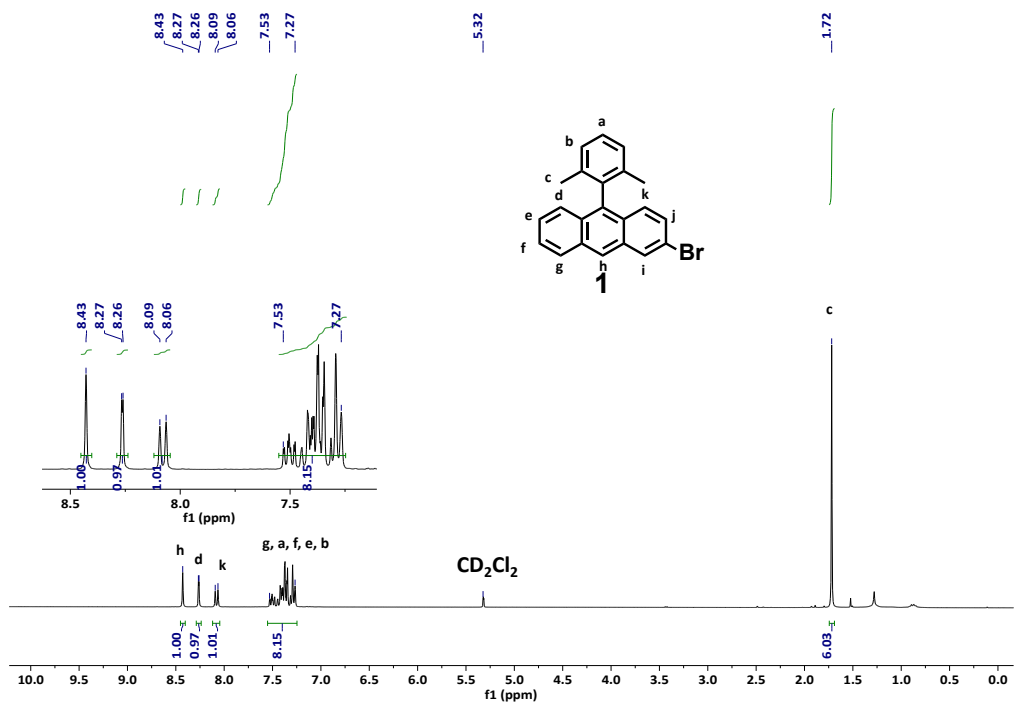

**Supplementary Fig. 36 |** ¹H-NMR spectrum of **1** dissolved in CD₂Cl₂, 300 MHz, 296 K.

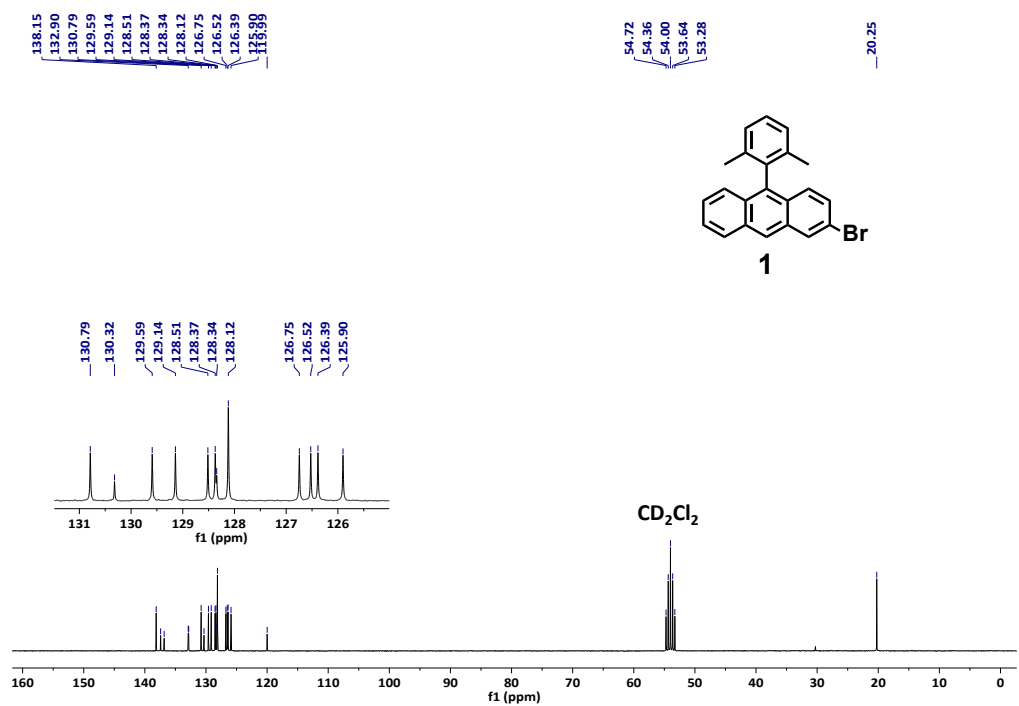

**Supplementary Fig. 37 |** ¹³C-NMR spectrum of **1** dissolved in CD₂Cl₂, 75 MHz, 296 K.

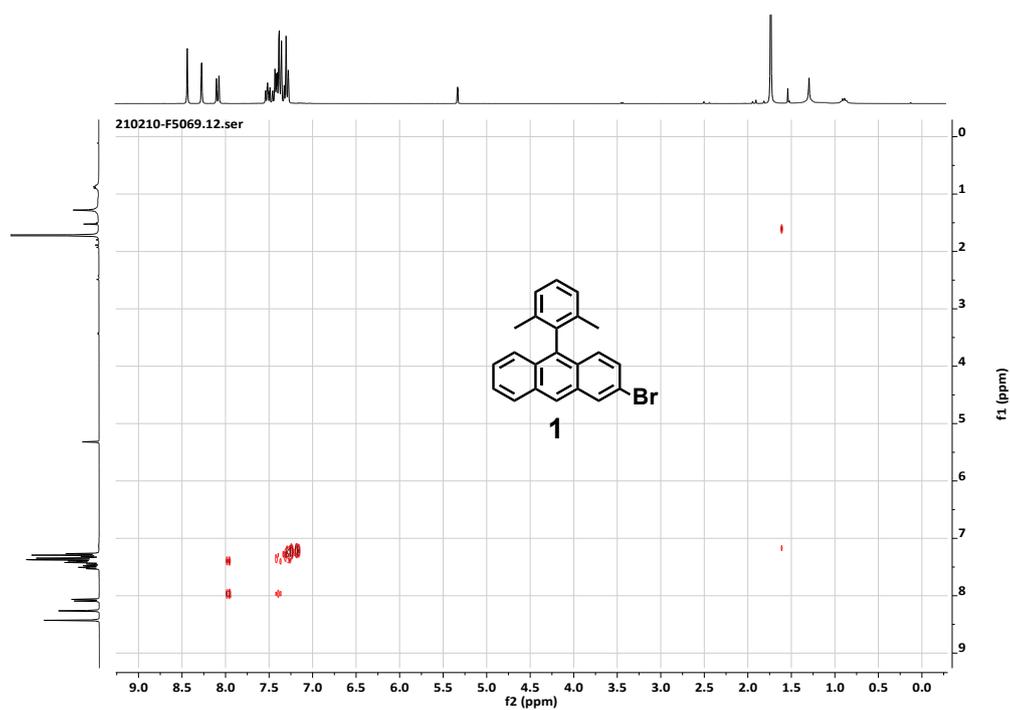

**Supplementary Fig. 38** | ¹H/¹H-COSY-NMR spectrum of **1** dissolved in dissolved in CD₂Cl₂, 300 MHz, 296 K.

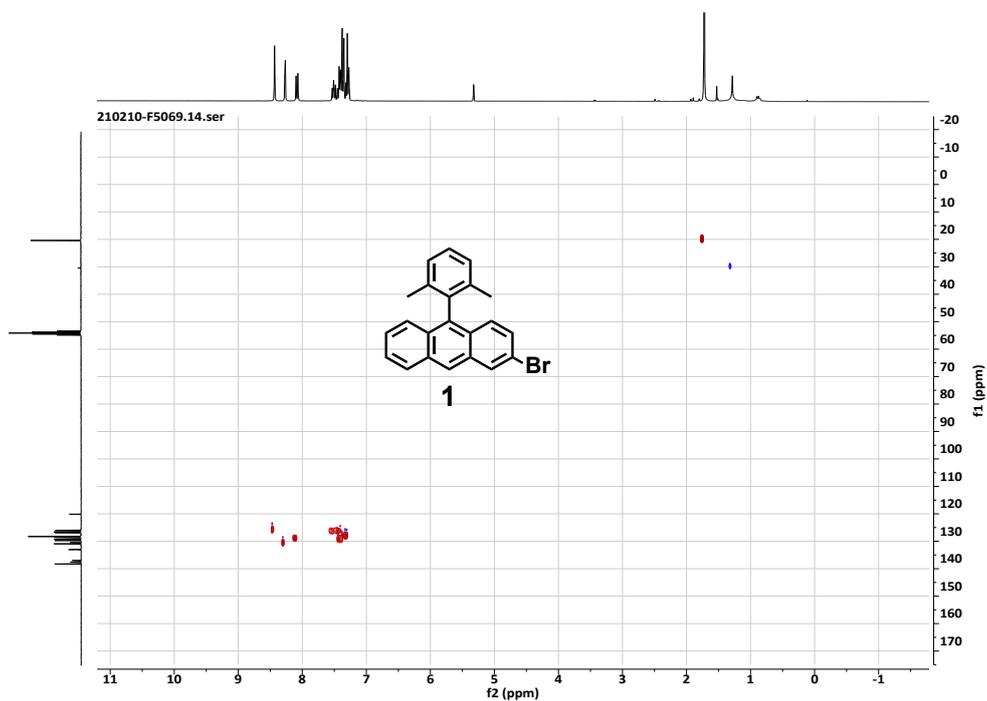

**Supplementary Fig. 39** | HSQC-NMR spectrum of **1** dissolved in CD₂Cl₂, 75 MHz, 296 K.

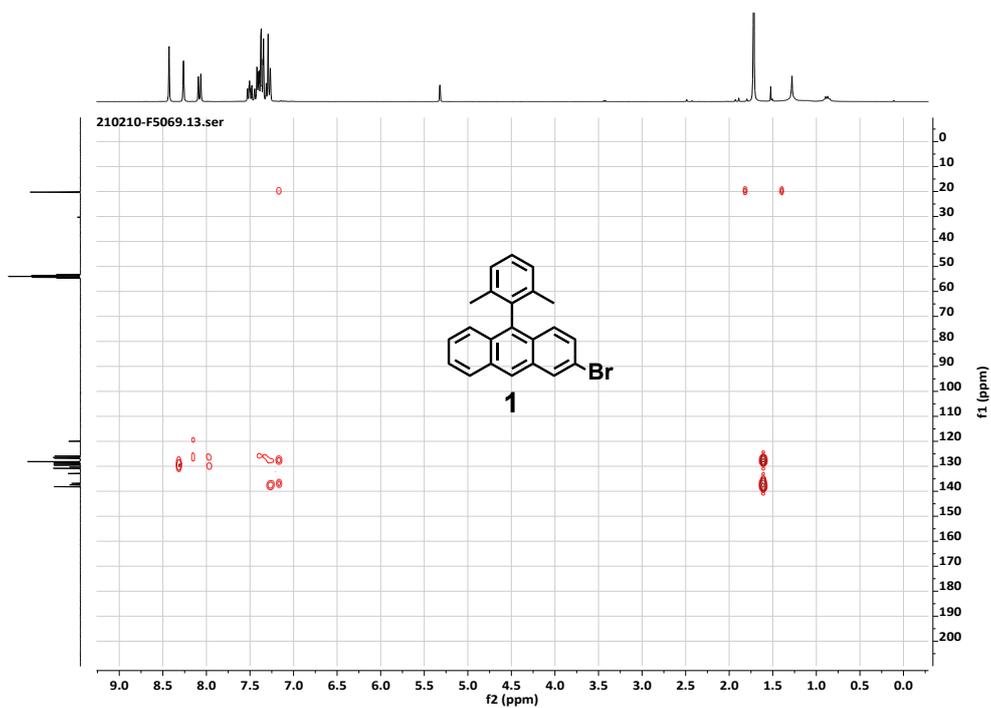

**Supplementary Fig. 40** | HMBC-NMR spectrum of **1** dissolved in CD₂Cl₂, 75 MHz, 296 K.

**Qualitative Compound Report**

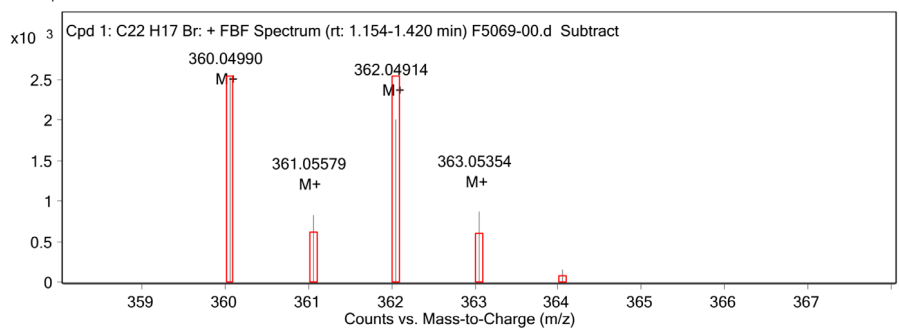

| Data File | F5069-00.d | | Sample Name | F5069 |
|---|---|---|---|---|
| Sample Type | Sample | | Position | Vial 34 |
| Instrument Name | Instrument 1 | | User Name | |
| Acq Method | ESI-pos-1-2-AcNi.m | | Acquired Time | 2/10/2021 8:15:02 PM |
| IRM Calibration Status | Success | | DA Method | default.m |
| Comment | C22H17Br | | | |
| Sample Group | | | Info. | 362.0496 |
| Stream Name | LC 1 | | Acquisition SW Version | 6200 series TOF/6500 series Q-TOF B.06.01 (B6172 SP1) |

Precursor 1   Precursor 1

**Compound Table**

| Compound Label | RT | Mass | Abund | Formula | Tgt Mass | Diff (ppm) |
|---|---|---|---|---|---|---|
| Cpd 1: C22 H17 Br | 1.171 | 360.05149 | 2522 | C22 H17 Br | 360.05136 | 0.35 |

MS Zoomed Spectrum

Cpd 1: C22 H17 Br: + FBF Spectrum (rt: 1.154-1.420 min) F5069-00.d  Subtract

360.04990 M+, 361.05579 M+, 362.04914 M+, 363.05354 M+

**MS Spectrum Peak List**

| m/z | z | Abund | Formula | Ion |
|---|---|---|---|---|
| 360.0499 | 1 | 2521.61 | C22H17Br | M+ |
| 361.05579 | 1 | 825.41 | C22H17Br | M+ |
| 362.04914 | 1 | 1998.44 | C22H17Br | M+ |
| 363.05354 | 1 | 863.15 | C22H17Br | M+ |
| 364.05805 | 1 | 153.97 | C22H17Br | M+ |

--- End Of Report ---

Agilent Technologies

**Supplementary Fig. 41** | Qualitative compound report of HR-ESI-MS (positive mode) for **1**.

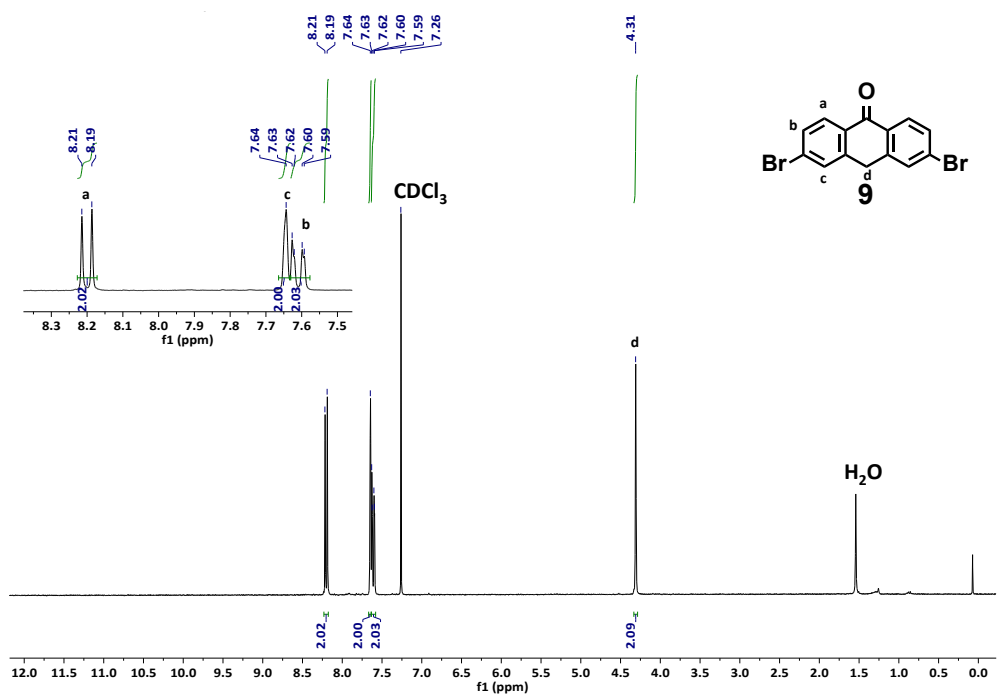

**Supplementary Fig. 42 |** $^{1}$H-NMR spectrum of **9** dissolved in CDCl$_3$, 300 MHz, 296 K.

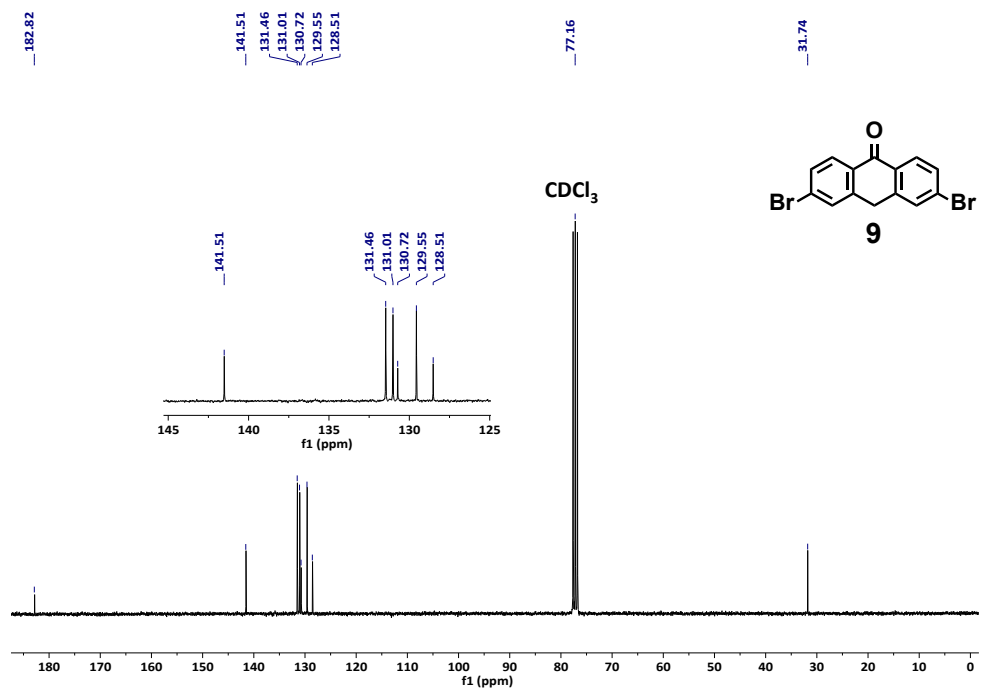

**Supplementary Fig. 43 |** $^{13}$C-NMR spectrum of **9** dissolved in CDCl$_3$, 75 MHz, 296 K.

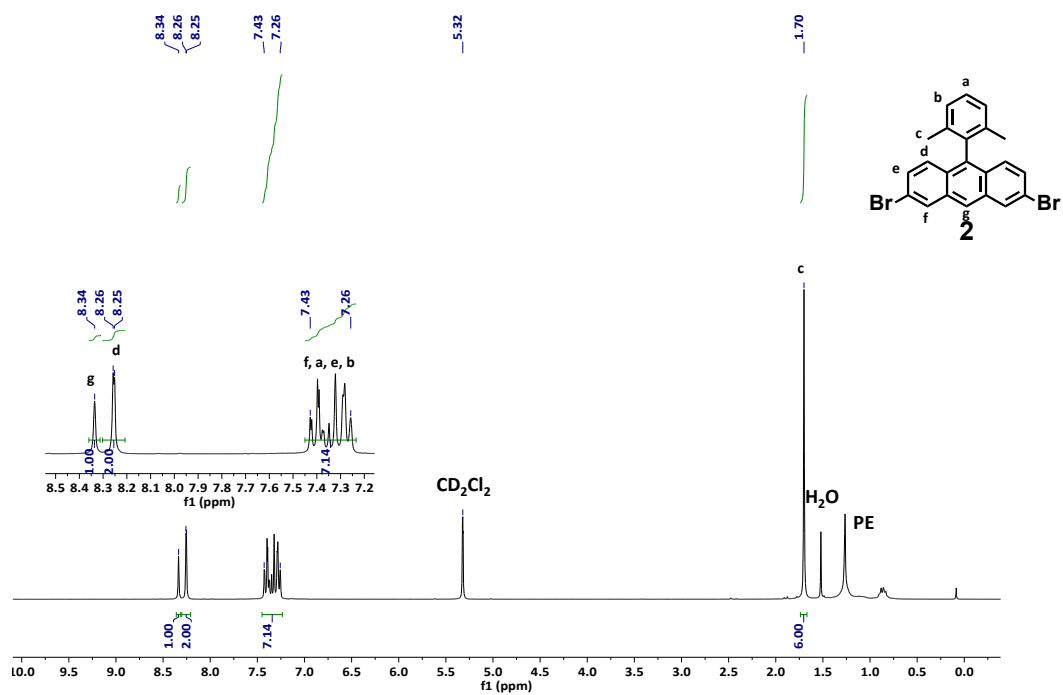

**Supplementary Fig. 44** | ¹H-NMR spectrum of **2** dissolved in CD₂Cl₂, 300 MHz, 296 K.

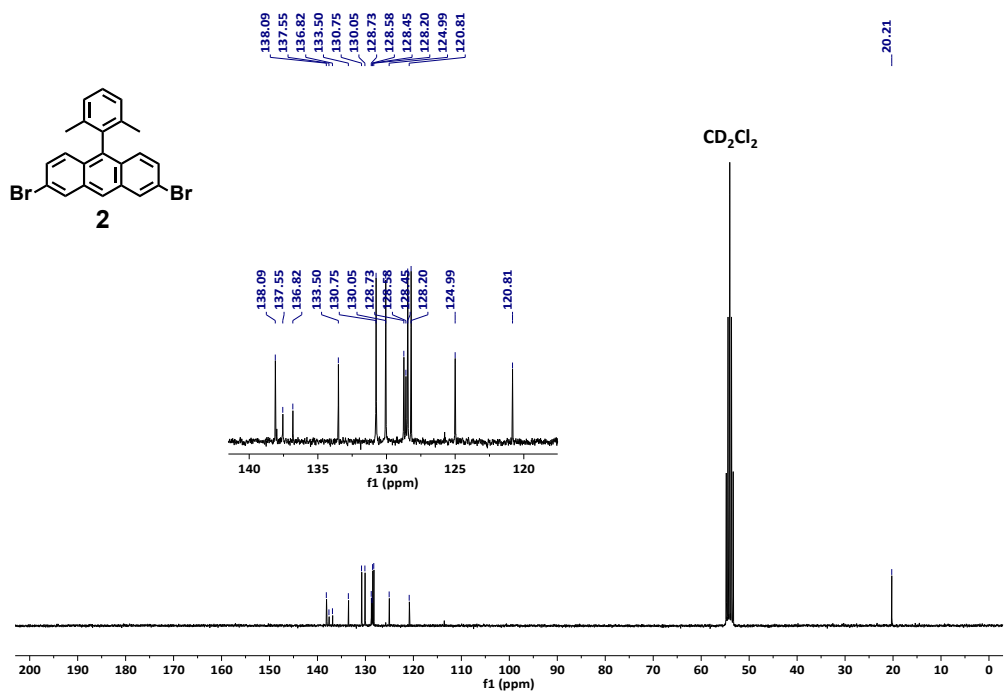

**Supplementary Fig. 45** | ¹³C-NMR spectrum of **2** dissolved in CD₂Cl₂, 75 MHz, 296 K.

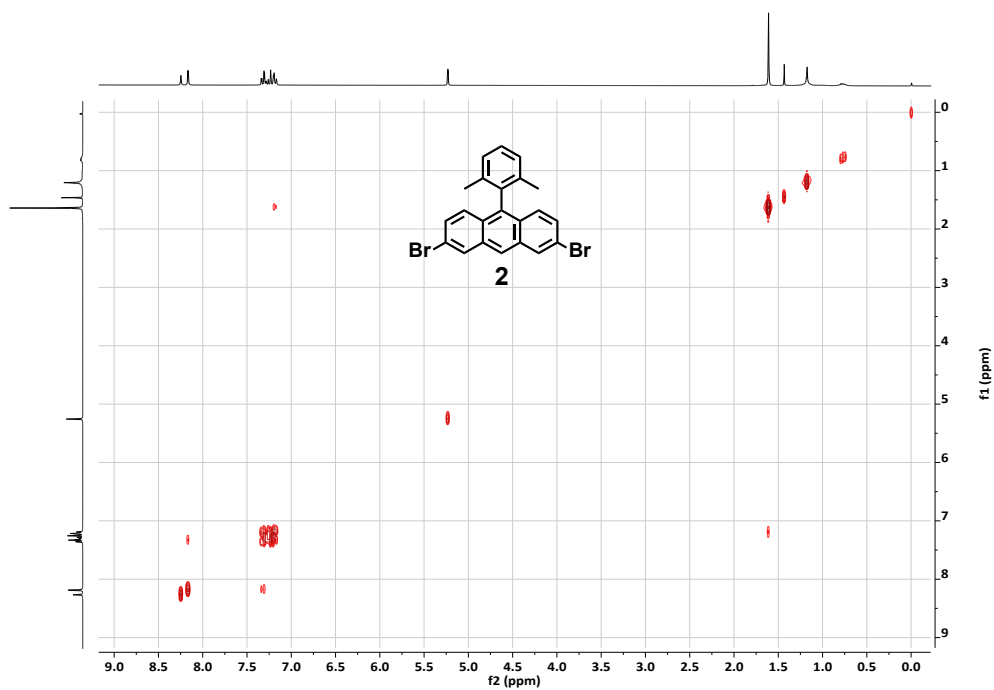

**Supplementary Fig. 46 |** ¹H/¹H-COSY-NMR spectrum of **2** dissolved in dissolved in CD₂Cl₂, 300 MHz, 296 K

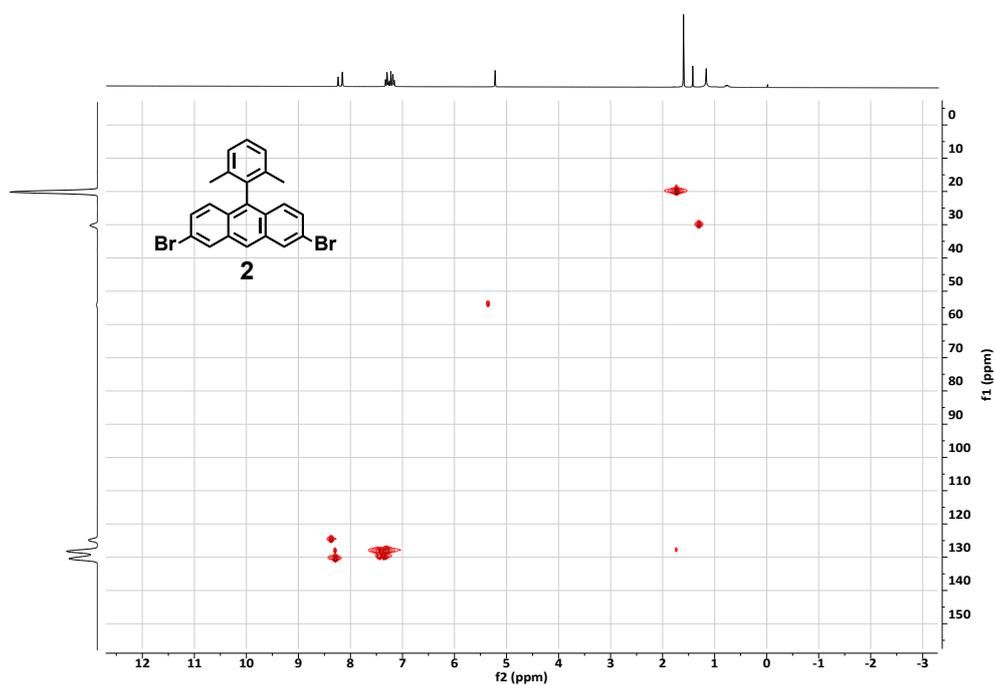

**Supplementary Fig. 47 |** HSQC-NMR spectrum of **2** dissolved in CD₂Cl₂, 75 MHz, 296 K

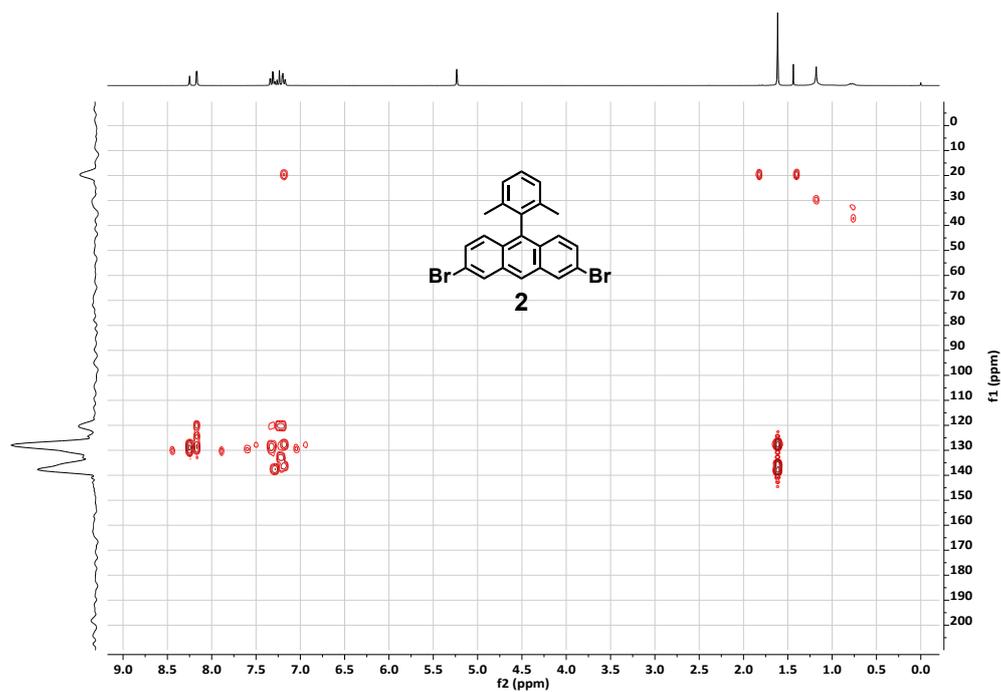

**Supplementary Fig. 48 |** HMBC-NMR spectrum of **2** dissolved in CD₂Cl₂, 75 MHz, 296 K.

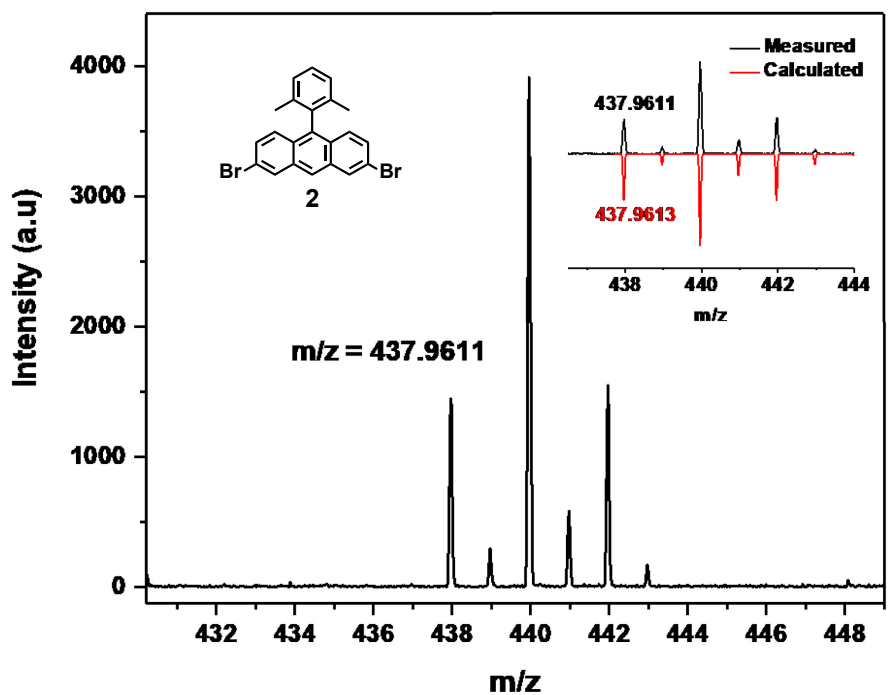

**Supplementary Fig. 49 |** Liquid-state HR-MALDI-TOF-MS of **2** (matrix: dithranol).

## 6. Supplementary references